\documentclass[twoside,12pt]{article}

\usepackage{graphicx}
\usepackage{epsfig}

\usepackage{amssymb}
\usepackage{dcolumn}  % Align table columns on decimal point
\usepackage{bm}       % bold math
\usepackage{amsmath}
\usepackage{sidecap}  % 

\newcommand{\bsigma}{\mbox{\boldmath$\sigma$}}
\newcommand{\btau}{\mbox{\boldmath$\tau$}}

\newcommand{\be}{\begin{equation}}
\newcommand{\ee}{\end{equation}}
\newcommand{\bea}{\begin{eqnarray}}
\newcommand{\eea}{\end{eqnarray}}
\newcommand{\nn}{\nonumber}
\newcommand{\half}{0.5}
\begin{document}

\title{The Skyrme Interaction in finite nuclei and nuclear matter}

\author{J. R. Stone$^{1,2,3}$ and P.--G. Reinhard$^{4,5}$\\
  $^1$
  Department of Physics, University Oxford,\\
  Oxford OX1 3PU / United Kingdom\\
  $2$
  Physics Division, Oak Ridge National Laboratory,\\
  P.O. Box 2008, Oak Ridge, TN 37831, USA\\
  $^3$
  Department of Chemistry and Biochemistry,\\
  University of Maryland, College Park, MD 20742, USA\\
  $^4$ 
  Institut fur Theoretische Physik, Universitat Erlangen,\\
  D-91054 Erlangen, Germany\\
  $^5$
  Joint Institute for Heavy Ion Research,
  Oak Ridge National Laboratory,\\
  P.\ O.\ Box 2008, Oak Ridge, TN 37831, USA\\
}
\date{}
%\author[JSaddress1,JSaddress2]{J. Stone}
%\ead{j.stone@physics.ox.ac.uk}
%\author[PGRaddress1,PGRaddress2]{P.--G. Reinhard}
%\ead{reinhard@theorie2.physik.uni-erlangen.de}
\maketitle

\begin{abstract}
Self-consistent mean-field models are a powerful tool in the
investigation of nuclear structure and low-energy dynamics.  They are
based on effective energy-density functionals, often formulated in
terms of effective density-dependent nucleon-nucleon interactions. The
free parameters of the functional are adjusted to empirical data. A
proper choice of these parameters requires a comprehensive set of
constraints covering experimental data on finite nuclei, concerning
static as well as dynamical properties, empirical characteristics of
nuclear matter, and observational information on nucleosynthesis,
neutron stars and supernovae. This work aims at a comprehensive survey
of the performance of one of the most successful non-relativistic
self-consistent method, the Skyrme-Hartree-Fock model (SHF), with
respect to these constraints. A full description of the Skyrme
functional is given and its relation to other effective interactions
is discussed. The validity of the application of SHF far from
stability and in dense environments beyond the nuclear saturation
density is critically assessed. The use of SHF in models extended
beyond the mean field approximation by including some correlations is
discussed. Finally, future prospects for further development of SHF
towards a more consistent application of the existing and promisingly
newly developing constraints are outlined.

\end{abstract}

% main text
%\section{Introduction}
%\label{sec:intro}
%\subsection{General overview of nuclear structure models}
%\label{sec:intrmf}
% this part --> PGR first, then JS
\section{Introduction}
\label{sec:intro}
%\subsection{Nuclear structure input to astrophysics}
%\label{sec:intrastro}
\subsection{General overview of nuclear structure models}
\label{sec:intrmf}

There is a great variety of nuclear structure models at very different
theoretical levels. At the most fundamental level are \textit{ab initio}
theories which start from a nucleon-nucleon force as input
\cite{Mac01aR} and compute the equation of state for nuclear matter by
diagrammatic techniques (for reviews see, e.g.,
\cite{Dic92aR,Hei00aR}) or from low-energy
quantum-chromodynamics (see e.g.  \cite{Lut00a,Kai02a,wei05,boh05}).  These
microscopic approaches have made considerable progress over the past
decades. One is even performing \textit{ab initio} calculations for finite
nuclei, e.g. in the no-core shell model \cite{Nav00a,for05,pie04},
coupled cluster calculations \cite{Dea05a}, or in the unitary
correlator method \cite{Rot05a}.  Nonetheless, the actual precision in
describing nuclear properties is still limited, typically 5\%
precision when \textit{ab initio} two-body forces are used\cite{Dea05a} (better
quality of about 1\% can be achieved when invoking effective
three-body forces). Moreover, the application to finite nuclei are
computationally very demanding, e.g. the largest no core shell model
calculations go at most up to A=20 \cite{Nav06a}. Thus fully
microscopic methods are presently not used for large-scale nuclear
structure calculations.

At the most phenomenological level are the macroscopic approaches
which are inspired by the idea that the nucleus is a drop of nuclear
liquid, giving rise to the liquid-drop model (LDM) or the more refined
droplet model, for a recent review see e.g. \cite{Mol97b}. With a mix
of intuition and systematic expansion in orders of surface effects (the leptodermous expansion) one can write down the corresponding
energy functional \cite{Mye82aR}.  There remain a good handful of free
parameters, i.e the coefficients for volume energy, symmetry
energy, incompressibility, or surface energy.  These have to be
adjusted to a multitude of nuclear bulk properties with the results that modern
droplet parameterizations deliver an excellent description of average
trends \cite{Nil95aB}.  Actual nuclei, however, deviate from the
average due to quantum shell effects, so that shell corrections have
to be added. They are related to the level density near the Fermi
surface and can be computed from a well tuned, empirical nuclear
single-particle potential. Macroscopic energy plus shell corrections
constitute the macroscopic-microscopic (mac-mic) approach which is
enormously successful in reproducing the systematics of known nuclear
binding energies \cite{Moe95aR}. One has to admit, though, that the
mac-mic method relies strongly on phenomenological input. This induces
uncertainties when extrapolating to exotic nuclei. Particularly
uncertain is the extrapolation of the single-particle potential used
in this model (Nilsson, Yukawa) because this is not determined
self-consistently but added as an independent piece of information.

An intermediate level of nuclear models consists in microscopic theories employing effective
interactions or effective energy-density functionals.  There exist two
basically different approaches. 
On the one hand, one has large scale shell model calculations
which aim at a fully correlated description of ground state and
excitations up to about 10 MeV, for a review see e.g. \cite{Koo97aR}.
These employ the wave functions of a single-particle basis from
empirical shell model potentials complemented by experimental
single-particle energies. The effective interaction in the valence
space (two-body matrix elements) is taken often from microscopic
information, namely the $G$-matrix from \textit{ab initio} calculations, with
a bit of final fine-tuning. Some calculations rely fully on a
phenomenological adjustment of the matrix elements.
On the other hand, there are the self-consistent mean-field models which aim
at a prejudice-free, self-consistent determination of the nuclear ground
state and low-energy collective dynamics. The present review con concentrates
on this latter class of models.

Self-consistent mean-field models, as compared to the mac-mic method,
take one big step towards a microscopic description of nuclei. They
produce the appropriate single-particle potential corresponding to the
actual density distribution for a given nucleus. Still, they cannot be
regarded as an \textit{ab initio} treatment because the genuine nuclear
interaction induces huge short-range correlations that are not
naturally included in a Hartree-Fock model. Self-consistent mean-field
models deal with effective energy functionals
\cite{Pet91aB,Rei92e}. The concept has much in common with the
successful density-functional theory for electronic systems
\cite{Dre90aB}.  The difference is, however, that electronic
correlations are well under control and that reliable electronic
energy-density functionals can be derived \textit{ab initio} because the
electron-electron interaction is well known.  The nuclear case is much
more involved because the nucleon as such is a composite particle and
a nucleon-nucleon interaction is already an approximate concept. Thus,
nuclear many-body theories, as discussed above, have not yet reached
sufficient descriptive power to serve as direct input for effective
mean-field models. They serve as guidance to develop the basic
features of the energy-density functionals \cite{Pet91aB}. The
open model parameters are adjusted phenomenologically. The three most
prominent schemes are the Skyrme energy functional, traditionally
called Skyrme-Hartree-Fock (SHF), the Gogny model, and the
relativistic mean field model (RMF), for a recent review see
\cite{Ben03aR}. The present review refers mainly to SHF whilst
establishing occasionally relations to the two other schemes (see also Section~\ref{sec:links}).

As said above, the self-consistent mean-field models lie between
\textit{ab initio} theories and the mac-mic method. The link between
\textit{ab initio} and SHF is still under development, as will be discussed
briefly in Section \ref{sec:links}.
The connection between SHF and the mac-mic method is much better
developed. There are attempts from both sides. The extended Thomas-Fermi
Skyrme interaction scheme (ETFSI) starts from SHF and derives an effective
mac-mic model by virtue of a semi-classical expansion
\cite{Bra81a,Bra85aR}. This has been exploited up to a quantitative level
\cite{Abo92a}.  From the macroscopic side there is an attempt to include more
self-consistency by virtue of a Thomas-Fermi approach
\cite{Mye98a,Mye00a}. The investigation of these links turns out to be useful
to gain more insight into the crucial constituents of either model.

One of the concerns in applying mean-field models is
uncertainty in the phenomenological determination of the free
parameters in the effective energy functionals \cite{rik03}. Many tens
of different parameter sets have evolved in the course of the
development. The main problem is that one does not have available a
one-to-one correspondence between individual parameters and a specific
piece of experimental information. It follows that there is a
continuum of parameter sets which would fit the experiment with a
comparable degree of accuracy. One of the important tasks in the
development of mean field models is to devise ways of constraining the
free parameters of interaction used. For example, new data from exotic
nuclei are an enormous help in that respect. One more recently identified
benchmark is the application of SHF to asymmetric nuclear
matter \cite{rik03}. This development is related to an increasing demand for
microscopic nuclear structure input into models of stellar matter,
occurring in neutron stars and core-collapse supernovae. Realistic
description of asymmetric nuclear matter under conditions of extreme
density, pressure and temperature poses a new constraint to nuclear
theory. It offers an extensive laboratory for testing the applicability
of effective mean field models, fitted to properties of finite nuclei
close to the beta-stability line, to the limits of nuclear binding at
the neutron and proton drip-line. The role of the SHF in providing
such input to nuclear astrophysics models and its implication is discussed in this work.

The present
review aims to survey the current status of SHF theory as applied in
both nuclear and astrophysics areas. The basic concepts and formalism
of the Skyrme approach, the choice of parameters and links to the
other models of nuclear interactions are detailed in
Section~\ref{sec:functional}. Section~\ref{sec:applfinite} is devoted
to applications in finite nuclei divided into subsections covering
static properties (\ref{sec:static}) (binding energies, spin-orbit
effects, neutron radii, super-heavy elements and fission barriers),
dynamic properties (\ref{sec:dynamic}) (giant resonances, Gamow-Teller
resonances, heavy ion collisions and rotational bands) and
nucleosynthesis (\ref{sec:nuclo}) (r-process and rp-process). These is
a final subsection devoted to discussion of collective correlations
beyond the mean field (\ref{sec:beyond}) (large amplitude collective
notion, soft modes and low energy quadrupole
states). Section~\ref{sec:infinite} focuses on nuclear matter and
astrophysical applications with subsections discussing the relevance
of SHF to construction of EOS for compact objects (\ref{sec:relshf}),
the key properties of nuclear matter, models of cold neutron stars
(\ref{sec:neutronstars}) and hot matter in type II (core-collapse)
supernova models. Constraints on parameters of SHF functionals are
summarized in Section~\ref{sec:summconstr} and conclusions are drawn
in Section~\ref{sec:concl}.

% \section{The Skyrme energy functional}
% \label{sec:functional}
%\subsection{A priori correlations}
%\subsubsection{On the description of correlations}
%\label{sec:corrbasic}
%\subsubsection{Motivation of the Skyrme functional}
%\label{sec:densmatexp}
%\subsection{Ingredients}
%\label{sec:ingredients}
%\subsubsection{The wavefunctions and densities}
%\label{sec:wavefun}
%\subsubsection{Basic formalism}
%\label{sec:basicform}
%\subsubsection{The center of mass correction}
%\label{sec:cmcorr}
%\subsubsection{Pairing interaction}
%\label{sec:pairing}
%\subsubsection{The mean field and pairing equations}
%\label{sec:mfeqs}
%\subsubsection{Computation of basic observables}
%\label{sec:observ}
%\subsection{Basic features}
%\label{sec:baiscfeat}
%\subsubsection{Fitting strategies}
%\label{sec:fitting}
%\subsubsection{Quality on gross properties}
%\label{sec:quality}
%\subsection{Links to other nuclear forces}
%\label{sec:links}
%\subsubsection{Relation to RMF}
%\label{sec:tormf}
%\subsubsection{Relation to the Gogny model}
%\label{sec:togogny}
%\subsubsection{Local Density Approximation and conformity with BHF}
%\label{sec:tobhf}
%\subsubsection{Nuclear effective forces and subnuclear degrees of freedom}
%\label{sec:qmc}
% this part --> PGR
\section{The Skyrme energy functional}
\label{sec:functional}

\subsection{\textit{A priori} correlations}

\subsubsection{On the description of correlations}
\label{sec:corrbasic}

The ultimate goal of a theory is, of course, an \textit{ab initio}
description which starts from the basic microscopic understanding of
the constituents of the system and predicts finally the properties of a
complex compound. The microscopic theory of nucleons and nuclei is
quantum chromodynamics (QCD) using quarks and gluons as constituents
\cite{Dis03aB}. However, QCD is a very involved theory and it is
already formidable work to compute the properties of single nucleons
or mesons. The description of a whole nucleus is presently far out of
reach. Approximate QCD at the level of a low-energy limit or chiral
perturbation theory allows approach finite nuclei and nuclear matter
and it has been applied with increasing success in recent years
\cite{Lut00a,Kai02a,wei05,boh05}. It should be noted, however, that
these \textit{ab initio} theories still allow for a few free parameters to
achieve fine-tuning of the results. But the field is still in a
developing stage and steady improvement may be expected.

Tradition nuclear \textit{ab initio} theories start from nucleons and
two-body interactions between them as the basic building blocks. The
interaction is determined from an exhaustive analysis of
nucleon-nucleon scattering, see e.g. \cite{Mac01aR}. The aim is a
microscopic description of the fully correlated nuclear state. It is
called correlated as opposed to the mean-field state (Slater
determinant or BCS state) which is the prototype uncorrelated state.
One distinguishes several types of correlations. Short-range
correlations are related to the hard repulsive core of the nuclear
interaction which develops at short inter-nucleon distances below 0.5
fm. Long-range correlations are mediated by the pronounced nuclear
resonance modes, the giant resonances; they mediate correlations over
distances because their propagator has large coherence length (longer
than any existing nucleus). Finally, there are the collective
correlations induced by the soft modes with large amplitude such as
center-of-mass motion, rotation, or low-lying quadrupole vibrations.
Each one of these types of correlation requires its own approach.
We will summarize them briefly here. For an extensive review of these
very different methods see \cite{Rei94aR}.

The family of conceptually simpler approaches is based on an obvious
ansatz for the correlated many-body wavefunction. Examples are the
Jastrow ansatz \cite{Pan97aR}, the exp$(S)$ method \cite{Kue78aR}, and
the unitary correlator technique \cite{Rot05a}. These three methods
have, in principle, much in common but differ in the details of the
practical evaluation. They start from a closed expression for the
wavefunction and evaluate the expectation values approximately by
elaborate expansion techniques (often called coupled cluster
expansion). They are designed to describe the short-range correlations
associated with the series of two-particle-two-hole ($2ph$)
excitations. A more general ansatz is used in the `no core shell
model' (NCSM) calculations which have been developed recently, see
e.g. \cite{Nav00a}. Here one aims at an expansion of the exact
many-body wavefunction in a basis of given shell model wavefunctions.
This method should in principle embrace all correlations, short as well as long
range and to some extent also collective ones.

Alternatively to the NCSM , one has the diagrammatic techniques which
are derived most often from the hierarchy of many-body Greens
functions \cite{Mat71aR}. The short-range correlations are related to
the series of ladder diagrams. The most widely used method is here the
Brueckner-Hartree-Fock (BHF) theory, for a recent review see
\cite{Dic92aR} and for its relativistic cousin,
Dirac-Brueckner-Hartree-Fock (DBHF) models \cite{Bro90a}.  The
long-range correlations stem from the series of bubble diagrams
(polarization propagators). They play the dominant role in electronic
systems \cite{Dre90aB}. There exist also noteworthy polarization
correlations in a nuclear environment, e.g., from the various giant
resonance modes \cite{Rei79a,Alb82a}. A combined treatment of short-
and long-range correlations has been tried in the theory of parquet
diagrams \cite{Jac82aR}. The enormous complexity of these methods has
so far hindered widespread applications of this approach. In practice,
diagrammatic \textit{ab initio} calculations in nuclear physics deal
only with the short-range correlations. That is to some extent
legitimate because they yield by far the dominant contributions due to
the huge short-range repulsion in the nucleon-nucleon interaction.

Collective correlations require very different methods. The effect
of zero-energy modes (center-of-mass, rotation, particle-number) is
handled by projection \cite{Rin80aB}. Low energy modes, as
e.g. quadrupole fluctuations, are usually treated by the
generator-coordinate method (GCM) \cite{Rei87aR} and approximations
therein \cite{Dan00aR}. Other methods to deal with large amplitude
motion in a wavefunction picture are re-quantized time-dependent
Hartree-Fock (TDHF) or path integrals \cite{Rei81e,Goe82a,Goe82b}.
Alternatively, one can describe a system in terms of operator algebra
and large amplitude motion is covered here by boson expansion
techniques \cite{Kle91aR}. 

The above short summary indicates that collective correlations differ
strongly from short- and long-range correlations and appear at a
different level of description. The collective correlations are
related to shape and surface of finite nuclei and are very sensitive
to shell effects which makes them vary strongly from nucleus to
nucleus. On the other hand, short- and long-range correlations are
fully active in the nuclear volume and persist up to nuclear
matter. Their effect is smoothly varying with proton and neutron
number. The latter feature justifies summarizing these correlations
implicitly in an effective energy-density functional, or effective
interaction respectively.  The way this is done and its consequences
is sketched in the following subsections. The main message to be kept
in mind is that these short- and long-range correlations have been \textit{a priori} built into the mean-field model and they should not be
computed again with these effective interactions (although there exist
techniques to recover carefully some of those correlations in the
$T$-matrix approach \cite{Toe88a}). The collective correlations, on
the other hand, cannot be transferred to a simple effective functional
because they do not obey smooth trends and rather vary strongly with
proton and neutron number. They need to be left for \textit{a posteriori}
treatment. The danger of double counting correlations is small because
strongly fluctuating effects cannot be included already in a smooth
energy functional. Examples for collective correlations will
be discussed in Section \ref{sec:beyond}.

Finally, we want to address again the limitations of \textit{ab initio} in
a nuclear context. The starting assumption of nucleons as basic
constituents is only approximately correct. It is then no surprise that all
strict \textit{ab initio} calculations starting from measured nucleon-nucleon
interactions fail to reproduce nuclear binding and radii at a
quantitative level (with typical errors of about 5\%
\cite{Fuc06a}). This fact has inhibited so far a parameter-free
deduction of effective energy-functionals.  The quality of \textit{ab initio}
calculations can be significantly enhanced by introducing effective
three-body forces. These, however, introduce free parameters in the
treatment which spoil the ideal of an \textit{ab initio} derivation. As a
consequence, the free parameters of nuclear energy-density functional
are directly fitted to empirical data, see Section
\ref{sec:baiscfeat}.

\subsubsection{Motivation of the Skyrme functional}
\label{sec:densmatexp}
%\label{sec:unitary}

The formally most transparent way to explain the construction of an
effective interaction is provided by the unitary correlator method.
The fully correlated state is written as a unitary transformation
$\hat{U}$ of a mean field state $|\Phi\rangle$, i.e.
\begin{equation}
  |\Psi\rangle
  =
  \hat{U}|\Phi\rangle
  \quad .
\end{equation}
At the lowest level, the unitary transformation is composed from exponentiated
two-body correlations 
\begin{equation}
\hat{U}=\exp{\left(\imath\sum c_{abcd}
   \hat{a}_a^+\hat{a}_b^+\hat{a}_c^{\mbox{}}\hat{a}_d^{\mbox{}}\right)}
\end{equation}
where $\hat{a}^+$, $\hat{a}$ are Fermion operators \cite{Rot05a}. The energy
expectation value is regrouped as
\begin{equation}
  E
  =
  \langle\Psi|\hat{H}|\Psi\rangle
  =
  \langle\Phi|\hat{U}^+\hat{H}\hat{U}|\Phi\rangle
  =
  \langle\Phi|\hat{H}_{\rm eff}|\Phi\rangle
  \quad.
\end{equation}
This defines 
the effective Hamiltonian 
\begin{equation}
  \hat{H}_{\rm eff}
  =
  \hat{U}^+\hat{H}\hat{U}.
\end{equation}
by associating the correlators to the Hamiltonian.  Note that the
expectation value with $\hat{H}_{\rm eff}$ is well defined only in
connection with mean-field states $|\Phi\rangle$ and for nothing
else. The price for the `effectiveness' is the restriction of the
legitimate Hilbert space. Such restrictions are best handled in
variational formulations and it is thus natural that effective
mean-field models come along in this fashion.  Thus far the principle
of building an effective interaction looks simple and is well
established in the derivation of effective charges (for a discussion
in diagrammatic terms see \cite{Kuo73a}). The practical evaluation,
however, is extremely tedious and not very enlightening.

More details about the desirable structure of a Skyrme-like effective
functional are disclosed by the technique of the density-matrix
expansion \cite{Neg72a}. Many \textit{ab initio} models deliver at the
end an effective two-body interaction as an involved integral operator
$\hat{G}$. This holds, in particular, for the Br\"uckner-Hartree-Fock
method (BHF) whose $G$-matrix serves finally as effective force for
the Hartree-Fock part, for a recent review see e.g \cite{Dic92aR}.  An
effective force as integral operator can also be extracted from many
other \textit{ab initio} models, as e.g. the above mentioned unitary
correlator method. In any case, the most general total energy becomes
\begin{equation}
  E_{\rm pot}
  = 
  \int\!dx\,dx'dy\,dy' 
  \rho({x},{x}')\,G({x},{x}';{y},{y}')\,\rho({y},{y}')
\end{equation}
where $\rho({x},{x}')$ is the one-body density matrix which falls off
quickly with increasing $|x-x'|$ with a typical range of $k_F^{-1}$
where $k_F$ is the Fermi momentum. More importantly, the integral
kernel $G$ is also non-zero only for small differences in all pairs of
coordinates with typical ranges mostly below $k_F^{-1}$. This suggests
an expansion in orders $(x-x')^n$ around $\bar{x}=(x+x')/2$ in the
form
\begin{equation}
  \rho({x},{x'})
  \approx
  {n}(\bar x)
  + i (x-x') \cdot j(\bar x)
  + \frac{1}{2}(x-x')^2 \big(\tau-\frac{1}{4}\Delta{n} \big)
  \quad,
\end{equation}
where ${n}$ is the local density, $j$ the local current, and $\tau$
the kinetic-energy density, see Section \ref{sec:wavefun}. It is
obvious that this expansion in $x-x'$ and similarly in $y-y'$ yields
properly the $\tau$ terms in the functionals. The $G$ matrix is also
well localized in the differences $x-y$ and $x'-y'$. An expansion of
the $G$ matrix with respect to these differences yields the gradient
terms $\propto\Delta{n}$ in the effective functional. Altogether, the
density-matrix expansion demonstrates how a zero-range effective
interaction emerges from the initially given involved operator
structure. We are dealing with a typical low-energy or low-$q$
expansion. It becomes apparent that this puts even more constraints on
the legitimate mean-field states. The spatial structure should be
sufficiently smooth. A safe estimate takes the pion wavelength as the
largest range in $G$. This complies with the spatial variations in the
wavefunctions and densities of order of $k_F^{-1}$ at normal nuclear
density. The validity of effective functionals exceeds often the range
of such safe estimates. But one should be warned about extension to
high densities. Sooner or later, effective functionals of zero range
will become inappropriate. There exist methods to extend the range to
a broader span of densities which have been worked out, e.g., for
$^3$He systems \cite{Wei92a} but not yet for nuclear systems.

Density functional theory provides the conceptually and technically
simplest way to derive an energy-density functional from \textit{ab initio}
calculations. This is the much celebrated local-density approximation
(LDA) \cite{Dre90aB}. We sketch it briefly for the case of one
component. A nuclear LDA deals, of course, with two components, for
protons and for neutrons separately. The basic steps can be summarized
as follows:
\begin{equation}
  {n}
  \quad\Longrightarrow\quad
  \frac{E}{N}\!=\!\epsilon({n})
  \quad\Longrightarrow\quad
  E
  =
  \int d^3r\,{n}\varepsilon\big({n}\big)
  \qquad\stackrel{{n}\rightarrow{n}(r)}{\Longrightarrow}\qquad
  E^{\rm(LDA)}
  =
  \int d^3r\,{n}(r)\varepsilon\big({n}(r)\big)
  \quad.
\end{equation}
One performs first a series of \textit{ab initio} calculations in bulk
matter for many different (homogenous) densities ${n}$. This yields an
energy per particle $E/N$ as function of ${n}$. It is rewritten as a
spatial integral which is still exact for homogenous densities. The
crucial step comes when allowing for spatially varying densities
${n}\longrightarrow{n}(r)$. This assumes that the finite and
inhomogeneous system can be treated as a succession of pieces of
bulk. This is an approximation for sufficiently smooth density
distributions. What ``sufficiently smooth'' means in practice remains
an open question as has been discussed already at the end of the
previous paragraph.  We will reconsider that question with a realistic
application in Section \ref{sec:tobhf}.

\subsection{The composition of the energy functional}
\label{sec:ingredients}

\subsubsection{The wavefunctions and densities}
\label{sec:wavefun}

The key idea in a mean field theory is that one seeks to describe the
many-body system exclusively in terms of a set of single-particle
wavefunctions $\left\{\varphi_\alpha,\alpha=1,...,A\right\}$. These
constitute the many-body state as a Slater determinant which is the
anti-symmetrized product of all occupied single-particle states. A
pure Slater state is simply applicable only to doubly-magic nuclei
which, however, are a small minority in the table of isotopes. All
others have partially open shells with a high density of almost
degenerated states which gives the residual two-body interaction a
chance to mix these states in order to produce a unique ground state
\cite{Rei84c}. This can be handled very efficiently and to a
reasonable approximation by a nuclear pairing scheme, see e.g.
\cite{Rin80aB,Ben03aR} for details. At the end, this amounts
associating an occupation amplitude $v_\alpha$ with each
single-nucleon state $\varphi_\alpha$. This amplitude can take values
continously in the interval $[0,1]$. The complementary non-occupation
amplitude is $u_\alpha=\sqrt{1-v_\alpha^2}$. Thus the typical
mean-field state is a BCS state
\begin{equation}
  |\Phi\rangle
  =
  \Pi_\alpha\big(
    u_\alpha^{\mbox{}}+v_\alpha^{\mbox{}}\hat{a}^+_\alpha{a}^{\mbox{}}_\alpha
  \big)|0\rangle
\label{eq:BCState}
\end{equation}
where $|0\rangle$ is the vacuum state. This BCS state comprises all the
information carried in the set
$\left\{\varphi_\alpha,v_\alpha;\alpha\!=\!1,...,\Omega\right\}$.  The
limit $\Omega>A$ denotes the size of the pairing-active space The
dynamical degrees-of-freedom are the single-particle wavefunctions
$\varphi_\alpha$ and the occupation amplitudes $v_\alpha$. The mean
field equations are obtained by variation of the total energy with
respect to these both quantities.  The expression for the total energy
is the key ingredient in the modeling. We will employ here the Skyrme
energy-density functional as given in Section \ref{sec:basicform}
together with a pairing functional, see Section \ref{sec:pairing}.
The equations for the Hartree-Fock mean field are obtained by
variation with respect to the wavefunctions $\varphi_\alpha$ while the
gap equations for the pairing mean field come from variation of the
$v_\alpha$.

Wavefunctions and occupations amplitudes uniquely define all one-body
densities. Energy-density functionals should require the knowledge of
only a few local densities and and currents, such as the local 
density ${n}({\bf
r})$. For nuclei, we have to distinguish proton and neutron density,
${n}_{\rm p}$ and ${n}_{\rm n}$. Note that we are using here the
symbol ${n}$ for the local particle number density. This differs from
the standard usage in the SHF community of the symbol $\rho$ for this
quantity. But it complies with the usage in nuclear astrophysics where
the symbol $\rho$ is reserved for the mass density (see Section
\ref{sec:relshf}), and it agrees also with the standards of electronic
density-functional theory \cite{Dre90aB}.  Furthermore, nuclear energy
functionals invoke also the kinetic-energy density $\tau$, the
spin-orbit density ${\bf J}$, the current ${\bf j}$, the spin density
${\bsigma}$, and the kinetic spin density ${\btau}$.  These read in
detail
\begin{subequations}
\label{eq:basdens}
\begin{eqnarray}
  {n}_q({\bf r})
  &=&
  \sum_{\alpha\in q}\big|v_\alpha^2\big|\,
  \big|\varphi_\alpha({\bf r})\big|^2
  \;\,\;
  q\in\{{\rm p,n}\}
  \quad,
\\
  \tau_q({\bf r})
  &=&
  \sum_{\alpha\in q}\big|v_\alpha^2\big|\,
  \big|\nabla\varphi({\bf r})\big|^2,
\\
  {\bf J}_q({\bf r})
  &=&
  -\imath
  \sum_{\alpha\in q}\big|v_\alpha^2\big|\,
  \varphi_{\alpha}^+({\bf r})\nabla\!\times\!\hat{\mathbf\sigma}
  \varphi_{\alpha}^{\mbox{}}({\bf r})
  \quad,
\\
  {\bf j}_q({\bf r})
  &=&
  -\frac{\imath}{2}
  \sum_{\alpha\in q}\big|v_\alpha^2\big|\,
  \left(
   \varphi_{\alpha}^+({\bf r}) \nabla\varphi_{\alpha}^{\mbox{}}({\bf r})
   -
   \nabla\varphi_{\alpha}^+({\bf r})\varphi_{\alpha}^{\mbox{}}({\bf r})
  \right)
  \quad,
\\
  {\bsigma}({\bf r})
  &=&
  \sum_{\alpha\in q}\big|v_\alpha^2\big|\,
  \varphi_{\alpha}^+({\bf r})
  \hat{\bsigma}\varphi_{\alpha}^{\mbox{}}({\bf r})
  \quad,
\\
  {\btau}({\bf r})
  &=&
  \sum_{\alpha\in q}\big|v_\alpha^2\big|\,
  \sum_{i\in\{xyz\}}
  \nabla_i\varphi_{\alpha}^+({\bf r})
  \hat{\bsigma}\nabla_i\varphi_{\alpha}^{\mbox{}}({\bf r})
  \quad.
\end{eqnarray}
It is often useful to recouple to sum and difference, e.g.
\begin{equation}
  {n}
  =
  {n}_{\rm p}+{n}_{\rm n}
  \quad,\quad
  \tilde{n}
  =
  {n}_{\rm p}-{n}_{\rm n}
  \quad,
\end{equation}
\end{subequations}
and similarly for all other densities and currents.  The sum plays a
role in the isoscalar terms of the energy functional and we will call
it the isoscalar density ${n}$. In a similar manner, the difference
plays the role of an isovector density $\tilde{n}$.

\subsubsection{Basic formalism}
\label{sec:basicform}

The Skyrme force was first introduced by Skyrme in \cite{sky56,sky59}
as an effective force for nuclear Hartree-Fock calculations. Its
widespread application started with the revival by Vautherin and Brink
in \cite{Vau70a,vau72}. It used to be formulated as an effective
interaction
\begin{eqnarray}
  \hat{V}_{\rm eff}({\bf r}_1,{\bf r}_2)
  &=&
  t_0(1\!+\!x_0\hat{P}_\sigma)\delta({\bf r}_1\!-\!{\bf r}_2)
  +
  \frac{t_3}{6}(1\!+\!x_3\hat{P}_\sigma)
  \delta\big(({\textstyle\frac{1}{2}}({\bf r}_1\!+\!{\bf r}_2)\big)
  \delta({\bf r}_1\!-\!{\bf r}_2)
\nonumber\\
  &&
  -\frac{t_1}{2}(1\!+\!x_1\hat{P}_\sigma)
  \big((\nabla_1\!-\!\nabla_2)^2\delta({\bf r}_1\!-\!{\bf r}_1)
       + h.c.\big)
\nonumber\\
  &&
  -{t_3}(1\!+\!x_2\hat{P}_\sigma)
  \big((\nabla_1\!-\!\nabla_2)\delta({\bf r}_1\!-\!{\bf r}_2)
       (\nabla_1\!-\!\nabla_2)\big)
\nonumber\\
  &&
  -\imath t_4(\nabla_1\!-\!\nabla_2)\!\times\!
  \delta({\bf r}_1\!-\!{\bf r}_2)(\nabla_1\!-\!\nabla_2)\big)
  \!\cdot\!(\hat\bsigma_1\!+\!\hat\bsigma_2)
  \quad,
\label{eq:skforce}\\
  \hat{P}_\sigma
  &=&
  \frac{1}{2}(1+\hat\bsigma_1\!\cdot\!\hat\bsigma_2)
  \quad,
\nonumber
\end{eqnarray}
where $\hat{P}_\sigma$ is the spin-exchange operator. It is a
zero-range interaction with kinetic and density-dependent terms. The
zero range is an idealization which is consistent with the smoothly
varying dependence of the mean field state on spatial
coordinates. Such a force makes sense only in connection with such smooth
wavefunctions and should never be used to regenerate short-range
correlations \cite{Vau70a}. The density dependent term $\propto t_3$
is of particular importance as it provides appropriate saturation and
thus secures the success of SHF in description of finite nuclei. It is
motivated by the concept of effective forces between nucleons in
nuclear environment in contrast to the forces between bare
nucleons. It can be seen as arising from either the concepts of
effective forces or energy functionals, resulting from the variation
with density of the underlying microscopic effective interaction or of
the energy (see \cite{Neg72a} and Section \ref{sec:densmatexp}). It
was originally formulated as a zero-range three-body force which is
equivalent to the density-dependent interaction (DDI) when used in
calculations of ground state properties. However, the two
interpretations of the $\propto t_3$ differ when excitations are
considered and the DDI approach is found to be more stable
\cite{Kre81a}.  On the other hand, the concept of a DDI leads to
inconsistencies in a variational formulation in that the variationally
derived two-body interaction is not identical to the initially given
interaction from which the energy was computed as expectation
value. What is done in practice, is to compute the
total energy as expectation value of the interaction first
(\ref{eq:skforce}) and then derive the Hartree-Fock equations and
solve them using the standard variational procedure. This is precisely
the concept of an effective energy-density functional. And this is the
line of development which we will pursue in the following.

Our starting point is then the most general Skyrme energy functional which
reads
\begin{subequations}
\label{eq:enfun}
\begin{eqnarray}
  E
  &=&
  \int d^3r\,{n}\left\{
    {\mathcal E}_{\rm kin}
    +{\mathcal E}_{\rm Skyrme}
    +{\mathcal E}_{\rm Skyrme,odd}
  \right\}
  +
  E_{\rm Coulomb}
  +
  E_{\rm pair}
  +
  E_{\rm cm}
  \quad,
\\
  {n}{\mathcal E}_{\rm kin}
  &=& 
  \frac{\hbar^2}{2m}\int d^3\,\tau 
  \quad, 
\label{eq:ekindens}
\\
  {n}{\mathcal E}_{\rm Skyrme}
  &=&
  \frac{B_0+B_3{n}^\alpha}{2}{n}^{2} 
  - 
  \frac{B_0^{\prime}+B'_3{n}^\alpha}{2}\tilde{n}^2
%\nonumber\\
%  &&
%  +\frac{B_3}{3}{n}^{\alpha+2}
%  -\frac{B_3^{\prime}}{3}\tilde{n}^2{n}^\alpha
\nonumber\\
  &&    
  +B_1 \big({n}\tau-{\bf j }^2\big)
  -B_1^{\prime}\big(\tilde{n}\tilde\tau-\tilde{\bf j}^2\big)
%\nonumber\\
%  &&
  -\frac{B_2}{2}{n}\Delta {n}
  +\frac{B_2^{\prime}}{2}\tilde{n}\Delta\tilde{n}
\nonumber\\
  &&
  -B_4{n}\nabla\!\cdot\!{\bf J} 
  -(B_4\!+\!B_4^{\prime})\tilde{n}\nabla\!\cdot\!\tilde{\bf J} 
  +
   \frac{C_1}{2}{\bf J}^2
   - \frac{C_1^\prime}{2}\tilde{\bf J}^2
  \quad, 
\label{eq:enfunsk}
\\
  {n}{\mathcal E}_{\rm Skyrme,odd}
  &=&
  -
  \frac{C_0+C_3{n}^\alpha}{2}{\bsigma}^{2} 
  +
  \frac{C_0^{\prime}+C_3^{\prime}{n}^\alpha}{2}{\tilde\bsigma}^{2} 
  +
  \frac{C_2}{2}{\bsigma}\!\cdot\!\Delta{\bsigma}
  -
  \frac{C_2^{\prime}}{2}{\tilde\bsigma}\!\cdot\!\Delta{\tilde\bsigma}
\nonumber\\
  &&
   -{C_1}\bsigma\!\cdot\!\btau
   +{C_1^\prime}\tilde{\bsigma}\!\cdot\!\tilde{\btau}
  -B_4 {\bsigma}\!\cdot\!(\nabla\!\times\!{\bf j})
  -(B_4\!+\!B_4^{\prime})
  \tilde{\bsigma}\!\cdot\!(\nabla\!\times\!\tilde{\bf j})
  \quad, 
\label{eq:enfunskodd}
\\
  {E}_{\rm Coulomb}
  &=& 
  e^2 \frac{1}{2} \int d^3r\,d^3r'
        \frac{{n}_{\rm p}({\bf r }){n}_{\rm p}({\bf r}')}
             {|{\bf r }-{\bf r }'|}
      - \frac{3}{4}e^2\left( \frac{3}{\pi} \right)^{1/3}
        \int d^3r\big[{n}_{\rm p}\big]^{4/3}
  \quad.
\label{eq:coulfun}
\end{eqnarray}
\end{subequations}
The $B$ ($B'$) parameters determine the strength of the isoscalar
(isovector) forces.
The spin-orbit term $\propto B_4, B'_4$ has only one free parameter
$B_4$ in the standard Skyrme functionals where one fixes $B'_4=0$.
The need for full isovector freedom in the spin-orbit term,
i.e. for a non-zero $B'_4$, was raised by the systematics of isotopic
shifts in Pb isotopes \cite{Rei95a}. 
The principle Skyrme functional ${\mathcal E}_{\rm Skyrme}$ contains
just the minimum of time-odd currents and densities which is required
for Galileian invariance \cite{Eng75a}, namely the
combinations ${n}\tau-{\bf j }^2$ and
${n}\nabla\!\cdot\!{\bf J} +{\bsigma}\!\cdot\!(\nabla\!\times\!{\bf j})$
. Further conceivable
time-odd couplings are collected in the optional terms in ${\mathcal
E}_{\rm Skyrme,odd}$. In case of a derivation from the zero-range
two-body force (\ref{eq:skforce}), they are related to the $B$
parameters of ${\mathcal E}_{\rm Skyrme}$ and the relation is given
implicitly in eq. (\ref{eq:bdef}). All these $C$-terms are optional
from the perspective of a pure energy-density functional and they are
not yet well determined.  The pure spin terms $\propto C_0, C_2$ play
a role only in truly time-odd situations such as, odd-$A$ nuclei or
the Gamow-Teller excitations (see Section \ref{sec:gamow-teller});
they are not yet well investigated.  The tensor spin-orbit terms
$\propto C_1, C'_1$ can contribute in most standard situations
although no specific consequence of their inclusion or need for them
has not yet been unambiguously identified.
The Coulomb functional (\ref{eq:coulfun}) depends only on the charge
density and stays outside this distinction. Its second term approximates
exchange in the Slater approximation \cite{Sla51}. Note that we use
the proton density in place of the charge density. This is a widely
used, more or less standard, approximation.
The center of mass (c.m.) term $E_{\rm cm}$ will be discussed in
Section \ref{sec:cmcorr} and the pairing energy $E_{\rm pair}$ in
\ref{sec:pairing}.

The energy-functional (\ref{eq:enfun}) leaves the time-odd part
(\ref{eq:enfunskodd}) open for free adjustment. The viewpoint of an
effective interaction as formulated in eq.  (\ref{eq:skforce})
establishes unique relations between the frequently used interaction parameters
$t_i,x_i$ and the functional parameters $B_i,C_i$:
\begin{subequations}{}
\label{eq:bdef}
\begin{alignat}{2}
B_0  & = \tfrac{3}{4}  t_0
         \quad , & \quad
B'_0 & = \tfrac{1}{2}  t_0 (\half + x_0)  \quad , \nn \\
B_1  & = \tfrac{3}{16} t_1 + \tfrac{5}{16} t_2 + \tfrac{1}{4} t_2 x_2
         \quad , & \quad
B'_1 & = \tfrac{1}{8} \Big[ t_1 (\half+x_1)-t_2 (\half+x_2)
                      \Big] \quad , \nn \\
B_2  & =   \tfrac{9}{32} t_1 
         - \tfrac{5}{32} t_2
         - \tfrac{1}{8}  t_2 x_2
         \quad , & \quad
B'_2 & = \tfrac{1}{16} \Big[ 3t_1 (\half+x_1)+t_2 (\half+x_2)
                      \Big] 
         \quad , \nn \\
B_3  & = \tfrac{3}{16} t_3
         \quad , & \quad
B'_3 & = \tfrac{1}{8} t_3 (\half+x_3)
         \nn \\
B_4  & = b_4 - \half b_4'
         \quad , & \quad
B'_4 & = \half b_4'
         \quad , 
\end{alignat}
\begin{alignat}{2}
C_0  & = - \half t_0 \big( \half - x_0 \big)
         \quad , & \quad
C'_0 & = \tfrac{1}{4} t_0 \quad , 
         \nn \\
C_1  & = \tfrac{1}{8} \Big[   t_1 \big( \half - x_1 \big)
                            - t_2 \big( \half + x_2 \big)
                      \Big]
         \quad , & \quad
C'_1 & = - \tfrac{1}{16} ( t_1 - t_2 )
         \quad , \nn \\
C_2  & = - \tfrac{1}{16} \Big[ 3 t_1 \big( \half - x_1 \big)
                               + t_2 \big( \half + x_2 \big) 
                         \Big]           
         \quad , & \quad
C'_2 & = \tfrac{1}{32} ( 3 t_1 + t_2 ) 
         \quad , \nn \\
C_3  & = - \tfrac{1}{8} t_3 (\half - x_3)
         \quad , & \quad
C'_3 & = \tfrac{1}{16} t_3
         \quad , 
\end{alignat}
\end{subequations}
%  \begin{equation}
%  \begin{array}{rclcrcl}
%     B_0 & = &   t_0 (1+\half x_0)
%   &\;\; , \;\; &
%     B'_0& = &   t_0 (\half+x_0)       \\
%     B_1 & = &  \frac{1}{4} \left[ t_1 (1+\half x_1)+t_2 (1+\half x_2)
%                \right]
%   &\;\; , \;\; &
%     B'_1& = &  \frac{1}{4} \left[ t_1 (\half+x_1)-t_2 (\half+x_2)
%                \right]                       \\
%     B_2 & = &  \frac{1}{8} \left[ 3t_1 (1+\half x_1)-t_2 (1+\half x_2)
%                \right]
%   &\;\; , \;\; &
%     B'_2& = &  \frac{1}{8} \left[ 3t_1 (\half+x_1)+t_2 (\half+x_2)
%                \right]                       \\
%     B_3 & = &  \frac{1}{4} t_3 (1+\half x_3)
%   &\;\; , \;\; &
%     B'_3& = &  \frac{1}{4} t_3 (\half+x_3)   \\
%     B_4 & = &  \half t_4               
%   &\;\; , \;\; &
%     B'_4& = &   0       \\
%     C_0 & = &   B_0
%   &\;\; , \;\; &
%     C'_0& = &   B'_0       \\
%     C_1 & = & \frac{1}{16}t_1-\frac{1}{8}t_1x_1
%                 -\frac{1}{16}t_2-\frac{1}{8}t_2x_2
%   &\;\; , \;\; &
%     C'_1& = & \frac{3}{16}t_1-\frac{1}{16}t_2
%  \\
%     C_2 & = &   ??
%   &\;\; , \;\; &
%     C'_2& = &   ???       \\
%  \end{array}
%  \label{eq:bdef}
%  \end{equation}
There is a one-to-one correspondence between the $B$ and the
$t,x$-parameterizations. The $C$-parameters are a consequence of the
interaction ansatz and can in this connection be considered as
function of the $B$s. At first glance, this feature looks like an
advantage of the interaction model. However, this imposed restrictions
may lead to conflicting requirements as worked out in Section
\ref{sec:spineffects} in connection withe isovector spin-orbit force.
The zero-range two-body spin-orbit interaction in the ansatz
(\ref{eq:skforce}) forces $B'_4=0$ while the energy functional takes
$B'_4$ as a free parameter which is found desirable to be adjusted
independently.

Empirical information for the other terms in the time-odd part
(\ref{eq:enfunskodd}) is much harder to obtain. These terms are not
active in the ground state of even-even nuclei. Odd nuclei could
supply useful information, however mixed with pairing properties
\cite{Rut99a}. This has not yet been explored in a systematic
manner. Excitations probe time-odd terms and the spin-couplings are
particularly crucial in modes with unnatural parity, see Section
\ref{sec:gamow-teller}. Again, this is an area where systematic
explorations are still ahead. Theoretical assistance can be obtained
from the RMF where spin properties are fixed by the Dirac equation. A
study of the non-relativistic limit suggests a minimalistic model
$C_i=0$ \cite{Sul03a}
which delivers the correct isovector structure for the spin-orbit
coupling \cite{Rei95a}.

\subsubsection{The center of mass correction}
\label{sec:cmcorr}

Electron clouds in an atomic or molecular system are bound by the
external Coulomb forces of the ionic background.  Nuclei are different
in that their center of mass moves freely through space without
confining fields (at least at nuclear scale). Thus the nuclear ground
state should be projected on center-of-mass (c.m.) momentum zero
\cite{Rin80aB}. The exact projection is tedious and a hindrance to
large scale calculations. Fortunately, the c.m. projection is well
behaved and can be approximated by simple expressions \cite{Schm91a}.
Two different approximate expression for the c.m. correction are in
use:
\begin{subequations}
\begin{eqnarray}
  E_{\rm cm}^{\mbox{}}
  &=&
  E_{\rm cm}^{\rm(full)}
  = 
  -\frac{1}{2mA}\langle\big(\sum_i\hat{p}_i\big)^2\rangle
  \quad,
\label{eq:cmfull}
\\
  E_{\rm cm}^{\mbox{}}
  &=&
  E_{\rm cm}^{\rm(diag)}
  = 
  -\frac{1}{2mA}\langle\sum_i\hat{p}_i^2\rangle
  \quad,
\label{eq:cmdiag}
\end{eqnarray}
\end{subequations}
where the brackets mean the expectation value over the BCS state
(\ref{eq:BCState}). The full correction (\ref{eq:cmfull}) is a second
order approximation to the exact c.m. projection and works very well for
nuclei with $A\geq 40$ and is still a fair approximation for lighter
nuclei \cite{Schm91a}. It encompasses effectively a two-body operator
which adds complications to the variational mean field equations. Thus
it is usually added \textit{a posteriori}, i.e. after the solution of the mean
field equations (projection after variation).
The correction (\ref{eq:cmdiag}) takes only the diagonal parts of the
two-body operator and so yields a one-body operator. This allows it to be
included in the mean field equations (variation after projection)
which is the standard for that variant of the theory. However, the form
(\ref{eq:cmdiag}) is a rather poor approximation to the full
correction (\ref{eq:cmfull}) and should be avoided.  An attempt to
cure the defects of this approach by a simple fit formula was made in
\cite{But84a} where the term (\ref{eq:cmdiag}) is augmented by a
factor \mbox{$f(A)=2/(t+1/3t)$} with \mbox{$t=(\tfrac{3}{2}A)^{1/3}$}.
There is also a center-of-mass correction for the nuclear density
distribution and quantities deduced therefrom. This will be discussed
in Section \ref{sec:observ}.

\subsubsection{Pairing interaction}
\label{sec:pairing}

Pure Slater states are distinguished by occupation weights
$n_\alpha\in\{0,1\}$. Such states are appropriate, however, only for
doubly magic nuclei. All others have partially open shells with a high
density of almost degenerated states which gives the residual two-body
interaction a chance to mix these states in order to produce a unique
ground state \cite{Rei84c}. This can be handled very efficiently and
to a reasonable approximation by a nuclear pairing scheme, see e.g.
\cite{Rin80aB,Ben03aR} for details. In the end, this amounts to
associating an occupation amplitude $v_\alpha\in{0,1}$ with each
single-nucleon state $\varphi_\alpha$. The complementing
non-occupation amplitude is $u_\alpha=\sqrt{1-v_\alpha^2}$.  Assuming
a zero-range residual interaction leads to the pairing energy
functional
\begin{subequations}
\begin{equation}
  E_{\rm pair}
  =
  E_{\rm pair}^{\rm(DI)}
  = 
  \frac{1}{4}\sum_{q\in\{{\rm p,n}\}}V_q^{\rm(pair)}\int d^3r\chi^2_q
  \quad,\quad
  \chi_q({\bf r})
  =
  \sum_{\alpha\in q}w_\alpha u_\alpha v_\alpha|\varphi_\alpha( {\bf r})|^2
  \quad,
\label{eq:ep1}
\end{equation}
where DI stands for $\delta$-interaction and $w_\alpha$ is some
phase-space weight. This pairing acts without prejudice throughout
the whole nuclear volume. There are good reasons to assume that most
of the pairing takes place near the nuclear surface
\cite{Bar99a}. That is accounted for by the density-dependent
$\delta$-interaction (DDDI) with the functional \cite{Ter97b}
\begin{equation}
  E_{\rm pair}
  =
  E_{\rm pair}^{\rm(DDDI)}
  = 
  \frac{1}{4} \sum_q v_{0,q} \int d^3r \chi^2_q
  \big[1 -\left(\frac{{n}}{{n}_0}\right)^\gamma\bigg]
  \quad,
\label{eq:ep2}
\end{equation}
\end{subequations}
where ${n}_0$ is the nuclear saturation density, typically
${n}_0=0.16\,{\rm fm}^{-3}$ and $\gamma$ models the surface profile of
the interaction. A standard value is $\gamma=1$. For a discussion of
the effects of varying $\gamma$ see, e.g., \cite{Dob01b}. The two
forms of the pairing functional (\ref{eq:ep1},\ref{eq:ep2}) are
practically standard in modern SHF calculations.  The strength
parameters $V_{\rm p}^{\rm(pair)}$ and $V_{\rm n}^{\rm(pair)}$ are
universal in that they can be adjusted to hold for the whole nuclear
landscape.
There are more elaborate treatments using the pairing part of the
Gogny force \cite{Dec80a} which is, of course, compulsory in Gogny
mean-field calculations and which is also often used in RMF approaches,
see e.g. \cite{Rin96aR}. 
Most older calculations use simpler schemes for pairing such as, a
constant pairing matrix element of a constant gap
\cite{Blo73a,Rin96aR}. These recipes depend very sensitively on the
pairing phase space and their parameters change substantially over the
nuclei. They do not meet the high standards of modern mean field
calculations.

The results of HFB or HFBCS calculations (see Section~\ref{sec:mfeqs})
depend on the space of single-nucleon states taken into account,
called here pairing phase space. The convergence with increasing phase
space is extremely slow \cite{Gon96a} so some cut-off procedure is
compulsory to obtain a manageable scheme. In fact, the cut-off is part
of the pairing description. It is provided by the phase-space weights
$w_\alpha$ in the above pairing functionals. The simplest way is to
use a sharp cut with $w_\alpha\in\{0,1\}$. This works well in
connection with large pairing spaces reaching up to 50 MeV above the
Fermi surface \cite{Dob84a}. Practicability often requires cuts at
lower energy.  The sharp cutoff then raises ambiguities if the level
density near the cutting edge is high. Much more elegant and stable is
a soft cutoff profile such as,
%\begin{equation}
$
  w_\alpha
  =
  \left[1+
    \exp{\left((\varepsilon_\alpha-(\epsilon_F+\epsilon_{\rm cut}))
            /\Delta\epsilon\right)}
  \right]^{-1}
$
%\label{eq:softcut}
%\end{equation}
where typically $\epsilon_{\rm cut}=5\,{\rm MeV}$ and
$\Delta\epsilon=\epsilon_{\rm cut}/10$ \cite{Bon85a,Kri90a}.  This
works very well for all stable and moderately exotic nuclei.  For
better extrapolability away from the valley of stability, the fixed
margin $\epsilon_{\rm cut}$ may be modified to use a band of fixed
particle number $\propto N^{2/3}$ instead of a fixed energy band
\cite{Ben00c}.

A basic problem with pairing is particle-number conservation.  The BCS
state (\ref{eq:BCState}) is adjusted to reproduce the correct {\em
average} number of protons and neutrons. But it comes with some
uncertainty in proton and neutron number as soon as $u_\alpha
v_\alpha\neq 0$. This uncertainty is embodied in the mean field
approximation. Nonetheless, one often sees a need for restoring exact
particle numbers. This can be achieved by particle-number projection
\cite{Rin80aB} although this is a bit cumbersome and, even worse, is
not necessarily consistent in connection with energy-density
functionals \cite{Ang01b}. As in the case of c.m. projection, one can
resort to approximate projection which is done usually in terms of the
Lipkin-Nogami (LN) prescription (for a detailed discussion with
comprehensive references see \cite{Ben03aR}).  Most of the results
presented here use simple BCS or HFB unless otherwise indicated.

\subsubsection{The mean field and pairing equations}
\label{sec:mfeqs}

The energy functional (\ref{eq:enfun}) once fully specified determines
everything. The mean field plus pairing equations are obtained in a
straightforward manner. Variation with respect to the single-nucleon
wavefunctions $\varphi_\alpha^+$ yields the mean-field equations and
variation with respect to the occupation amplitudes $v_\alpha$
provides the Bogoliubov equations, both together comprising the
Hartree-Fock-Bogoliubov scheme (HFB).  In practice, the feedback of
$v_\alpha$ variation on the mean field is neglected, yielding the
Bardeen-Cooper-Schrieffer (BCS) approximation to pairing.  We skip the
detailed mean field and pairing equations for reasons of brevity. A
comprehensive description is found in the recent review \cite{Ben03aR}.

Numerical implementation of the HFB(CS) equations requires further
consideration. There are two different ways to represent wavefunctions
and fields.  On the one hand are the basis expansions which usually
employ harmonic oscillator wavefunctions (for a recent detailed
application see \cite{Dob00d}).  On the other hand are grid techniques
where the wavefunctions and fields are represented on a grid in
coordinate space. Depending on the symmetry, this can be a radial 1D,
an axial 2D, or a Cartesian 3D grid.  Coordinate-space grids are well
suited for self-consistent mean fields which employ local densities.
For the kinetic energy, various approximations are available: Fourier
representation or finite difference formulae at various orders. The
optimum choice depends on the dimensionality, see
e.g. \cite{Blu92a}. A detailed example of implementation on a
spherical one-dimensional grid is presented in \cite{Rei91b}.

\subsubsection{Computation of basic observables}
\label{sec:observ}

The most prominent observable is, of course, the total energy.  It is
the starting point of the description (see Section
\ref{sec:basicform}) and thus naturally results from any mean field
calculation. The best possible reproduction of the energies is, of
course, the most important feature in the development of the model,
see Section \ref{sec:fitting}. A more refined view emerges when
looking at energy differences which may reveal information on the
underlying shell structure (see Section \ref{sec:etrends}) or lead to
effects beyond a mere mean-field description (see Section
\ref{sec:beyond}). A further energy-related quantity is the single
particle spectrum. It is a natural outcome of mean field
calculations. There is, however, no direct relation to experimental
observables. One usually deduces single particle energy from
neighboring odd-A nuclei which means that the results are masked by
polarization and correlation effects. But differences of
single-particle energies, in particular spin-orbit splittings, can be
extracted  fairly well \cite{Rut98a}.

Next to the energy, the density distribution is a key feature of
nuclear structure. Elastic electron scattering allows a more or less
model free experimental determination of the nuclear charge formfactor
$F_{\rm ch}({\bf k})$ \cite{Fri82a}. The formfactor is related to the
density by a simple Fourier transformation
$
F_q (\vec{k})  
= \int \! d^3r \; \exp{\imath\vec{k}\!\cdot\!\vec{r}}{n}_q(\vec{r})
$
where the variable index $q\in\{\mbox{ch,p,n}\}$ indicates that this
one-to-one relation holds for any distribution, charge, proton or
neutron like.  It depends only on $k\!=|\vec{k}|$ for spherical
systems.  We will base the following discussion on the less intuitive
but formally simpler formfactor and assume spherical symmetry.  Mean
field calculations yield proton and neutron formfactors.  The charge
formfactor is obtained by multiplying these with the intrinsic proton
and neutron formfactors (equivalent to folding of the densities)
\cite{Fri75a}
\begin{equation}
F_{\rm ch} (k) 
= \sum_q [ F_q G_{E,q} + F_{ls,q} G_M ]
\exp{\left(\frac{\hbar^2 k^2}
                {8\langle \vec{\hat{P}}_{\rm cm}^2 \rangle}\right)}
\label{eq:FormC}
\end{equation}
where $F_{ls,q}$ is the form factor of $\nabla \cdot \vec{J}_q$
augmented by a factor $\mu_q/4m^2$ with $\mu_q$ being the magnetic
moment of the nucleon, $G_{E,q}$ is the electric form factor and $G_M$
the magnetic form factor of the nucleons (assumed to be equal for both
species). The overall exponential factor takes into account the
center-of-mass correction for the formfactor complementing the
corresponding energy correction as discussed in Section
\ref{sec:cmcorr}. It employs the same variance of the total momentum
$\langle\hat{\bf P}_{\rm cm}^2 \rangle$ and its physical
interpretation is an unfolding of the spurious vibrations of the
nuclear center-of-mass in harmonic approximation \cite{Schm91a}.  The
nucleon form factors $G_{E,q}$ and $G_M$ are taken from nucleon
scattering data \cite{Sim80aE,Wal86aPC}, for details see
\cite{Ben03aR}.

Energy density functionals should, in principle, provide a reliable
description of all density distributions, or,
equivalently, formfactors. Actual functionals employ analytically
simple forms which are smooth functions of the densities as motivated
by a local-density approximation. It has been shown that this limits
the predictive value to the regime $k<2k_{\rm F}$ in the formfactor
where $k_{\rm F}$ is the Fermi momentum \cite{Rei92c}.  Fortunately,
the salient features of the nuclear shape are determined in that low
$k$ range. They can be characterized in terms of three form
parameters: r.m.s. radius $r_{\rm ch}$, diffraction radius $R_{\rm
ch}$, and surface thickness $\sigma_{\rm ch}$ \cite{Fri82a}.  They are
computed as \cite{Fri82a}
\begin{subequations}
\begin{eqnarray}
  r_{\rm ch} 
  &=&
  \frac{3}{F_{\rm ch}(0)} 
  \frac{d^2}{dk^2} F_{\rm ch} (k) \bigg|_{k=0} 
  \quad,
\\
  R_{\rm ch}  
  &=& 
  \frac{4.493}{k_0^{(1)}}
  \quad,\quad
  F_{\rm ch} (k_0^{(1)}) = 0
\label{eq:Rdif}
\\
  \sigma_{\rm ch}
  &=& 
  \frac{2}{k_m} \log{\left(\frac{F_{\rm box}(k_m)}{F_{\rm ch}(k_m)}\right)} 
  \quad,\quad
  F_{\rm box}(k) 
  = 
  3 \, \frac{j_1(k_mR_{\rm ch})}{k_mR_{\rm ch}} 
  \quad,\quad  
  k_m = 5.6/R_{\rm ch}
\label{eq:surf}
\end{eqnarray}
\end{subequations}
The diffraction radius $R_{\rm ch}$ parameterizes the overall
diffraction pattern which resembles these of a filled sphere
of radius $R_{\rm ch}$. It is called
the box equivalent radius \cite{Fri82a}. The actual nuclear form
factor decreases faster than the box form factor $F_{\rm box}$ due to
the finite surface thickness $\sigma$ of nuclei which is thus
determined by comparing the height of the first maximum of the box
equivalent form factor and of the mean-field result $F_{\rm ch}$.
The simple simple combination $\sqrt{\frac{3}{5}R_{\rm ch}^2 + 3
\sigma^2_{\rm ch}}-r_{\rm ch}$ of these three form parameters serves
as a nuclear halo parameter which is found to be a relevant measure of
the outer surface diffuseness \cite{Miz00a}.

The charge distribution is mostly sensitive to the proton
distribution. Useful complementing information is contained in the
neutron distribution which is discussed briefly in Section
\ref{sec:neutronradii}.

\subsection{Basic features}
\label{sec:baiscfeat}

The form of the energy functional as presented in Section
\ref{sec:functional} is based on a low momentum expansion of many-body
theory \cite{Neg72a} which leads naturally to local point couplings
with some density dependence. The parameters of the functional are
universal in the sense that they apply unmodified to all conceivable
nuclei however they have to determined by fit to experimental data 
since nuclear many-body
theories are not yet precise enough to allow an \textit{ab initio} computation
of nuclear properties with the precision which we need for practical
applications. Thus it is still state of the art to adjust the
parameters of the functional phenomenologically. Typical strategies
for the adjustment will be discussed in Section \ref{sec:fitting} and
its practical consequences subsequently. 
Thereby we include also the RMF in the considerations as the
relativistic cousin of SHF because similarities and differences
between SHF and RMF are often quite instructive.
The connection to
BHF and the importance of each
parameter is discussed briefly in \ref{sec:tobhf}.

\subsubsection{Fitting strategies}
\label{sec:fitting}

The parameters of the functional are adjusted such that a
certain set of observables is optimally reproduced.  Different groups
have different biases in selecting these observables. Fits are usually
restricted to a few semi- or doubly-magic spherical nuclei (an
exception is a recent large-scale fit to all known nuclear masses
\cite{Sam02a}). All fits take care of binding energy $E_{\rm B}$ and r.m.s.\
charge radii $r_{\rm ch}$ after which different tracks are pursued.
SHF fits include extra information on spin-orbit splittings while RMF
incorporates the spin-orbit interaction automatically as a
relativistic effect, see Section \ref{sec:tormf}.  Pairing properties
are usually adjusted to the odd-even staggering of binding energies,
see e.g. \cite{Sat98a,Ben00c}.  Some fits add information on nuclear
matter, the series \cite{Cha98a} even on neutron matter.  Others make
a point to include information from the electromagnetic formfactor
\cite{fri86a} (see Section \ref{sec:observ}). Differences exist also
in the bias and weight given to the various observables; e.g., the
force BSk1 fits exclusively to binding energies ($\equiv$ atomic
masses).  Recently, attention has been brought to possible corrections
for proton-neutron correlations in $N=Z$ nuclei, often called the
Wigner energy \cite{Sat97b} (see e.g. the fits of \cite{Sam02a}). A
detailed discussion of fitting strategies can be found, e.g., in
\cite{Ben03aR}.

\begin{table}
\begin{center}
\begin{tabular}{l|c|cccc|cl}
\hline
  force & ref. & c.m. & l*s T=1 & $\eta_{\rm ls}$ & Coul.ex.& nuc.mat. &comments \\
\hline
  SkM$^*$ & \cite{Bar82a}& diag & no & no & yes & yes & fission barriers\\
  SkP &\cite{Dob84a} &  diag & no & no & yes & yes & pairing\\
  SLy6 & \cite{Cha98a}& full & no & no & yes & yes & neutron matter\\
  SkI3/4 &\cite{Rei95a} &full & yes & no & yes & no & isotopic shift Pb\\
  BSk1 & \cite{Sam02a}& diag & no & no & yes & no & deformed nuclei\\
\hline
  PC-F1 &\cite{Bue02a} & full & & & no & no &\\
  NL-Z &\cite{Ruf88a} & diag & & & no & no &\\
  NL3 &\cite{Lal97a} & guess & & & no & yes & isovector bias\\
\hline
\end{tabular}
\end{center}
\caption{\label{tab:options}
The selection of parameterizations used throughout this paper
with their actual options
(`ref.' = first publication, `c.m.' = center-of-mass correction, `l*s T=1' = isovector
spin-orbit,
$\eta_{\rm ls}$ = switch for spin-orbit tensor term,
`Coul.ex.' = Coulomb exchange in Slater approximation,
`nuc.mat.' = some nuclear matter properties had been included in the
fit).
The entry `guess' for NL3 means that the c.m. correction
had been used in the simple form as $E_{\rm cm}=30.75\,A^{1/3}$.
The diagonal c.m. correction  for BSk1 is augmented with the recipe 
from \cite{But84a}.
}
\end{table}
In view of these different prejudices entering the fits, there exist
many different parameterizations for SHF as well as RMF. We confine
the discussion to a few well adjusted and typical sets.  They are
summarized together with their citations, actual options, and fit bias
(last two columns) in Table \ref{tab:options}. A few further
explanations are in order. They expand mainly the key words in the
column `comments'. SkM$^*$ belonged to the second generation of Skyrme
forces which delivered for the first time a well equilibrated
high-precision description of nuclear ground states. It was developed
with an explicit study of surface energy and fission barriers in
semiclassical approximation. SkP aimed at a simultaneous description
of the mean field and of pairing. The set SLy6 and its cousins (see
\cite{Cha98a}) have been developed with a bias to neutron rich nuclei
and neutron matter aiming at astrophysical applications.  SkI3 and
SkI4 exploit the freedom of an isovector spin-orbit force to obtain an
improved description of isotopic shifts of r.m.s. radii in neutron
rich Pb isotopes which posed a severe problem to all conventional
Skyrme forces, see Section \ref{sec:spineffects}. BSk1 (and subsequent
variants) fit exclusively to binding energies (computed in HFB), but
employ an considerable pool of even-even nuclei including the majority
of those with deformation. As for the c.m. correction in spherical
nuclei, a simple correction for the rotational projection had been
employed in deformed systems and an ad-hoc correction for the Wigner
energy in $N$=$Z$ nuclei was applied \cite{Sat97b}.
The RMF parameterizations need no adjustment for the spin-orbit force
because that is implied in the four-spinor structure of the
relativistic wavefunctions \cite{Rei89aR}. Coulomb exchange is rarely
included in the RMF, probably for reasons of formal similarity with
the meson-field terms in the model. PC-F1 is a point-coupling model
\cite{Bue02a} whereas NL3 \cite{Gam90a} and NL-Z \cite{Ruf88a} use the
standard RMF with finite range couplings. PC-F1 as well as NL-Z are
fitted with the same strategy and data pool as SkI3/4 \cite{Rei95a},
i.e. information of the electro-magnetic formfactor is included.  NL3
takes care of the incompressibility and puts weight on correct
isovector trends.

\subsubsection{Quality on gross properties}
\label{sec:quality}

\begin{figure}
\centerline {\includegraphics[angle=0,width=15cm]{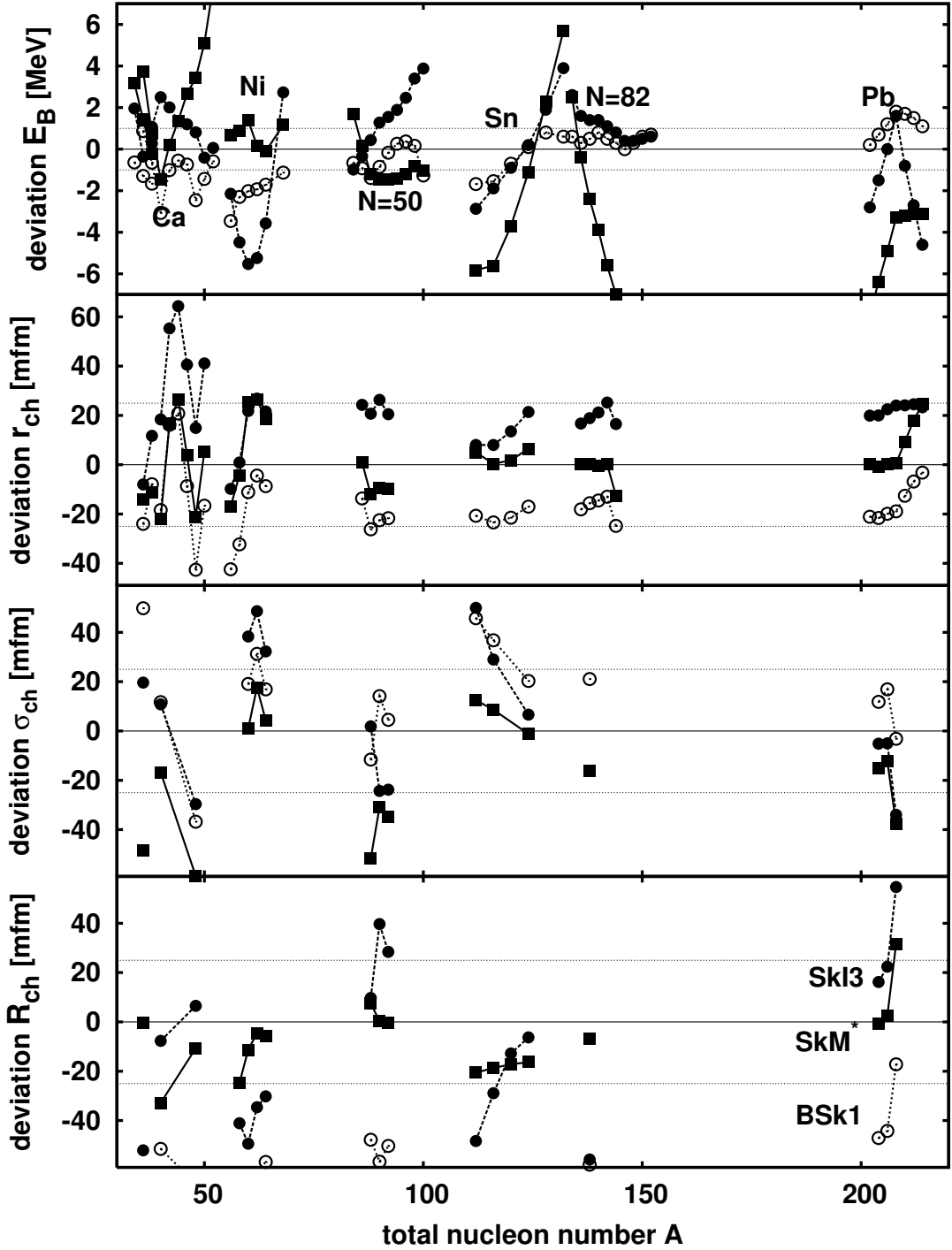}}
\caption{\label{fig:scanERS_abs}
Distributions of deviations
${\mathcal O}^{\rm(exp)}-{\mathcal O}^{\rm(model)}$
for the four key observables in the description of nuclear ground
states: binding energy $E_B$, charge r.m.s. radius $r_{\rm ch}$,
charge surface thickness $\sigma_{\rm ch}$, and charge diffraction
radius $R_{\rm ch}$.  Results are shown for three forces: SkM$^*$
(filled squares), SkI3 (filled circles), and BSk1 (open circles). The
typical deviations which can be achieved by good fits are indicated by
the band of dashed horizontal lines.}
\end{figure}
Figure \ref{fig:scanERS_abs} shows the quality of the
reproduction of energy, radii and surface thickness as delivered by
three forces from our selection and for a chains of spherical
semi-magic nuclei spread over the nuclear chart. The dashed horizontal
lines indicate the typical average deviation which can be achieved,
being approximately $\pm 1$ MeV for the energy and $\pm 0.025$ fm for
radii and surface thickness.  Note that these limits are demanding
requiring, e.g., for $^{208}$Pb a relative deviation from
experiment of 0.006\% for the energy and of
0.003\% for the radii. Larger relative deviations are natural for
lighter nuclei because correlation effects become relatively
larger. The figures shows indeed a nearly constant
absolute deviation
throughout all sizes, even with a slight tendency to larger errors for
small systems (particularly for the r.m.s. radii).  The errors are not
statistically distributed but show unresolved trends. The large
isotopic and isotonic trends, for example for SkM$^*$ results, may
suggest missing correlations effects. But BSk1 resolves those trends
to a large extent. Thus we see here largely limits of a functional,
probably with a remaining smaller part from irreducible correlations.
The example indicates that the capabilities of nuclear energy
functionals are not yet fully exhausted and that improvements can
still be expected. For example, BSk1 was fitted with strong bias on
atomic masses and performs best in that respect.  The forces from the
BSk family could achieve an average error of 0.7 MeV for the energies
\cite{Sam02a} which is the best result amongst the self-consistent
models.
Even though, the present performance of the mean field models is
already very satisfactory as can be seen from the overall quality
shown in the figure. 
Moreover, the models provide also a reliable
description of many other observables as we will see later in
Section \ref{sec:infinite}. 

The extrapolation of gross properties, energies and radii, to infinity
relates these to the key properties of nuclear matter as they are
saturation point, incompressibility, and symmetry energy. There is no
direct experimental access to these ``observables'' and thus the
quality of different parameterizations cannot be discussed as directly
as done here for finite nuclei. However, compliance with other
extrapolations to bulk matter, particularly those within the
liquid-drop model \cite{Rei06a}, is usually checked.  However, some
empirical evidence can be obtained from astro-physical considerations
such as, properties of neutron stars. We thus have placed the
discussion of nuclear matter properties in Section
\ref{sec:infinite}. The discussion of basic properties of nuclear
matter in relation to SHF is found in particular in Section
\ref{sec:keyprop}.

\subsection{Links to other nuclear forces}
\label{sec:links}

\subsubsection{Relation to RMF}
\label{sec:tormf}

The relativistic mean-field model (RMF) was developed to give a
competitive description of nuclei in the seventies
\cite{Bog77a,Ser79a}, at about the same time as SHF, for reviews see
e.g. \cite{Ser86aR,Rei89aR,Rin96aR,Vre05aR}. The RMF is conceived as a
relativistic theory of interacting nucleonic Dirac fields and mesonic
mean fields with the anti-particle contributions in the nucleon
wavefunctions being suppressed (`no--sea' approximation). However, the
mean-field approximation would not be valid in connection with the
true physical meson exchange fields. Thus the meson fields of the RMF
are effective fields at the same level as the forces in SHF are
effective forces. The RMF is the relativistic cousin of SHF, and the
same strategy applies: the model is postulated from a mix of intuition
and theoretical guidance with parameters to be fixed
phenomenologically.
For the modeling, one chooses the most basic meson fields, one scalar
($\sigma$) and one vector ($\omega$) field in the isoscalar channel, and a
vector-isovector field (${n}$). The pion does not contribute at Hartree level
because the pseudo-scalar density vanishes. A second isovector meson
($\delta$) is conceivable but was found to be ineffective \cite{Ruf88a}.  The
necessary density dependence is introduced through non-linear terms (cubic and
quartic) in the scalar meson field \cite{Bog77a}. This leads to a model with
about the same descriptive power as SHF \cite{Ben03aR}. The indirect modeling
of density-dependence was originally motivated by the aim to maintain
renormalizability of the theory \cite{Ser86aR}. There are variants of the RMF
which proceed differently in this respect. The point-coupling RMF (PRMF) employs
covariant zero-range couplings plus non-linear terms in the density, in a way 
similar to SHF \cite{Hoc90a} and it reaches a competitive descriptive
power \cite{Bue02a}. Much more elaborate density dependences are sometimes
discussed to enhance the flexibility of the model, see
e.g. \cite{Typ99a,Hof01a,Nik04a}.

It is possible to draw straight connections between RMF and SHF by
considering a twofold expansion, a non-relativistic ($v/c$) and a
zero--range limit of RMF \cite{Rei89aR,Thi86a}. The ($v/c$) expansion
applies to the scalar density ${n}_s$ and reads
\begin{equation}
    {n}_s
    =
    {n}
    +
    \frac{1}{2m^*}\left\{
      \tau
      - \vec{j}^2/{n}_0
      + \nabla \cdot \vec{J}
    \right\}
\label{eq:vclim}
\end{equation}
where $m^*$ is the nucleon effective mass. It is obvious that a
kinetic term $\propto\tau$ and a spin-orbit term $\propto\vec{J}$
emerge naturally, appearing as a specific combination.  The one extra
parameter for the scalar coupling in RMF is equivalent to the one
extra parameter for the kinetic term in the SHF functional
(\ref{eq:enfunsk}). We reiterate that the
spin-orbit coupling is built into the RMF and needs no extra
adjustment. This `spin-orbit for free' is the major difference between
RMF and SHF.  Moreover, the RMF-like spin-orbit force is predominantly
isoscalar which differs from the conventional SHF form
($B_4^{\prime}=0$ in the SHF functional). An example of measurable
consequences of this difference is discussed in Section
\ref{sec:spineffects}. The ($v/c$) expansion suffices to map the PRMF
into SHF. The conventional RMF needs as a further step an expansion of
the finite range of the meson field into a leading zero-range coupling
and subsequent gradient correction. This provides formally the
structure of the surface terms $\propto\Delta{n}$ in the functional
(\ref{eq:enfunsk}).  Thus far the equivalence is straightforward.  The
density dependence, however, is hard to map.  It is built into
conventional RMF by using a non-linear meson coupling and this
mechanism is very different from the SHF where a term with higher
power (than two) in the density is used. PRMF with its
straightforward expansion in powers of density ${n}$ is much closer to
SHF in that respect.

\subsubsection{Relation to the Gogny model}
\label{sec:togogny}

The breakthrough of self-consistent nuclear models in the seventies
has generated not only the SHF and the RMF, but also the Gogny model
as a third and equally powerful option \cite{Gog70a,Dec80a}. The Gogny
model is an effective interaction as SHF or RMF and its parameters are
adjusted to empirical data. It is a non-relativistic model as SHF, but
employs a finite range two-body interaction with all exchange terms
correctly treated. The density dependence and spin-orbit force are
added in the same functional forms as in the Skyrme functional
(\ref{eq:enfunsk}), namely as zero-range effective
interactions. It could
perhaps be more properly named Gogny-Hartree-Fock. The Gogny
functional can be mapped formally to SHF with the general techniques
of the density-matrix expansion \cite{Neg72a} which was discussed in a
more general context in the second paragraph of Section
\ref{sec:densmatexp}.

\subsubsection{Local Density Approximation and conformity with BHF}
\label{sec:tobhf}

 The formal aspects of the
derivation of an energy-density functional from \textit{ab initio} input by
means of the LDA were outlined in the third paragraph of Section
\ref{sec:densmatexp}. We will discuss here its performance in a
realistic example. Starting from from BHF calculations of nuclear
matter \cite{Bal04a} we calculate binding energies and r.m.s radii of
finite nuclei across the nuclear chart in LDA.
\begin{SCfigure}[0.5]
{\includegraphics[angle=0,width=11cm]{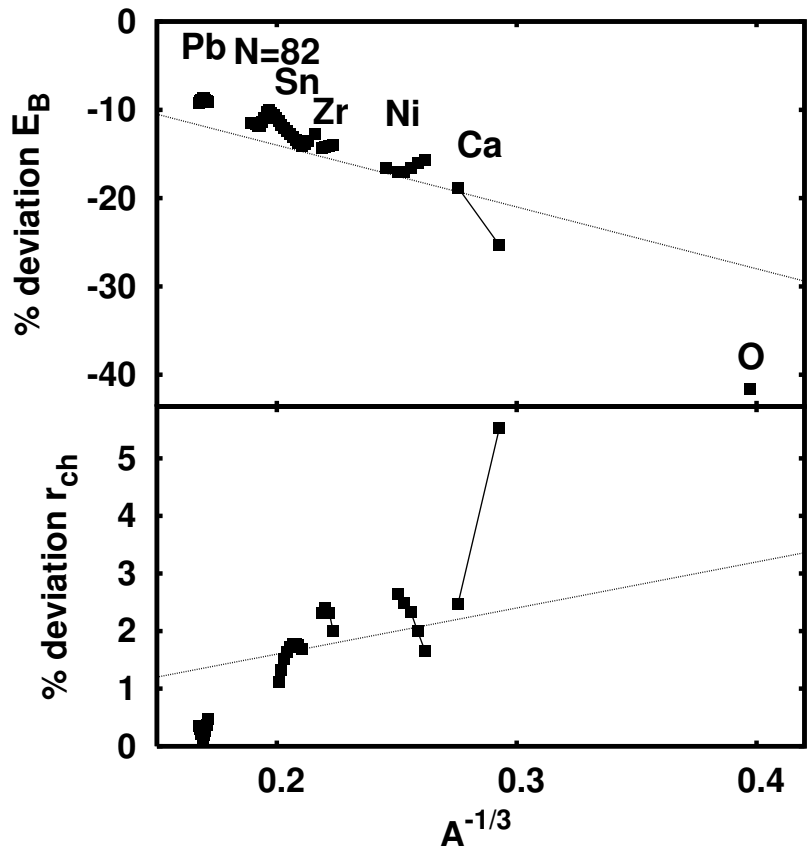}}
\caption{\label{fig:scanER_bhfpure}
Distributions of deviations 
${\mathcal O}^{\rm(exp)}-{\mathcal O}^{\rm(model)}$
for charge r.m.s radii and energies as computed from LDA using an
equation of state for nuclear matter from BHF calculations in
\cite{Bal04a}. Beyond strict LDA, a pairing and a spin-orbit force have been
added, but they have a negligible effect on the results.  The fine
dotted lines indicate an average trend $\propto A^{-1/3}$.  }
\end{SCfigure}
Figure
\ref{fig:scanER_bhfpure} shows the average deviation from experiment
that can be achieved in this model. The good news is that LDA provides
at once a basically pertinent view of nuclear size and energy. At
second glance, however, one realizes that the deviations are much too
big for practical applications, e.g., in astrophysical scenarios.
That failure is partially due to the fact that strict \textit{ab initio}
calculations (using only nucleon-nucleon interactions) never reach the
proper saturation point of nuclear matter, leaving a principle error
of about 5\% on energies (see Section \ref{sec:intrmf}). However, the
errors in the LDA calculation presented here exceed the deviation in
energy by much more, particularly for smaller nuclei. This is a clear
insufficiency of the LDA applied to finite nuclei. The key for
quantitative success of LDA lies in the gradient terms which allow
modeling of the nuclear surface energy. These terms are by far the
most important terms on the way to the excellent description
documented in Figure \ref{fig:scanERS_abs}. The situation is to some
extent similar to electronic density functional theory where gradient
corrections enhance the descriptive power to chemical precision
\cite{Per96,Dre90aB}. Recall, however, that the crucial first step
remains LDA which delivers a sound basis for further refinement. It is
conceivable to develop a partially \textit{ab initio} determination of the SHF
functional where the density dependent terms are derived with LDA from
given microscopic calculations. The problem however is that the
microscopic understanding is limited so that there exists an
uncomfortably large variety of predictions for nuclear matter
properties, see e.g. \cite{Fuc06a}.
A study in somewhat similar spirit is found in \cite{Cao06a}. It goes
beyond the strict grounds of the LDA in that it also adjusts the
effective nucleon mass, and with it the kinetic terms
$\propto\tau$. This yields ground state binding energies
within 5\% precision, better than above but still far from the goal of
 0.5\% reliability. Once more the need to include gradient term is clear.

For completeness, we mention that there exist also
several attempts to map relativistic Brueckner-Hartree-Fock
calculations onto an effective RMF, for most recent examples see
e.g. \cite{Ser05a,Ron06a}.  The formal relations are to some extend
more direct than in nonrelativistic models. The basic obstacle,
however, remains the fact that all nuclear ab-initio models up to now
have a limited descriptive value. 

\subsubsection{Nuclear effective forces and subnuclear degrees of freedom}
\label{sec:qmc}

Although nuclear models of point-like nucleons interacting via
effective forces have proven quite successful in the description of nuclear
properties, the predictive power of all nuclear models is not
sufficient as yet and new, more fundamental approaches are sought for,
that would improve the situation . Application of QCD to nucleon
systems is more frequently examined in dense (hot) matter such as the
cores of compact stars where the physical conditions for a partial or
full quark deconfinement are satisfied and quark effects, at least to
some approximation, must be included, (see e.g. \cite{mai04,agu03,and02}).

 The question of the effect of quark degrees of freedom in finite
nuclei is still treated as tentative. Their consideration in nuclear
systems at low energy challenges the application of non-perturbative
QCD, but progress has been made.  A recent example of a model aiming
towards understanding the role of subnucleonic degrees of freedom in
many-body effective Hamiltonians \cite{gui04} utilizes the quark meson
coupling model (QMC) in non-relativistic approximation
\cite{gui96}. The main idea of the model is to express the response of
the quark structure of the nucleon to the nuclear environment. The
interaction between the quarks of \textit {different} nucleons
(assumed to be non-overlapping bags) in the nuclear medium is
represented by exchange of $\sigma$, $\omega$ and $\rho$ mesons, with
coupling constants treated as free parameters. In the zero-range limit
of the model, the effective Hartree-Fock QMC Hamiltonian can be
written in a form which is directly comparable with the Hartree-Fock
Hamiltonian corresponding to the Skyrme effective force. The three
free parameters of the QMC model, the coupling constants $G_\sigma$,
$G_\omega$ and $G_\rho$ were fitted to reproduce correctly the binding
and symmetry energy and the saturation density of nuclear matter. The
remaining four parameters, the meson rest masses and the bag radius R
are fixed to experimental masses for the $\omega$ and $\rho$ mesons,
taking R=0.8 fm as an empirical value. The mass of the scalar $\sigma$
meson $m_\sigma$, not known well enough from experiment, is allowed to
vary in the expected region 500--600 MeV. Comparing corresponding
terms in the QMCHF and SHF Hamiltonians, the Skyrme parameters $t_0,
x_0, t_3, 5t_2-9t_1$ and $t_4$ can be expressed in terms for the
parameters of the QMC model. The parameters show a close resemblance
to those of the SIII Skyrme force that is rather impressive.

The QMC model in the form defined in \cite{gui04} did not give good
properties of nuclear matter at densities above nuclear saturation
density, which limited its applications to the physics of nuclear
matter and compact objects. This deficiency has been removed in the
current version of the model (\cite{gui06} and references therein)
where a density-dependent, effective nucleon-nucleon force of the
Skyrme type has been derived using a new version of the QMC model that
leads to an effective Hamiltonian with a density-dependent two-body
force for finite nuclei and allows a treatment of high density nuclear
matter consistent with relativity. It has been demonstrated that,
following a procedure similar to that in \cite{gui04}, the Skyrme
parameters $t_0, t_1,t_2,t_3,t_4$ and $x_0$ are calculated from the
QMC formalism in quite close agreement with those obtained from
fitting to experimental data in the SHF model with essential just one
adjustable parameter, the mass of the $\sigma$ meson. When the QMC
Hamiltonian is used in the HF approximation for doubly-closed-shell
nuclei, it yields binding energy per particle, charge and neutron
radii and spin-orbital splittings for $^{16}$O, $^{40}$Ca, $^{48}$Ca
and $^{208}$Pb in a very good agreement with experiment considering
that the model depends only on three adjustable parameters (meson
coupling constants). For application of the model far from stability,
the QMC-HFB approach was used with density dependent contact
interaction acting in the particle-particle channel. As an example, it
yielded the position of the neutron drip-line at around N=60 for Ni
and N=82 for Zr, similar to predictions provided by the SLy4 Skyrme
parameterization. Calculation of two-neutron and two-proton separation
energies suggests strong shell quenching at N=28 around Z=14 (proton
drip-line region) and Z=32 (neutron drip-line region) as also
predicted by HFB calculations with the SLy family \cite{Cha98a} of
Skyrme forces. Detailed application of the new QMC model to nuclear
matter and neutron stars is expected in a forthcoming
publication. This development can be seen as an interesting attempt to
relate the Skyrme interactions to an effective model based on
subnuclear degrees of freedom.

We mention for completeness that another many-body model of the
nucleus, inspired by QCD quantum field theory, where the strong
coupling regime is controlled by a three-body string-type force and
the weak coupling regime is dominated by a pairing force, has been
recently proposed \cite{boh05}. This model has however so far more of
conceptual than practical interest as, although it yields reasonable
results for the surface density of finite nuclei, and correct
properties of symmetric nuclear nuclear matter, it needs more
development before it is applicable to a wide range of properties of
finite nuclei.

%\section{Applications to finite nuclei}
%\label{sec:applfinite}
%\subsection{Static properties}
%\label{sec:static}
%\subsubsection{Trends of ground state binding energies}
%\label{sec:etrends}
%\subsubsection{Spin-orbit effects}
%\label{sec:spineffects}
%\subsubsection{Neutron radii}
%\label{sec:neutronradii}
%\subsubsection{Super-heavy elements}
%\label{sec:she}
%\subsubsection{Fission barriers}
%\label{sec:fission}
%\subsection{Dynamic properties}
%\label{sec:dynamic}
%\subsubsection{Giant resonances}
%\label{sec:giantres}
%\subsubsection{Gamow-Teller resonances}
%\label{sec:gamow-teller}
%\subsubsection{Heavy ion collisions}
%\label{sec:collisions}
%\subsubsection{Rotational bands}
%\label{sec:rotat}
% this part --> PGR
\section{Applications to finite nuclei}
\label{sec:applfinite}
\subsection{Static properties}
\label{sec:static}

Basic bulk properties such as ground state binding energies and radii
of magic and only a few semi-magic nuclei have been used as input to
the adjustment of the parameterizations. The quality achieved in
calculations of these properties in a broad range of nuclei is very
good, as discussed in Section \ref{sec:quality}. Thus experience shows
that SHF and RMF have a good interpolating power sin the valley of
stability but extrapolations are much less certain. It is the great
challenge for the further development of mean field models to
accommodate simultaneously more data in order to enhance the
predictive power into unknown regions such as nuclei appearing in
various processes of nucleosynthesis or super-heavy elements. One
option is to investigate more detailed ground-state observables as a
part of this effort. This will discussed in this Section. Going deeper
into details is a very critical test of mean field models and it often
turns out that one needs to go beyond a mere mean-field
description. This will be sketched briefly in Section \ref{sec:beyond}
dealing with correlations from soft modes.  Another option is to look
for excitation properties in a regime still accessible to mean-field
models. That will be addressed in Section \ref{sec:dynamic}.

\subsubsection{Trends of ground state binding energies}
\label{sec:etrends}

Ground state binding energies as such are very well described in
properly adjusted effective functionals, as discussed in Section
\ref{sec:baiscfeat}.  A closer look at Figure \ref{fig:scanERS_abs}
reveals that the remaining discrepancies between theory and experiment
are not statistically distributed but show systematic and as yet not
well understood trends. It is to be expected that energy $\textit
{differences}$ constitute a much more critical observable. In fact, we
will see in Section \ref{sec:beyond} that correlations can play a role
there. In the present Section, we concentrate on the basic mean field
effects. The observed variations will remain typical, even if
polarization and correlation are to be added later.
 
Energy differences give indirect access to underlying pairing and
shell structure. The odd-even staggering of binding energies carries
information on the pairing gap although care has to be taken to sort
out effects due to interferences with shape fluctuations \cite{Sat98a,Ben00c}.
Differences between even-even nuclei are related to the shell
structure at the Fermi energy. The first order difference yields the
two-nucleon separation energies,
$S_{\rm 2n}=E(Z,N)-E(Z,N-2)$ and $S_{\rm 2n}=E(Z,N)-E(Z-2,N)$.
Shell gaps are associated with jumps in the separation energies (as long as
shape fluctuations are absent, see Section \ref{sec:beyond}). These jumps
appear in a more pronounced manner as steep spikes in the second differences,
the two-nucleon shell gaps
\begin{equation}
  \delta_{\rm 2n}
  =
  E(Z,N+2)-2E(Z,N)+E(Z,N-2)
  \quad,\quad
  \delta_{\rm 2p}
  =
  E(Z+2,N)-2E(Z,N)+E(Z-2,N)
  \quad.
\end{equation}
These differences play a crucial role in identifying magic shell closures
\cite{Ben99a} and `can serve as indicators for waiting points in
the $r$-process \cite{pfe01}. However, they have to be viewed with care
for two reasons: first, the relation to the underlying single-particle
spectrum can be masked by shape effects, and second, they are a measure
of a single shell gap and not of the level density in a whole region
around the Fermi energy.  Thus a study of two-nucleon shell gaps
should be combined with complementary considerations of shapes and
level structure.  An example for shape fluctuations is discussed in
Section \ref{sec:beyond}.  The computation of level densities is
discussed in Section \ref{sec:she} with the example of shell
stabilization of super-heavy elements.

\begin{SCfigure}[0.5]
{\includegraphics[angle=0,width=11cm]{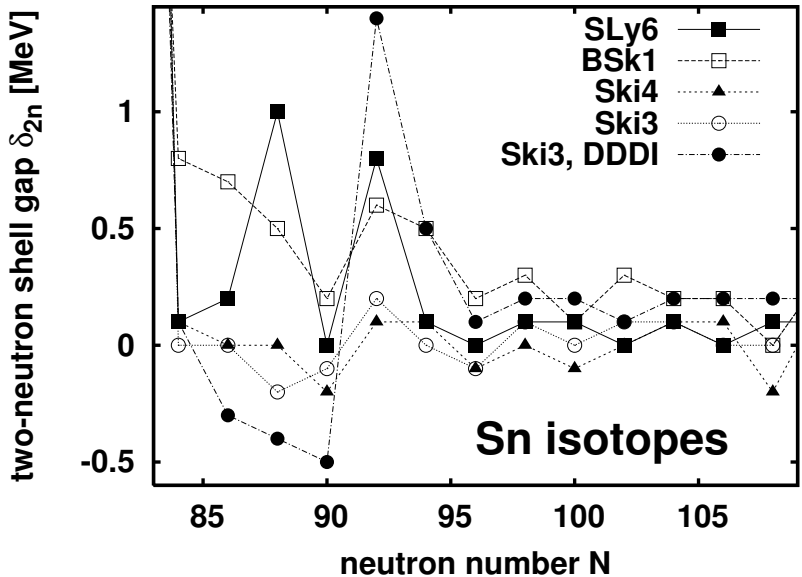}}
\caption{\label{fig:d2n_Z50}
The two-neutron shell gap $\delta_{\rm 2n}$ along the chain of neutron rich
exotic Sn isotopes computed in spherical mean field calculations for various
SHF parameterizations.  
}
\end{SCfigure}
Shell effects play a crucial role for estimating the abundance of elements as
produced in the $r$-process and traditional mean field models seem to call for
corrections \cite{Che95a}. Before going that far, one has first to explore the
predictive power or, more properly, the variance of existing models. Figure
\ref{fig:d2n_Z50} shows the two-neutron shell gaps  $\delta_{\rm 2n}$ 
along the neutron rich Sn isotopes for a variety of SHF parameterization. This
chain plays a crucial role in the $r$-process and $\delta_{\rm 2n}$ is a
useful indicator, particularly for differences between the parameterizations.
One realizes immediately that the different SHF parameterizations yield very
different predictions. Not only that, the variation of the pairing treatment adopted (see
Section \ref{sec:pairing}) has an equally dramatic effect. This example shows
that predictions of waiting points from mean-field model are still rather
vague. It is a great challenge to improve the predictive power of the
models. This calls in particular for a better control of shell structure as
well as of pairing. The strategy for improvement is to take  more and more
observables into consideration in fitting of the Skyrme parameters. Potential additions will be addressed
briefly in the following parts of the present Section.

\subsubsection{Spin-orbit effects}
\label{sec:spineffects}

\begin{SCfigure}[0.5]
{\includegraphics[angle=0,width=11cm]{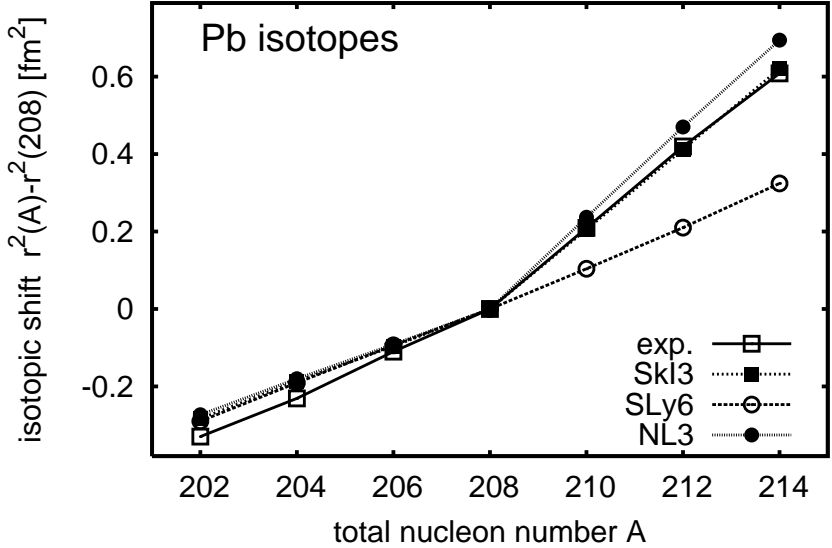}}
\caption{\label{fig:Pb_isoshifts_ls}
The isotopic shifts of charge r.m.s. radii, $r^2$($^{A}$Pb)-r$^2$($^{208}$Pb), for
the chain of Pb isotopes around double magic
$^{208}$Pb$_{82}^{126}$. Results for the SHF parameterizations SLy6,
SkI3, and the RMF parameterization NL3 are compared with experimental
data \cite{Ans86a}.  
}
\end{SCfigure}
New data from exotic nuclei reveal often hitherto unexplored features
of self-consistent mean-field models. One prominent example is
provided by the isotope shifts of charge radii in neutron rich Pb
isotopes \cite{Ans86a}. The trend shows a pronounced kink at doubly
magic $^{208}$Pb, see Figure \ref{fig:Pb_isoshifts_ls}. It was found that
SHF fails to reproduce this kink but continues uninterrupted with the
slope from the isotopes below $^{208}$Pb \cite{Taj93b}.  The example
of SLy6 in Figure \ref{fig:Pb_isoshifts_ls} is typical for all
conventional SHF parameterizations, i.e. those with fixed $B'_4=0$ in
the SHF functional (\ref{eq:coulfun}). All RMF parameterizations, on
the other hand, deliver the observed discontinuity \cite{Sha93b}, see
the example of NL3 in Figure \ref{fig:Pb_isoshifts_ls}.  A comparison
of the models in detail revealed the mechanisms underlying this
systematic difference \cite{Rei95a,Sha95a}. It was found that the
spin-orbit splittings of high lying neutron shells are substantially
larger in SHF than in RMF. As a consequence, the $2g_{9/2}$ neutron
level which is gradually filled when adding neutrons to $^{208}$Pb, is
more deeply bound and has a smaller radius in SHF. This is the level
which determines the outer tail of the density for $N>126$. Thus the
RMF produces larger neutron radii in isotopes above $^{208}$Pb than
SHF. 
The proton radii, in turn, are stretched due to the large nuclear
symmetry energy which is, in fact, particularly large in the RMF
(see Figure \ref{fig:nucmat_params}). And that is what produces in the
RMF results the kink above $^{208}$Pb, in agreement with the data. Now
the source of the difference between SHF and RMF is traced back to the
difference in the spin-orbit splitting. This difference, in turn, can
be associated with the different isovector structure of the spin-orbit
coupling in RMF and conventional SHF, see Section
\ref{sec:basicform}. The RMF corresponds to a spin-orbit functional
without isovector terms. for which $B_4\!+\!B_4^{\prime}\approx 0$
whereas the conventional SHF form has $B_4^{\prime}=0$. A minor
extension of the SHF to the full form (\ref{eq:coulfun}) allows to
reproduce the kink also with SHF. The Figure \ref{fig:Pb_isoshifts_ls}
shows the force SkI3 as example in which the relativistic variant
$B_4\!+\!B_4^{\prime}=0$ was chosen \cite{Rei95a}.
It seems that the isotopic shifts in the Pb region are the by far the
most sensitive observable for the isovector spin-orbit force. Many
other observables which are, in principle, sensitive to shell effects,
as e.g. level densities in super-heavy elements (see Sections
\ref{sec:she} and \ref{sec:fission}) did not yet show such specific
evidence.

\subsubsection{Neutron radii}
\label{sec:neutronradii}

\begin{SCfigure}
{\includegraphics[angle=0,width=11cm]{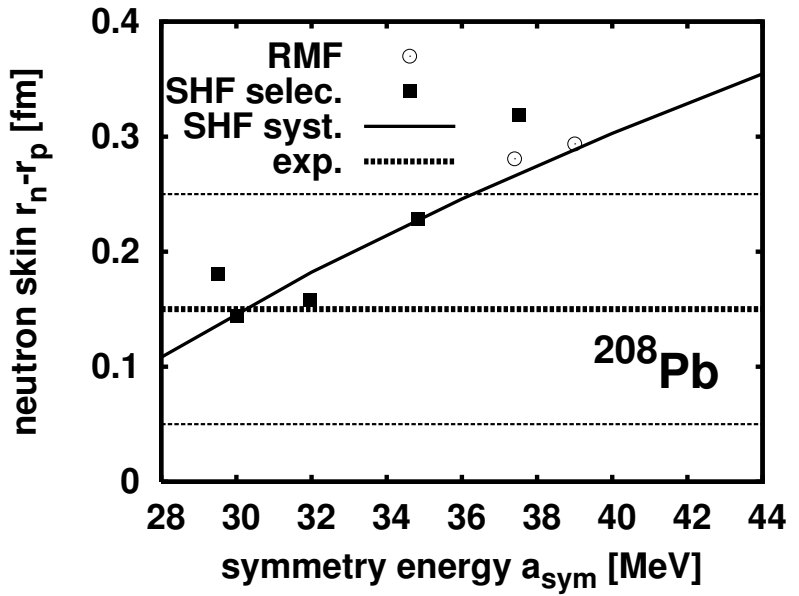}}
\caption{\label{fig:rneut_pb}
Neutron skin, $r_{\rm neut}\!-\!r_{\rm prot}$ for $^{208}$Pb for various
parameterizations, drawn versus the symmetry energy $a_{\rm
sym}$. Results for selected parameterizations
(RMF and `SHF selec.')  and the trend from systematically fitted SHF
forces (`SHF syst.') are shown. The experimental value is indicated
by heavy dotted horizontal line \cite{Cla03}. 
The assumed error on the data is indicated by faint dotted lines. 
}
\end{SCfigure}
Neutron radii provide additional very valuable information complementing
the rich pool of data from charge radii. Unfortunately, their
experimental determination is more model dependent than for charge
radii because the strong interaction is involved \cite{Bat89aER}.
%There is hope that a clean
%tool will be available soon from parity-violating electron nucleus
%scattering experiments \cite{Hor01b}.
%
More experimental information on neutron radii would improve models in several
respects. For example, there is a close connection between the
equation of state of neutron matter and the neutron r.m.s. radius of
$^{208}$Pb, see \cite{Bro00b} for the SHF and \cite{Typ01a} for the
RMF.
Other interesting phenomena related to the neutron density are the
neutron skins defined as the difference between neutron and proton
radii (or even neutron halos developing towards the drip lines
\cite{Miz00a}). One can establish a direct relation between isovector
forces and the neutron skin. A systematic variation of SHF
parameterizations with respect to all conceivable bulk properties has
shown that there exists a unique one-to-one relation between neutron
skin and symmetry energy $a_{\rm sym}$ \cite{Rei99a}.  This is
demonstrated in Figure \ref{fig:rneut_pb} for $^{208}$Pb. The symbols
show results for a variety of mean-field parameterizations. They line
up in a way which indicates a systematic increase of the skin
thickness with increasing $a_{\rm sym}$.  In order to check that trend
thoroughly, a set of SHF forces was fitted all to the same set of
ground state data but with an additional constraint on $a_{\rm sym}$
that was systematically varied \cite{Rei99a}. The result corroborates
the trend and shows an almost linear connection between neutron skin
and $a_{\rm sym}$. Such a unique correspondence can be established for
all nuclei. The trends are particularly pronounced in neutron rich
exotic nuclei. Furthermore, neutron radii have strong influence on the
reaction cross section of exotic nuclei. This indicates that a correct
symmetry energy and its density dependence, as discussed in Section
\ref{sec:infinite}, is crucial for understanding nucleosynthesis and
some other astrophysical phenomena dependent on reaction rates
involving neutron rich nuclei.

A different and complementing access to neutron radii has been brought
up in \cite{Vre03a} by exploiting a combined 
knowledge about the position of the  Gamow-Teller resonance and of the
isobaric analogue state.

 \subsubsection{Super-heavy elements}
\label{sec:she}

\begin{figure}
\centerline{\includegraphics[angle=0,width=17cm]{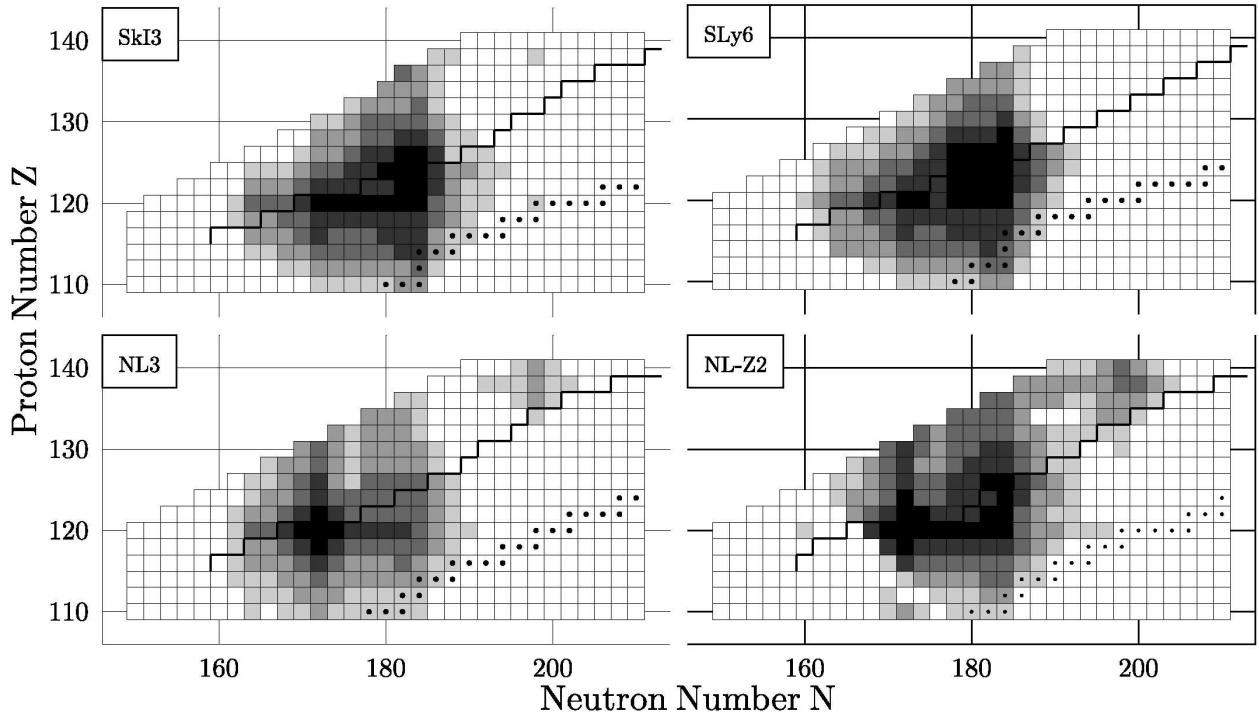}}
\caption{\label{fig:SHshells}
The shell correction energies (\ref{eq:shellcor}) of super-heavy
elements for various SHF and RMF parameterizations as indicated.
There are five gray values for five intervals of width 3 MeV.
The black boxed stand for corrections below -12 MeV, the next
gray value for -9 to -12 Mev and the lightest gray for 0 to -3 MeV.
(Adapted from \cite{Kru00a}.)  
}
\end{figure}
The key question in the study of super-heavy elements is their
stabilization, through shell effects, against spontaneous fission. A
first estimate of where stability may be found can be drawn from the
shell correction energy given approximately as
\begin{equation}
 E_{\rm shell}
 =
 \sum_\alpha w_\alpha\varepsilon_\alpha-\int
 d\varepsilon\,g(\varepsilon)
\label{eq:shellcor}
\end{equation}
where the $\varepsilon_\alpha$ are the single particle energies,
$w_\alpha$ the associated multiplicities, and $g(\varepsilon)$
represents a smooth energy distribution generated from the set
$\{\varepsilon_\alpha\}$ by an appropriate shell averaging procedure
which has to include continuum states in case of SHF spectra (for
details of the prescription see e.g. \cite{Kru00a}).
Figure~\ref{fig:SHshells} shows a summary of shell correction energies
for the landscape of super-heavy elements.
Note that the results are based on spherical mean-field calculations
which somewhat underestimates stabilization in case of deformed
nuclei. One finds broad islands of shell stabilization rather than
narrow and deep valleys as they are typically found for lighter
nuclei. This means that the notion of magic nucleon numbers fades away
for super-heavy nuclei. This is due to the fact that the level density
increases with $A^{-1/3}$ and with it the gaps in the
spectrum which are the key features for defining magic shells. Seen
from that perspective, it seems rather surprising that such large
shell correction energies still emerge. The reason is a separation of
states with low from those with high multiplicity. There arise broad
spectral regions which carry only a few low-multiplicity states. These
inhibit large spectral gaps but still allow for large shell
corrections in a broad range of isotopes.  The emergence of large
regions of stable nuclei is good news for the potential experimental
accessibility as it should not be necessary to hit precisely a single
longer-lived final isotope. Rather there is a broad choice of more
stable final isotopes in a heavy-ion fusion experiment which enhances
the chances for finding successful collision combinations to produce
them.
\begin{SCfigure}[0.5]
\caption{\label{fig:rho_120_180}
The proton (right) and neutron (left) density distributions
for the super-heavy element $Z$=120, $N$=180 for 
different Skyrme forces and one RMF parameterization as indicated.  }
{\includegraphics[angle=0,width=11cm]{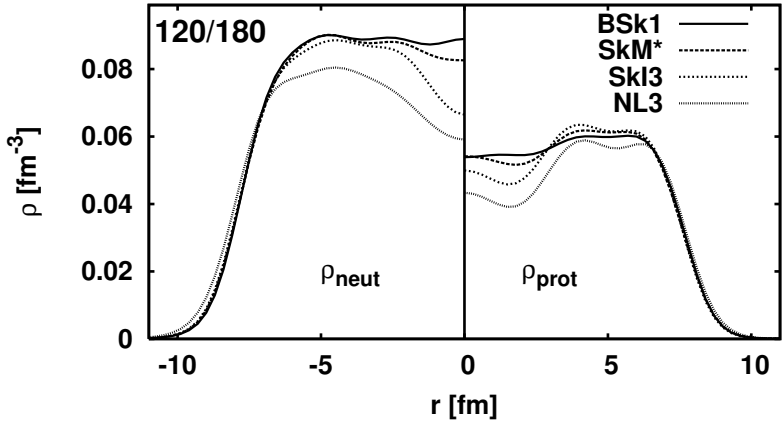}}
\end{SCfigure}
We are still a long way from assessing the super-heavy elements and
even farther away from measuring their detailed properties. Let us,
nonetheless, speculate a bit about their density distribution. Strong
Coulomb repulsion should reduce the proton density in the interior
producing a more or less deep dip as indicated in Figure
\ref{fig:rho_120_180} for the fictive nucleus
$^{300}$XX$_{120}^{180}$. where proton densities show indeed all some
dip at the center. However, this dip differs very much between the
different parameterizations. It is interesting to note that the
neutron distributions show the same succession of dip depths. The
depth increases with decreasing effective mass and BSk1, with
effective mass $\approx 1$, has practically a flat distribution. This
indicates that the variation of dips is a shell effect. Forces with
low effective mass tend to cut a hole in the nuclear center. For
proton densities, the situation is less clear with the simple Coulomb
effect probably well seen for BSk1 and amounting to about 10\%
reduction at the center.  Larger reductions are induced by shell
effects. The results of figure \ref{fig:rho_120_180} are typical for
region of nuclei with A$\sim$ 300 with similar effects predicted and
discussed for $^{292}$XX$_{120}^{172}$ in \cite{Ben99a}. Taking the
predictions bit further, the next region of shell stabilization,
around $^{480}$XX$_{160}^{320}$ \cite{Ben01a}, opens the possibility
of bubble nuclei which are almost proton-empty at the center.

\subsubsection{Fission barriers}
\label{sec:fission}

Spontaneous fission becomes a crucial decay channel for actinides and
super-heavy nuclei. There exists a wealth of information about fission
in actinide nuclei, for a review see \cite{Spe74aER}. The most
remarkable feature is the double-humped barrier. The ground state is
usually prolate deformed with a typical deformation of $\beta_2\approx
0.3$.  As deformation increases, the energy passes a first (inner)
barrier which leads to an elongated fission isomer at around
$\beta_2\approx 0.9$ and finally goes to fission through a second
(outer) barrier. The first barrier explores triaxial deformation which
lowers the barrier by about 3 MeV as compared to the axial value. The
second barrier explores reflection-asymmetric shapes.
%The potential landscapes of adjacent
%actinide nuclei are similar, though for some nuclide there might
%appear a second isomeric state \cite{Zha86aE}.
The double-humped fission barrier of $^{240}${Pu} has often served as
a benchmark for mean-field models, see Ref.\ \cite{Flo74a} for results
obtained using Skyrme interactions, \cite{Ber84a} using Gogny forces
and \cite{Blu94a,Rut95a} for the RMF, see also \cite{Ben00b}.
Collective correlations (see Section \ref{sec:beyond}) modify the
fission path significantly \cite{Rei76a,Rei87aR}. The typical
corrections are a lowering by 0.5--1 MeV of the first barrier and
about by 2 MeV for the second. The triaxial shape lowers the first
barrier by another 1--2 MeV such that both (axial) barriers are
subject to correction by about 2--3 MeV.

\begin{SCfigure}[0.5]
{\includegraphics[angle=0,width=11cm]{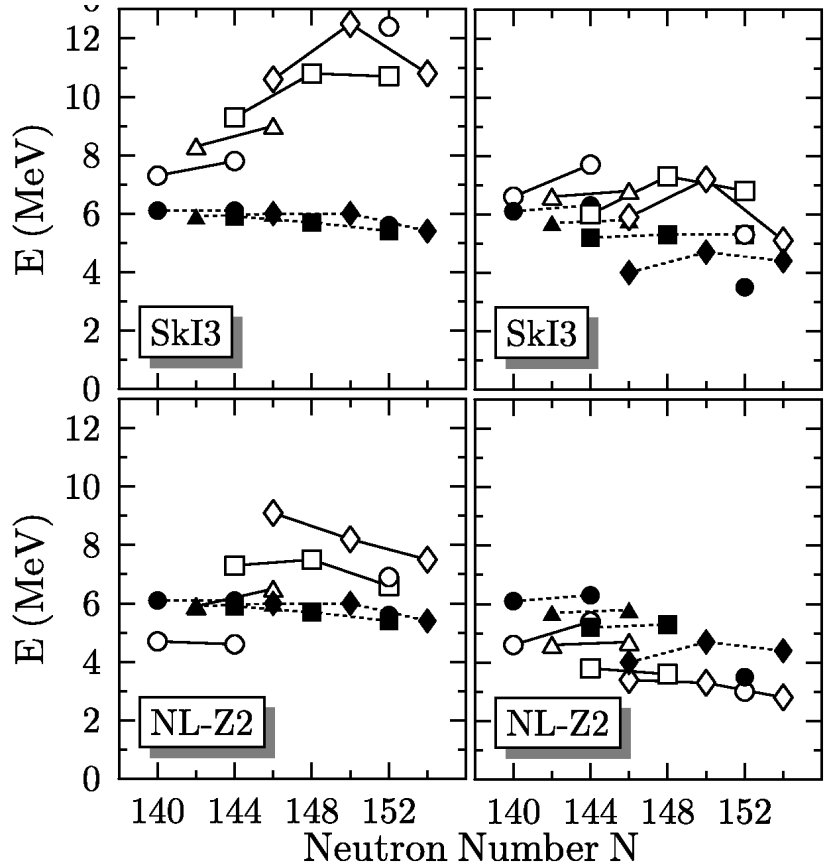}}
\caption{\label{fig:actinides_barrier}
The first (left panels) and second (right panels) fission barriers from 
axial and reflection-symmetric calculations with SkI3 or NL-Z2. 
Th (\mbox{$Z=90$}), U (\mbox{$Z=92$}), Pu (\mbox{$Z=94$}), 
Cu (\mbox{$Z=96$}), and Cf (\mbox{$Z=98$}) isotopes are denoted by 
open circles (for \mbox{$N=140$}, 142), open triangles, open squares, 
open diamonds, and again open circles (\mbox{$N=152$}), respectively. 
Experimental data (full symbols) are taken from \protect\cite{Mam98a}.
Data points for the same element are connected by lines.
}
\end{SCfigure}
Figure \ref{fig:actinides_barrier} compares calculated (SHF and RMF)
and experimental heights of the barriers. One has to take into account
that the theoretical values are deduced from axially symmetric shapes
without triaxial and correlation corrections, however several
conclusions can be safely drawn.  At first glance, we see that both
models yield the correct order of magnitude for the barriers, which is
not a trivial achievement in view of the various counteracting effects
(shell effects versus collective trends from symmetry and surface
energy). On closer inspection, we see that the RMF yields
systematically lower barriers than SHF. Both are generic features
which persist when comparing a broader selection of parameterizations
\cite{Bue04a}. Accounting for typical reductions of about 2--3 MeV,
the RMF tends to underestimate fission stability. The SHF looks more
successful for the first barrier but also predicts second barriers
somewhat too low.

\begin{figure}
\centerline{\includegraphics[angle=0,width=14cm]{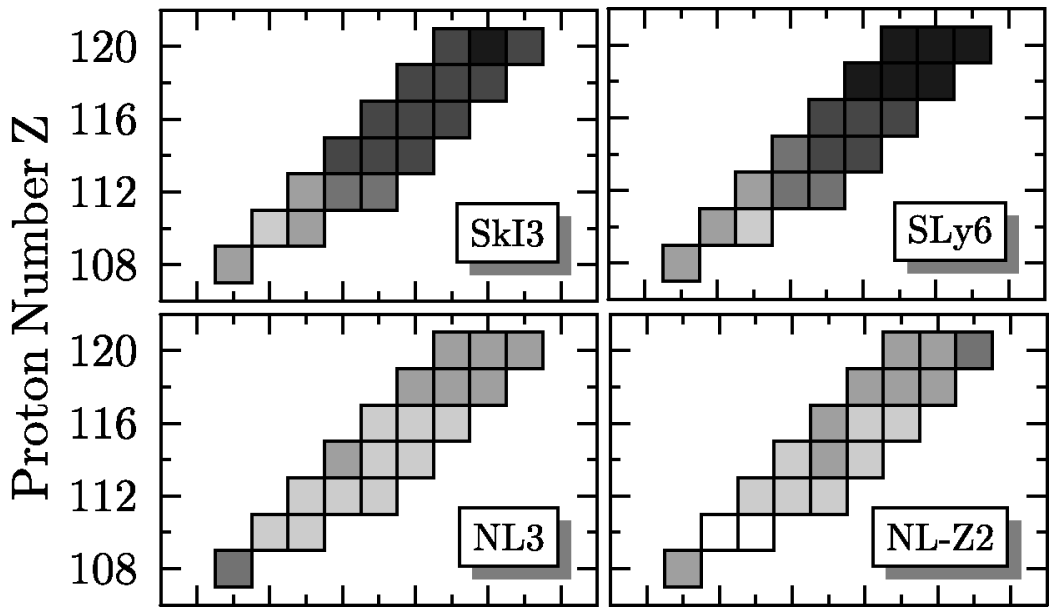}}
\caption{\label{fig:allforces_barrier_review}
Fission barriers of super-heavy
elements for various SHF and RMF parameterizations.
The grey scale proceeds in bins of 2 MeV width. The black boxes indicate
barriers higher than 12 MeV and the lightest grey denotes barriers between
2--4 MeV.  
(Adapted from \cite{Bue04a}.)  
}
\end{figure}
The fission landscape of super-heavy nuclei differs from that in the
actinides through the disappearance of the fission isomer and the second
barrier. On the prolate side, only the first barrier needs to be
considered. For the heaviest nuclei there may be an alternative
fission  channel which goes through oblate and then triaxial shapes
\cite{Ben04d}. We confine the discussion here to the axially symmetric
first barrier at the prolate side.
The systematics of barriers for the landscape of super-heavy elements
and for two mean field models is shown in Figure
\ref{fig:allforces_barrier_review}. The results are typical for a
broader variety of forces \cite{Bue04a}.  All models and forces agree
that there is a regime of low fission barriers around \mbox{$Z=110$}
and that the (axial) barriers increase again when going towards
\mbox{$N=184$}. There are again significant differences in that RMF
yields systematically lower barriers than SHF. Taking into account a
possible lowering by triaxiality and correlations, we conclude that
the RMF predicts fission instability practically everywhere in that
landscape. The SHF provides a much more optimistic outlook for
ultra-heavy element stability, qualitatively in accordance with
experiments which indeed managed to identify nuclei up to $Z=116$
\cite{Oga99aE,Oga99cE,Oga01aE}. The reason for this systematic
difference has not yet been identified uniquely. It comes probably
from several sources, amongst them the much higher symmetry energy in
the RMF and the somewhat different shell structure.

\subsection{Dynamic properties}
\label{sec:dynamic}

Self-consistent mean-field models allow also dynamical applications.
From the given energy functional, one can equally well derive
time-dependent mean-field equations. The SHF functional \ref{eq:enfun}
provides already the time-odd couplings involving current ${\bf j}$
and spin-density $\bsigma$ which are now activated in dynamical
situations with non-vanishing net flow. The dynamic extension of SHF
is often called time-dependent Hartree-Fock (TDHF) and it has enjoyed
great attention three decades ago as a tool for analyzing the
principle mechanisms of heavy-ion collisions, for reviews see
\cite{Neg82aR,Dav85a}. But TDHF covers much more. Many excitation
properties can be derived from TDHF. An important class of excitations
deals with small amplitude motion, i.e. in the limit of harmonic
oscillations. This applies to the basic nuclear resonance modes (e.g.,
giant resonances or Gamow-Teller resonances), in fact, any excitation
in the spectral range from 2--30 MeV. The small amplitude limit of
TDHF yields the much celebrated random-phase approximation
(RPA). Examples in that regime will be discussed in Sections
\ref{sec:giantres} and \ref{sec:gamow-teller}. Low-energy excitations
are associated with soft modes (quadrupole surface oscillations,
rotation, c.m. motion) and explore large amplitudes. These modes
proceed very slowly and can be dealt with in various adiabatic
approximations. Examples will be discussed in Section \ref{sec:rotat}
and \ref{sec:lampl}, and the example of fission was already sketched
in Section \ref{sec:fission}. And there is finally the regime of
heavy-ion collision which employs large amplitudes and is not
necessarily slow. This will be discussed in Section
\ref{sec:collisions}. In all cases, one should consider, in principle,
an inclusion of pairing in a TDHF with BCS states (TDHFB).  This is
rarely done at the level of full TDHFB. It is a standard choice in the
adiabatic situations. The small amplitude limit of TDHFB yields the
quasi-particle RPA (QRPA) which is the method of choice for computing
excitations in  non-magic nuclei.

\subsubsection{Giant resonances}
\label{sec:giantres}

Prominent features of the nuclear excitation spectrum are the giant
resonances of which the most striking is the isovector dipole
resonance. The isovector dipole couples strongly to photons and thus
provides the largest photo-reaction cross sections. A correct
description of these is crucial for estimating astrophysical reaction
chains. The isovector dipole excitations can be described
within a mean field theory utilizing a time-dependent mean field in
the limit of small amplitudes, often called the random-phase
approximation (RPA). There exists a large body of RPA
calculations in the context of self-consistent models, for SHF see 
recent compilations \cite{Col95a,Rei99a}, for Gogny forces
\cite{Bla76a,Bla77a}, and for the RMF \cite{Ma01a,Rin01b}.
The dipole spectra in heavy nuclei show one prominent resonance peak
with some small broadening due to neutron escape, Landau damping and
two-body correlations \cite{Ber83aR,Rei86c,Mar05a}. This peak is 
well described by most of the existing mean-field parameterizations.
A small part of dipole strength is found at lower energies giving rise
to what is called a pygmy resonance \cite{Vre01a}. The pygmy peak is
sensitive to shell structure and depends strongly on the particular mean field
model. The spectral fragmentation increases towards lighter nuclei and
when going towards the drip lines. 
\begin{SCfigure}[0.5]
\caption{\label{fig:gdr_140Sn}
The photo-absorption strength for $^{140}$Sn computed with RPA using
four different Skyrme parameterizations as indicated. The
self-consistent RPA techniques from \cite{Rei92a,Rei92b} have been
used. The spectra are folded with a Gaussian width of
$\Gamma=(E_{\rm n}-\varepsilon_{F,\rm neutr})*0.2$ MeV to simulate
escape and collisional broadening.
}
{\includegraphics[angle=0,width=11cm]{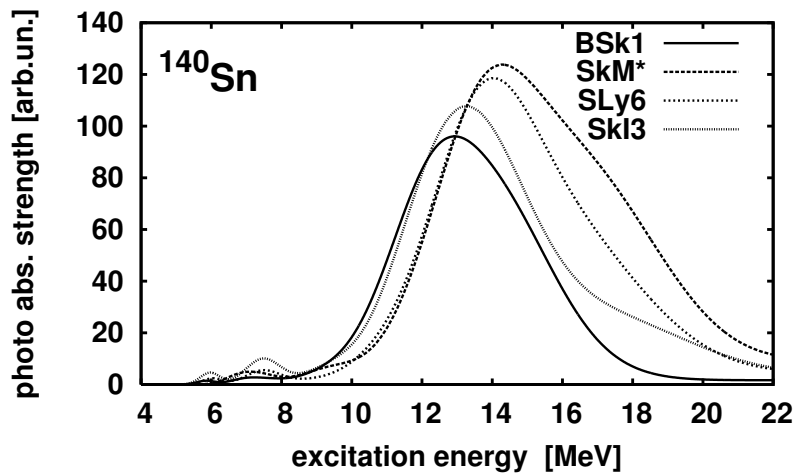}}
\end{SCfigure}

\begin{SCfigure}[0.5]
\caption{\label{fig:GDRomeg_asym}
Upper panel: Energy of the giant dipole resonance peak in $^{208}$Pb
as function of the symmetry energy $a_{\rm sym}$. The resonances
were computed with a set of Skyrme forces where
$a_{\rm sym}$ was systematically varied \cite{Rei99a}.
Lower panel: The slope of density dependence for symmetry energy.
  }
{\includegraphics[angle=0,width=11cm]{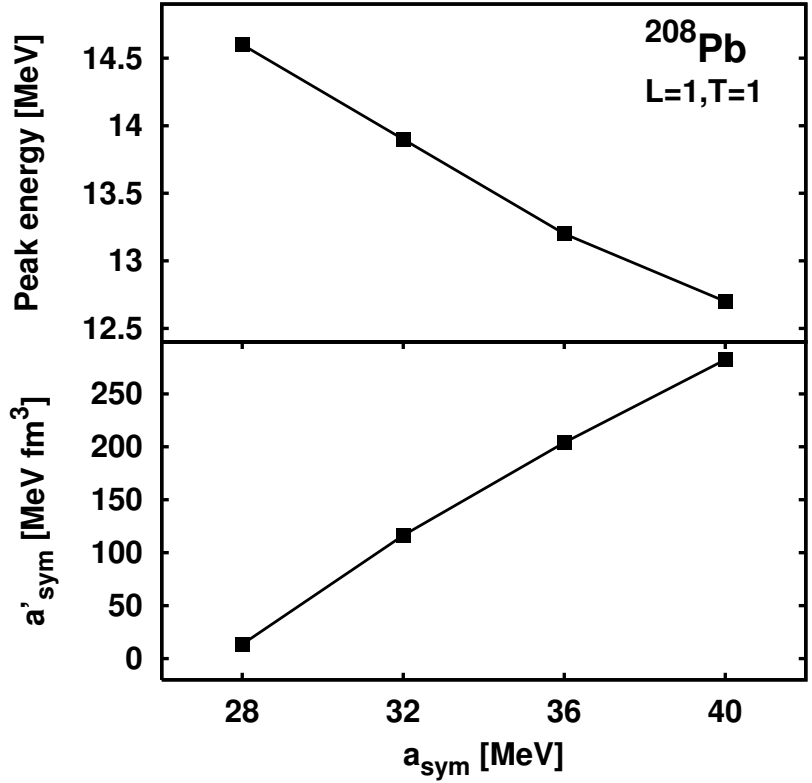}}
\end{SCfigure}
The giant dipole resonance in $^{208}$Pb shows one clear, although broad, peak
and can be well characterized by the peak position. It is interesting to see
how this peak position depends on the bulk properties of the model. One
expects naturally a strong relation to the symmetry energy $a_{\rm sym}$ which
is associated with a given parameterization (see Section \ref{sec:keyprop}).
For such studies, we have developed a set of SHF parameterizations where one
bulk parameter was fixed in addition to the elsewise same fitting to gross
properties of a selection of finite nuclei \cite{Rei99a}. This systematics
had already been used for the discussion of neutron radii in Figure
\ref{fig:rneut_pb}. It is now applied to a study of the giant dipole
resonance. Figure \ref{fig:GDRomeg_asym} shows the trend of the peak energy of
the resonance in $^{208}$Pb with varied symmetry energy. The result is at
first glance surprising. The energy decreases with increasing $a_{\rm sym}$.
One would have expected the opposite trend because $a_{\rm sym}$ represents
the spring constant for isovector motion. Note, however, that $a_{\rm sym}$
represents a feature at bulk equilibrium density. The actual nuclear density
varies about that value and it significantly lower than that in the nuclear
surface where the transition density ($\propto\nabla\rho$ \cite{Rei85a}) of
the resonance mode has its maximum. We have to consider the density
dependence of $a_{\rm sym}$.  It can be characterized by the slope at bulk
equilibrium, i.e.
\begin{equation}
  a'_{\rm sym}
  =
  \frac{d}{dn}a_{\rm sym}\Big|_{n=n_0}
  \quad.
\end{equation}
The trend of $a'_{\rm sym}$ is shown in the lower panel of Figure
\ref{fig:GDRomeg_asym}. The slope is basically positive and it increases
dramatically with increasing $a_{\rm sym}$.  The subsequent decrease of the
effective symmetry energy with decreasing density more than counterweights the
increase of the bulk-equilibrium value. Thus we find at the end this curious
counter-trend observed in the upper panel of Figure \ref{fig:GDRomeg_asym}.
As an example, we show in Figure \ref{fig:gdr_140Sn} the predicted
photo-absorption spectra for the neutron rich exotic nucleus
$^{140}$Sn. Its neutron emission threshold $\varepsilon_{F,\rm neutr}$
is about 4 MeV. The spectra show the dominant giant dipole
resonance with several well separated pygmy resonances a few MeV above
threshold. The four chosen Skyrme parameterization all reproduce
stable ground state properties well and yield
comparable giant dipole resonance peaks in $^{208}$Pb.  However the
predictions in $^{140}$Sn differ visibly, particularly what
the pygmy strength is concerned.  The proper description of these
details in fragmented spectra and for low-lying dipole strength
remains a great challenge to mean-field models and requires further
thorough exploration.

\begin{SCfigure}[0.5]
\caption{\label{fig:GDR_trends_comp}
The average peak position of the giant dipole resonance in a broad
range of spherical nuclei drawn versus $A^{-1/3}$ (which is
proportional to the inverse radius). Results are shown for two
different Skyrme forces as indicated and compared with experimental
data where available.  }
{\includegraphics[angle=0,width=11cm]{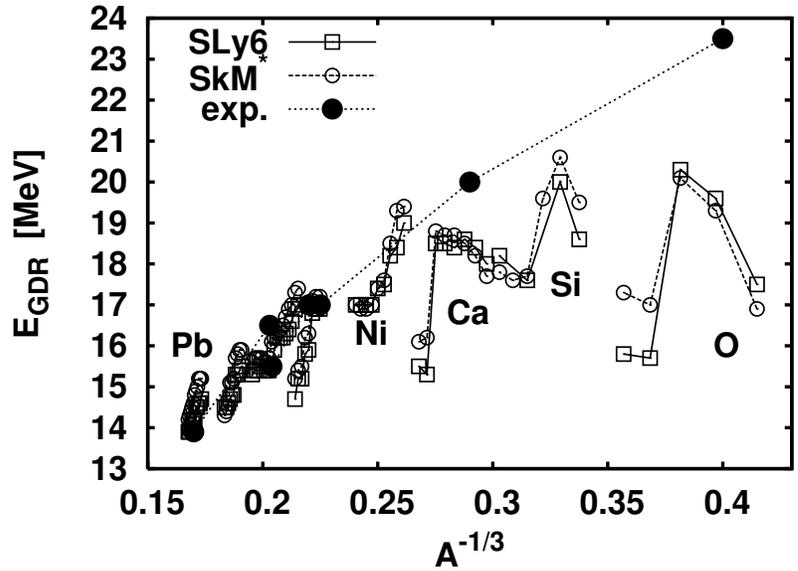}}
\end{SCfigure}
Despite the good agreement with experiment in heavy nuclei, a most
puzzling feature appears for the average peak position in the giant
resonance region on moving to light nuclei. The average resonance
energy is underestimated dramatically and systematically for light
nuclei, such as e.g. $^{16}$O and as yet no SHF parameterization can
describe adequately the resonance both in $^{208}$Pb and $^{16}$O
\cite{Rei99a}.
As a first step towards an understanding of this problem, we
illustrate in Fig.~\ref{fig:GDR_trends_comp} the trend of the average
resonance frequency over a broad range of nuclei from $^{16}$O to
$^{208}$Pb. Results for two conventional SHF forces, typical for any
other force investigated so far, are shown. They are drawn versus
$A^{-1/3}$ which helps to spot trends.  The Goldhaber-Teller model
predicts an $A^{-1/3}$ law while the Steinwedel-Jensen model indicates
an $A^{-1/6}$ dependence \cite{Bra83a} while experiment is closer to
the $A^{-1/6}$ low. The SHF theoretical predictions deviate from both
these estimates and find a constant resonance energy for the light
nuclei (large $A^{-1/3}$). The clear deviation for $Z<28$ nuclei is
presently unexplained, thus SHF predictions can only be considered
acceptable for $Z\ge28$.

So far, we have discussed the spectra of small-amplitude modes.  The
lowest excitations in nuclei are often related to large amplitudes,
e.g. rotations and soft quadrupole vibrations. Time-dependent
mean-field theory is also capable to describe these large amplitude
modes. The case of heavy-ion collisions is discussed in the now
following Section. Low-energy spectra require some additional steps
for proper re-quantization of the semi-classical mean-field dynamics.
This will be discussed in Section \ref{sec:beyond}, particularly in
the part \ref{sec:lowQ}.

\subsubsection{Gamow-Teller resonances}
\label{sec:gamow-teller}

Giant resonances (see Section \ref{sec:giantres}) and low-energy
quadrupole modes (see Section \ref{sec:lowQ}) access mainly the flow
part of the time-odd functional in subequation
(\ref{eq:enfunsk}) of the Skyrme functional.  This part is basically
fixed by Galileian invariance. Excitations with unnatural parity
($\Pi=(-1)^{L+1}$) probe the other time-odd parts in subequation
(\ref{eq:enfunskodd}) which are not yet so well fixed. Thus unnatural
parity states carry a lot of new information.
In this family, the dominant isoscalar excitations have magnetic dipole ($M1$) 
structure. Orbital $M1$ strength is related to the nuclear 
scissors mode which has been observed by \cite{Boh96aE}, 
see \cite{LoI97aR} for a review.
The most prominent isovector mode is the Gamow-Teller (GT) resonance,
for a review see \cite{Ost92aR}. The spectrum in the GT channel plays
a role in computing the probabilities for $\beta$-decay which, in
turn, is a crucial ingredient in the reaction chains of the
$r$-process. Furthermore, the energy distribution of GT strength is a
vital information needed for calculation of electron capture rates at
densities and temperatures appropriate for core-collapse supernovae
models. Existing shell model calculations (\cite{lan03,hix03} and
references therein) are not applicable for all needed neutron-heavy
nuclei and extrapolation techniques have to be applied. 
Self-consistent SHF models are not subject to this limitation
and may prove to be a versatile tool in this context. Thus a proper
modeling of these modes is a desirable feature in a self-consistent
description.

While many time-odd terms contribute to isoscalar $M1$ excitations, GT
excitations are determined exclusively by the time-odd isovector
spin-spin part of the effective interaction, see the terms $\propto
C'_0,C'_2$ in eq. (\ref{eq:enfunskodd}). These terms are only
loosely fixed in usual parameterizations at present. Studies of GT
resonances may supply useful data for the further development of the
forces. Fully self-consistent calculations in the SHF framework are
still rare, for a thorough discussion of previous work see
\cite{Ben02b}.  The volume properties of the residual interaction in
the GT channel are characterized by the Landau parameter $g'_0$
\cite{Rin80aB}.  The lack of predictive power becomes obvious from the
fact that different SHF parameterizations shows a large spread just
for this parameter. The $g'_0$ becomes, e.g., for the present
selection of forces:
SkM* $\leftrightarrow  0.31$,
SkP  $\leftrightarrow  0.06$,
SLy6 $\leftrightarrow -0.04$,
SkI3 $\leftrightarrow  0.20$,
SkI4 $\leftrightarrow  1.38$,
BSk1 $\leftrightarrow  0.22$.
Note that negative Landau parameters signify an instability of the
ground state, for $g'_0$ a spin-isospin instability.  It is to be
reminded, however, that the volume parameters serve only as a first
guidance. The actual situation in finite nuclei is additionally much
influenced by the surface terms, in the GT channel then by the $C'_2$
term in eq. (\ref{eq:enfunskodd}). There exist already a few attempts
to look for more suitable forces. The force SGII has been tailored for
this particular purpose \cite{Gia81a} and a Skyrme force SkO' was
found to perform fairly well in the GT channel \cite{Eng99a}.  Large
scale systematic studies and a proper adjustment of these features
are necessary and have still to come.
There exist also recent studies of GT states within the RMF, see e.g.
\cite{Paa05b}.
The advantage is here that RMF has less uncertainties in the spin
channel. It seems, however, that the modeling of the density-dependent
couplings leave some freedom yet to be fixed.

\subsubsection{Heavy ion collisions}
\label{sec:collisions}

As mentioned above, fully fledged TDHF with Skyrme forces helped a lot
to eludicate the principles of heavy-ion collisions
\cite{Neg82aR,Dav85a}. It gave information on basic properties such as
as the bifurcation between fusion and inelastic scattering at a
certain critical impact parameter and the typical pattern of a
Wilczynski plot (double differential cross section with respect to
scattering angle and energy). The early studies were all hampered by
restrictions and approximations due to numerical limitations.  One of
the open problems was that too little dissipation emerged from these
calculations thus underestimating fusion cross sections and energy
loss in inelastic collisions. A large part of dissipation could be
regained including the spin-orbit force properly
\cite{Lee87a,Rei88a}. Explicit dissipation mechanisms through dynamic
two-body correlations beyond mean-field have been intensively
investigated, as reviewed in \cite{Abe96aR}.  But the full degrees of
freedom of a pure TDHF description had not been explored so far. Only
recently has ever-increasing computing power allowed fully fledged
TDHF calculations without compromises to be explored. Glimpses of a
new generation of TDHF studies are emerging, offering in assessing
more precisely the borderline between mean field and correlation
effects. \cite{Mar05a,Uma06a}.

\subsubsection{Rotational bands}
\label{sec:rotat}

Nuclear spectra exhibit extended rotational bands which disclose a
variety of interesting phenomena, such as the breakdown of pairing due
to the coriolis force, or the phenomenon of backbending caused by the
crossing of two rotational bands with very different moment of
inertia.  It is no surprise then that one can find a large body of
literature on that topic, for some experimental reviews see
\cite{Jan91aER,Bak95aER,Han99aER}.

 Rotation is a
large amplitude motion for which TDHF is the appropriate
tool. However, rotational symmetry allows to simplification of the
treatment.  A transformation from the intrinsic to the laboratory
frame shows that the rotating wavefunction is correctly described by
constrained Hartree-Fock (CHF) calculation in which the mean-field
Hamiltonian $\hat{h}$ is augmented by a constraint for each angular
momentum $\hat{J}_i$, i.e.
\begin{equation}
  \hat{h}
  \;\longleftrightarrow\;
  \hat{h}-\omega_i\hat{J}_i
\end{equation}
where $\omega_i$ is the Lagrange parameter (corresponding to the
associated rotational frequency) and $\hat{J}_i$ is a component of the
total angular momentum orthogonal to one of the principle axes of the
deformed nucleus. One considers usually axially symmetric nuclei and
the rotation axis has then to be chosen orthogonal to the symmetry
axis. This then yields unavoidably a fully triaxial problem.
Moreover, time-reversal symmetry is broken by the rotation which
complicates the treatment of pairing. Although conceptually
straightforward, the technical side of these CHF calculations for
rotational bands is extremely involved, for details see e.g.
\cite{Rin80aB}. Nonetheless, the high interest in rotational dynamics
has motivated several groups to admirable efforts in that direction
leading subsequently to the large body literature as mentioned above.
A detailed report would go far beyond the limits of the present review.

Mean-field theories are well suited to deal with that phenomenon since
the high quantum number involved validate the semi-classical limit
which is implied in TDHF and self-consistency eliminates the need for
further free parameters.  SHF was employed in early studies on
rotating nuclei, see e.g. \cite{Fle79a}, and since then a widespread
literature about rotational bands on the grounds of mean-field models,
mainly SHF emerged, for reviews see e.g. \cite{Afa99aR,Fra01aR}. As
discussed above, the Skyrme functional is somewhat uncertain in the
time-odd domain, particularly concerning the spin couplings. This
carries over to the rotation studies, because rotation breaks
time-reversal symmetry and thus accesses the time-odd components of a
model. The RMF method, is thought to be better suited to analyzing
spin properties since they are automatically included within the lower
components of the Dirac wavefunctions, has been used also, however the
computations are much more involved than in SHF, see for example
\cite{Afa96a}.

%\subsection{Nucleosynthesis}
%\label{sec:nuclo}
%\subsubsection{The r-process} 
%\label{sec:rprocess}
%\subsubsection{The rp process}
%\label{sec:rpprocess}
% this part --> JS
\subsection{Nucleosynthesis}
\label{sec:nuclo}
It is the present understanding that all elements in nature, apart from $^1$H, $^2$H, $^3$He, $^4$He and $^7$Li which originated from the Big Bang \cite{thi01}, are made during stellar evolution and stellar explosions. The main influence of nuclear physics on element synthesis is given in calculation and predictions of cross sections of nuclear reactions and decay rates in the stellar environment. This environment is not stable but changes on a timescale much longer than the typical time scale of the strong and week interactions, dominating nuclear transmutation. The environment, characterized by density, temperature and properties of nuclei present in the medium is crucial for type, rate and products of nucleosynthesis. This Section reviews the role of the Skyrme interaction in simulations of the two major ways of creation of heavy elements, the r- and rp- processes.
\subsubsection{The r-process} 
\label{sec:rprocess}
 Very neutron rich environments at high temperature permit neutron captures, much faster than competing $\beta$-decay, to occur up to the neutron drip-line and are sites of the rapid neutron capture (r-process) which is thought to be responsible for creating about half of all nuclei with A$\ge$70. Observational data on elemental and isotopic solar abundances of nuclei (see \cite{kra93} and references therein) serve as a stringent test of the r-process models concerning both the characteristics of the stellar environment and the nuclear physics input. Details of the r-process models are beyond the scope of this review and are given elsewhere \cite{kra93,pfe01,thi01}. Here we only point out that the current understanding is that in the high temperature and high density environment, leading to the r-process, chemical equilibrium between neutron capture and the reverse photodisintegration develops within each isotopic chain (for a given Z) and the flow along the r-process path is governed by $\beta$-decay from one isotopic chain to another. It follows that in equilibrium conditions the  r-process path is uniquely determined by the neutron number density $n_{\rm n}$, temperature $T$ and neutron separation energy  $S_{\rm n}(n_{\rm n},T)$ (Q-value) \cite{pfe01} and cross sections of individual capture processes are not important. The process is slowed down when it reaches nuclei with magic number of neutrons (N=50, 82 and 126) and $\beta$-decay becomes comparable or faster than neutron capture. Then a series of $\beta$-decays and (n,$\gamma$) processes takes place each changing Z and A by one \cite{kra88}. The nuclei in this series are called the waiting point nuclei. As these nuclei approach stability, $\beta$-decay lifetimes get longer and the process gradually breaks off and proceeds with more favored neutron capture (competing at some cases with $\beta$-delayed neutron emission) towards the next neutron closed shell. 
   
The nuclear masses (in particular neutron separation energies $S_{\rm n}$) and gross $\beta$-decay properties (half-lives and $\beta$-delayed neutron emission probabilities)  along the contour line of constant $S_{\rm n}$ are the key nuclear inputs into r-process modeling. Another important piece of information comes from investigations of shell structure far from stability, in particular the strength of shell gaps which affect r-process abundances in the vicinity of magic numbers.  At the endpoint of the r-process path when fissionable nuclei are reached, $\beta$-delayed fission becomes important. The fission fragments may alter heavy element abundances and also increase general r-process abundances \cite{thi01}. The important nuclear physics input to modeling of this final phase of the r-process includes fission probabilities  of nuclei at highly excited states and precise calculation of fission barriers. 

\begin{figure}
\centerline {\includegraphics[angle=90,width=15cm]{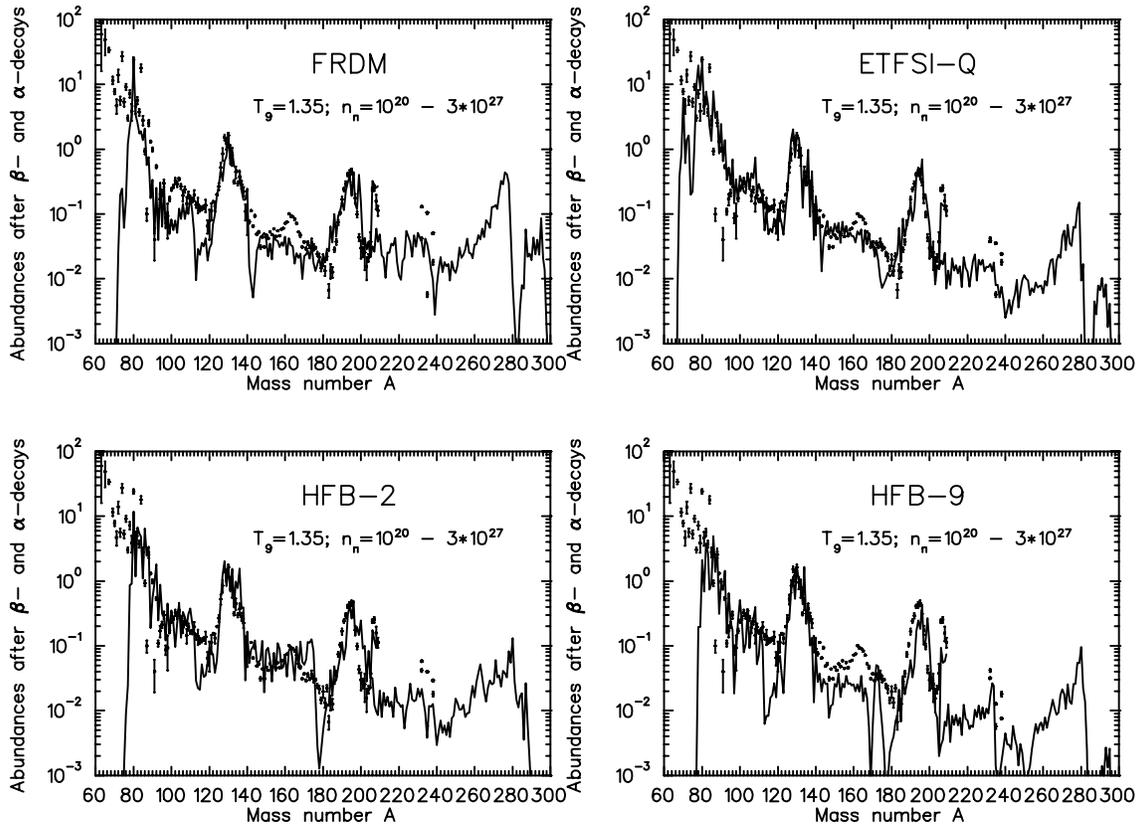}}
\caption{\label{fig:abund} Observational solar abundance distribution N$_{\rm r,\odot}$ (dots) in comparison with calculation \protect\cite{krpf06} N$_{\rm r,calc}$ (full line), using four different mass models, FRDM \protect\cite{Mol97b}, EFSI-Q \protect\cite{pea96}, HFB-2 \protect\cite{gor01} and HFB-9 \protect\cite{gor05}. Except for the $S_{\rm n}$ energies, all other nuclear physics and astrophysics parameters we kept the same in the calculations.
}
\end{figure}

Calculation of nuclear masses has been of major interest for decades. The major contribution to this field has been achieved in the Finite Range Droplet Model (FRDM) of Moller et al. \cite{Mol97b}. An extensive attempt to employ microscopic models using the Skyrme interaction in the Extended Thomas-Fermi Strutinsky Integral \cite{abo95,pea96}, HF+BCS \cite{ton00} and HFB \cite{gor01,sam04,gor05} mean-field models in precise ground state mass calculations has been made in recent years. A number of new Skyrme parameterizations have been developed for this purpose (SkSC, MSk and BSk families) by optimizing fitting to all known mass-data only, not taking into account any other experimental constraints. Examination \cite{sto05} of the results  that these microscopic calculations did not improve on the best r.m.s error of 0.633 MeV achieved by FRDM in fitting the latest table of experimental masses \cite{aud95} and did not show an improved predictive power, especially in the region of neutron-heavy nuclei. We display results of calculations of the r-abundance distribution N$_{\rm r,calc}$ using two representative HFB models, the FRDM model and ETFSI-Q model (with SkSC4 Skyrme parameterization) in Fig.~\ref{fig:abund} in comparison with observational data N$_{\rm r,\odot}$. It is seen that in particular the HFB-9 predictions  give serious differences from observational data, especially in the A=120 and A=140-180 regions. 
It was concluded \cite{sto05} that is very unlikely that the present HF mass models will ever yield atomic masses with the precision required by the r-process and related applications, unless a very different approach is explored. It is important to realize that not only atomic masses but derived observables, dependent on mass {\it differences}, like one- and two-nucleon separation energies and Q$_\beta$ values are also required. Very careful exploration of the real sensitivity of the mean field Hamiltonian to particular aspects of available experimental information on nuclei is needed. Then the most sensitive data should be fitted together to find truly optimized Skyrme parameter sets. Furthermore, it is clearly desirable that the correlations between parameters of the Skyrme functional, individually or in groups, are systematical examined and their role unambiguously understood. This may provide a very powerful and decisive constraint in the search for meaningful Skyrme functionals.

Gross $\beta$-decay properties have been treated in a variety of models (see, e.g. \cite{kra06}) that range from phenomenological to large scale microscopic calculations. At this point all models have some deficiency and it is not yet possible to calculate the whole r-process. The limitations are mainly in restriction to spherical shapes, Gamow-Teller (GT)-transitions only, or too small shell model spaces. One of the key issues is as-correct-as-possible treatment of nucleon-nucleon correlations in a model. The only fully self-consistent HFB+QRPA calculation of  $\beta$ decay rates of the r-process waiting-point nuclei utilized the Skyrme parameterization SkO$^\prime$ \cite{eng99}. An important extension of the Skyrme-HFB model employed here was the incorporation of time-odd terms in the Hamiltonian, needed for the correct treatment of the GT excitation. These terms contribute to the energy of polarized states, i.e. those with non-zero angular momentum, including the 1$^+$ states populated in the $\beta$-decay. They are usually neglected in Skyrme-HF calculations which are then strictly speaking applicable only to even-even nuclei. The consequence of consideration of the T-odd terms is the introduction of additional variable parameters and their relationship with the parameters of the T-even part of the Hamiltonian has to be established \cite{dob95}. The results of this calculation include a study of the GT-strength distribution over the spectrum of excited states and the effect of the strength of the $pn$ particle-particle interaction. It was shown that the HBF-QRPA model \cite{eng99} in all but very heavy nuclei calculates half-lives that are shorter then predictions of the FRDM+QRPA \cite{mol90} and EFTSI+QRPA \cite{bor97} in some cases in improved agreement with experiment. Unfortunately, a full study of the impact of the calculation on the r-process could not be completed because not all waiting point nuclei were calculated and the no deformation effects were included in the model. It would be desirable to follow up this issue and establish the origin of the difference in predictions of $\beta$-decay half-lives.   
      
 The presence and strength of neutron shell closures affects the rate of the r-process flow through the waiting point nuclei. This information comes from mean field or other mass model calculations and, to large extent, from experiment (see e.g. \cite{kra06} and references therein). It has been shown \cite{chen95} that, for example, abundance deficiencies around A=120 and A=140, calculated using mass predictions from global macroscopic-microscopic models, indicated too strong N=82 shell closure. When the masses were calculated locally in the vicinity of the shell closure using an HFB model with SkP force which predicted quenching of the N=82 gap, the agreement between calculated and observed solar r-process abundances seriously improved. At the present time it is clear that existence of yet unknown changes in the shell structure at the neutron drip-line region may have a profound impact on the flow of the r-process.

\subsubsection{The rp-process}
\label{sec:rpprocess}
Hot environments with a large surplus of hydrogen (protons) permit proton capture on seed nuclei close to the stability line up to proton drip-line - in a process called the rapid proton capture (rp-process) \cite{wal81}. It originates in explosive hydrogen burning at temperatures high enough that a break-out of the hot CNO cycle is possible. It proceeds by a series of proton capture reactions and $\beta^+$-decays and at sufficiently high density and temperature can get well beyond A=64 and Z=32 \cite{sch98} until the rp-process reaches the proton drip line. The rp-process is thought to be the dominant source of type I X-ray bursts and responsible for the nuclear composition of the crust of an accreting neutron star \cite{sch99}. The key nuclear input includes proton separation energies $S_{\rm p}$ to determine proton capture and reverse photodisintegration, alpha separation energies $S_\alpha$ for treatment of back-processing via $\alpha$ emission or ($\gamma,\alpha$) disintegration  and $Q_{\beta^+}$ = $Q_{\rm EC}$ - 2$m_{\rm e}c^2$ for calculation of the rate of weak processes along the rp-process path.

As for the r-process, the lack of experimental data in important regions of the rp-process path forces us to turn to theoretical models. The mass models play, as usual, a key role. Weak interaction rates are usually known experimentally for nuclei with $Z\le34$. For $\beta$-decay of isotopes heavier than A = 20 the decay from thermally populated excited states has to be taken into account and the temperature dependence of the decay rate has to be calculated. The standard approach is to calculate these rates within QRPA \cite{mol90} and, wherever possible, in the shell model. 

An interesting attempt to calculate properties of nuclei along the rp-process path close to the proton drip-line has been reported by Brown et al., \cite{bro02} who calculated one-proton and two-proton separation energies in the spherical Skyrme HF model. They calculated Coulomb displacement energies of mirror nuclei and utilized known masses of the neutron rich partner of a mirror pair. The newly developed SkX Skyrme parameterization with a especially added charge symmetry breaking (CSB) term was used \cite{bro00}. The results in the region of A = 41--75, given with the error  dependent on the experimental error of the neutron-rich nucleus and assumed 100 keV theoretical error, were extensively tested in rp-process simulation and models for the x-ray burst. It has been clearly shown that the x-ray burst tails are sensitive to nuclear masses at and beyond the $N=Z$ line between Ni and Sr. The new calculation leads to a tighter constraint on proton capture Q values as compared to extrapolation of experimental masses \cite{aud95}. The authors connect this success with the special parameterization of the SkX$_{\rm CBS}$ force to account for charge-symmetry breaking. The results are indeed better in comparison with the original SkX force. However the question remains how unique is this extension of the SkX force and, whether some other parameterizations than SkX would behave similarly if the CSB term was added to them.

%\subsection{Beyond the mean-field: Collective correlations}
%\label{sec:beyond}
%\subsubsection{Large amplitude collective motion}
%$\label{sec:lampl}
%\subsubsection{Correlations from soft modes}
%\label{sec:collcorr}
%\subsubsection{Low energy quadrupole states}
%\label{sec:lowQ}
% this part --> PGR
\subsection{Beyond the mean-field: Collective correlations}
\label{sec:beyond}

\subsubsection{Large amplitude collective motion}
\label{sec:lampl}

As outlined in Section \ref{sec:corrbasic}, correlations cover
anything beyond mean field and embrace very different
mechanisms. Short-range and long-range correlations are assumed to
be incorporated into the effective energy-density functional. Collective correlations, however, cannot be accounted for in a simple
energy-density functional and need to be considered as explicit
correlations in addition to the mean field calculations.

The effective energy functionals can still be used for that task since
low-energy (and thus large amplitude) collective motion can be derived
as the adiabatic limit of time-dependent mean fields. This applies
exactly to center of mass motion which perfectly decouples from
internal excitations but to a good approximation also. This applies to
rotation whose ground state has zero energy and where the excitation
energies are by far the smallest in the nuclear spectrum. The approach
has been usually applied to low-energy modes associated with surface
vibrations of quadrupole type, sometimes also of octupole nature. The
zero-energy modes (center of mass in Section \ref{sec:cmcorr},
rotation in Section \ref{sec:rotat}) are related to symmetry
breaking through the mean field and restoration through
correlations. In the same spirit, this includes also particle-number
projection in connection with BCS states although the associated
collective `deformation' corresponds to a motion through phase
angles and does not have any geometrical interpretation.

Somewhat in analogy to the Born-Oppenheimer picture in molecular
physics, the adiabatic motion is thought to evolve along a collective
path which is set of mean-field states $\{|\Phi_{\rm q}\rangle\}$ where $q$
labels the collective deformation (e.g., rotation angle or quadrupole
momentum). In the ideal case, the path is generated by adiabatic TDHF
(ATDHF) \cite{Bar78a,Goe78a,Rei87aR}. In practice, one approximates
that by a constrained Hartree-Fock (CHF) calculation in which a wanted deformation
$q=\langle\Phi_{\rm q}|\hat{Q}|\Phi_{\rm q}\rangle$ 
is achieved by adding a constraining operator to the mean field
Hamiltonian, i.e. $\hat{h}\longrightarrow\hat{h}-\lambda\hat{Q}$ where
$\lambda$ is the Lagrange multiplier. The dynamical properties are
explored by self-consistent cranking taken over from the ATDHF scheme.
The collective spectrum (correlated ground state and excitations) is
usually described within generator-coordinate method (GCM) as a coherent
superposition of these deformed mean-field states i.e.
\begin{equation} 
  |\Psi_{\rm n}\rangle
  =
  \int dq\,|\Phi_{\rm q}\rangle f_{\rm n}(q)
  \quad.
\label{eq:gcmansatz}
\end{equation}
The equations for the superposition functions $f_{\rm n}$(q) are obtained by
variation. The machinery of the GCM and the various approximations
invoked in actual calculations are a large topic, far beyond
the scope of this review. We refer to summaries in
\cite{Rei87aR,Bon90b,Rin80aB,Ben03aR} and mention only the two basically
different solution schemes: some groups go for a direct numerical attack on a
finite grid of $|\Phi_{\rm q}\rangle$, other groups deduce a collective Hamiltonian
in terms of the $q$ and solve that.

The deformation coordinates $q$ embrace all the typical low energy
modes. The three nuclear center of mass coordinates are included among
these and their approximate treatment leads to the c.m. correction as
explicitely included in the energy functional and as discussed in
Section \ref{sec:cmcorr}. The next important considerations are the
five coordinates for vibration-rotation, either as the five quadrupole
momenta or recoupled to total quadrupole coordinate $\beta$,
triaxiality $\gamma$, and three Euler angles
\cite{Gre11B}). The GCM, or collective Hamiltonian, in these coordinates
yields the low-energy quadrupole spectra and consistent with it also
the correlated ground state associated with the lowest solution
$f_{\rm 0}$. An example of spectra will be discussed in Section
\ref{sec:lowQ} and of ground-state correlations in \ref{sec:collcorr}.
It is to be noted that octupole modes can also become very soft and
accessible to the adiabatic picture, for a typical application see
e.g. \cite{Ska93b}.

What is meant by `beyond mean field' clarifies when we consider in GCM ansatz
(\ref{eq:gcmansatz}) expressed sums of mean field states. The superposition is
necessary for a fully quantum mechanical treatment of the system. TDHF
as such can also describe large amplitude motion and it does it in
terms of a trajectory $|\Phi(t)\rangle$ which remains at every instant
a mean-field state and never goes `beyond'. One can say that TDHF
represents the classical limit of many-body dynamics. The classical
trajectory belongs to the mean-field approach and GCM serves to
`re-quantize' the description. There are regimes of collective
motion where a semi-classical description suffices, such as
fission which had been discussed in Section \ref{sec:fission} and
heavy-ion reaction which were sketched in Section
\ref{sec:collisions}. A particularly subtle case are the small
amplitude vibrations discussed in Section \ref{sec:giantres} and
\ref{sec:gamow-teller}. TDHF allows computation of the frequency spectrum
of classical vibrations. The re-quantization is performed
unconsciously through the equivalence of classical and
quantum-mechanical picture for the harmonic oscillator. That simple equivalence fails
for non-harmonic motion which is usually encountered in soft modes and
this, in turn, requires the elaborate GCM techniques.

\subsubsection{Correlations from soft modes}
\label{sec:collcorr}

\begin{SCfigure}[0.5]
%{\includegraphics[angle=0,width=9cm]{Sn_d2p_ski3.eps}}
{\epsfig{figure=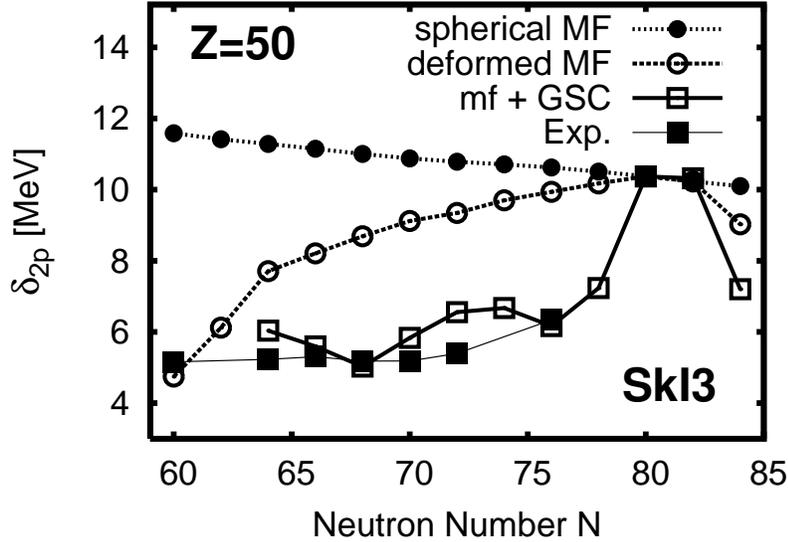,width=11cm}}
{\caption{\label{fig:Sn_d2p_ski3}
  The two-proton shell gap $\delta_{\rm 2p}$ 
  calculated for the interaction SkI3 at various levels
  of the model: spherical mean field,
  deformed mean field, and with collective ground state correlations
  from quadrupole vibration and rotation.
  The experimental data are taken from \protect\cite{Rad00a}.
}}
\end{SCfigure}
The impact of collective correlations on bulk properties of semi-magic
nuclei is generally very small. That is why these nuclei are
preferable used in the fitting of forces. The situation may change,
however, when considering the much more sensitive isotopic or isotonic
differences. The crucial influence of collective correlations on isotopic
differences of radii, the isotopic shifts
\cite{Ott89aER}, has been long recognized in many places, e.g.
\cite{Gir82a,Bon91a}. The effect on energy 
differences, though less in focus up to now, is not less important. As
an example, Figure \ref{fig:Sn_d2p_ski3} shows the second energy
difference, the two-proton shell gap,
\begin{equation} 
  \delta_{2p}
  =
  E(Z\!+\!2,N)-2E(Z,N)+E(Z\!-\!2,N),
\end{equation}
in Sn isotopes. The results from all spherical calculations agree very
nicely with twice the spectral gap at the magic $Z=50$ shell
closure. This confirms the idea behind using the shell gap as a
measure of the spectral gap. However, the spherical results are far
from the measured values, in both size and trend. As a next step, we
allow for (axially) deformed ground states. This lowers the
$\delta_{\rm 2p}$ gap for the isotopes with low neutron number because
the neighboring isotones (Cd and Te) gain binding energy through
deformation while semi-magic Sn remains spherical.  The deformation
effect is still insufficient to match experimental gaps. We are
obviously in a transitional regime between spherical and deformed
shapes where we expect large shape fluctuations and consequently
substantial collective correlations. And indeed, the $\delta_{\rm 2p}$
gaps obtained using correlated ground states match experimental values
very nicely. (These calculations were done within the Gaussian Overlap
Approximation to GCM, for details see
\cite{Fle04b}.) This example shows that correlation can have a dramatic
effect on isotopic (or isotonic) differences. It need not always 
be that dramatic -- a similar study in the Pb isotopes has shown that
there the deformation effect suffices to reproduce the data
\cite{Ben02a}. Nonetheless, a careful check is advisable in any case.

\subsubsection{Low energy quadrupole excited states}
\label{sec:lowQ}

There is a widespread literature on the computation of
low-energy spectra on the basis of SHF or the Gogny model using
various variants of the GCM. These deal mainly with low-lying
quadrupole excited states, see e.g.
\cite{Rod00c,Hee93a,Fle04a}, sometimes also octupole excitations, see e.g.
\cite{Ska93b,Hee94a}. Well adjusted modern Skyrme parameterizations
provide usually a pertinent description of these soft modes.
\begin{SCfigure}[0.5]
%{\includegraphics[angle=0,width=9cm]{Sn_chain_rev.eps}}
{\epsfig{figure=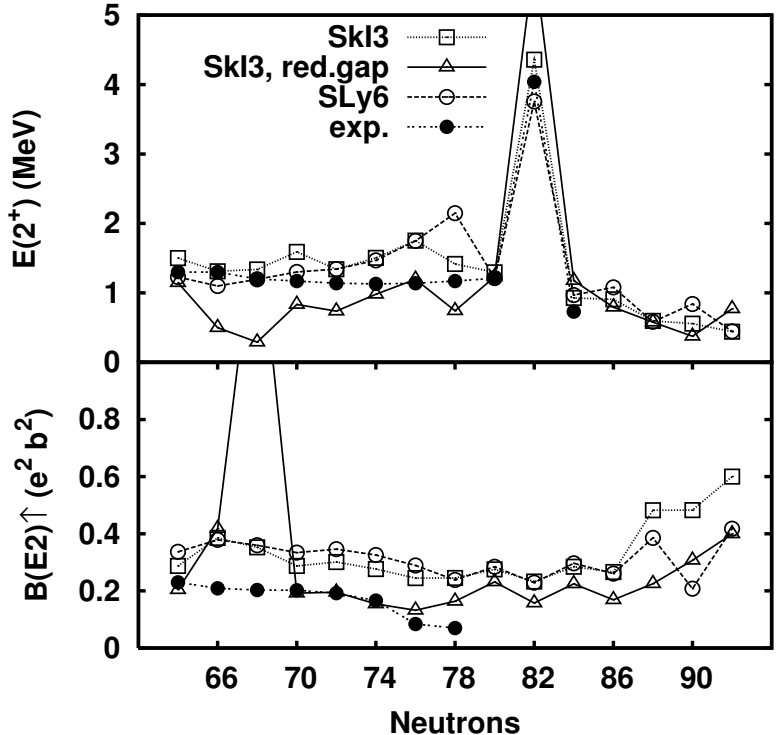,width=11cm}}
{\caption{\label{fig:Sn_chain_rev}
The energy (upper panel) and B(E2) transition moments (lower panel)
of the lowest quadrupole state ($2^+$) in Sn isotopes
computed with two different forces as indicated and compares with
experimental data. For the force SkI3, we also consider 
the case of a 75\% reduced pairing strength.
}}
\end{SCfigure}
Their typical performance is demonstrated in Figure
\ref{fig:Sn_chain_rev} taking Sn isotopes as example. Doubly magic
$^{132}$Sn is clearly distinguished by its large excitation energy.
Pairing is inactive in a doubly-close-shell nucleus and the energy
spectrum is closely related to shell structure, in particular the
shell gap in the single-nucleon spectra. The low lying quadrupole mode
in all other isotopes is dominated by the pairing gap. It is no
surprise then that many different SHF parameterizations yield very
similar results provided that the pairing strength has been calibrated
properly for each case. The crucial influence of pairing is
demonstrated for SkI3 by reduction of the pairing strength by to 75\%
of the standard strength. The particularly strong effect seen around
neutron number $N=68$ is related to a semi-shell closure at a point at
which the BCS scheme runs into a critical regime close to
breakdown.

As discussed above, the results for the excitation energy are
generally very satisfying.  The transition moments, however, are not
yet so well described as is true of most models. This is because the
moments are much more sensitive to the collective mass which, in turn,
is extremely sensitive to the choice of the collective path
\cite{Rei84b}. We suspect that moment calculations could be much
improved by using the full ATDHF instead of approximating it through
CHF.  In spite of this unresolved detail, we want to emphasize that
self-consistent mean-field models with well adjusted bulk and pairing
properties yield directly a pertinent description of large amplitude
collective dynamics without thee need of additional re-tuning.

%\section{Infinite matter and astrophysical applications}
%\label{sec:infinite}
%\subsection{Relevance of nuclear properties to physics of compact objects 
%}
%\label{sec:comp}
%\subsection{Relation to SHF and the Equation of State}
%\label{sec:relshf}      
%\subsubsection{Key properties of nuclear matter} 
%\label{sec:keyprop}
%\subsection{Neutron Stars}
%\label{sec:neutronstars}
%\subsubsection{Cold non-rotational neutron star models}
%\label{sec:coldnonrot}
%\subsubsection{Validity of the Skyrme model in high density matter: Hyperons and quark matter in neutron stars?}
%\label{sec:validity}
%\subsubsection{Inhomogeneous matter.}
%\label{sec:inhommatter}
%\subsubsection{Hot matter and type II (core-collapse) supernova models.}
%label{sec:hotmatter}
% this part --> JS
\section{Infinite matter and astrophysical applications}
\label{sec:infinite}
The concept of infinite nuclear matter has been introduced to nuclear
physics as an extrapolation of the almost homogenous conditions in the
center of heavy nuclei. In this fully homogeneous matter the
difference between neutrons and protons is neglected and no single
particle effects are considered. These conditions lead to constant
density and the absence of Coulomb energy, pairing and surface
effects. These features make it an attractive medium for testing and
comparing models of bulk nuclear properties. Infinite nuclear matter
is, however, not an observable system and the quantities
characterizing its properties are of empirical nature (see
Section~\ref{sec:keyprop}).  There are two limiting states of infinite
nuclear matter, symmetric nuclear matter (SNM) with equal number of
protons and neutrons and pure neutron matter (PNM) with no protons
present. These two states have fundamentally different properties. The
energy per particle in SNM reaches a negative minimum value (i.e. it
saturates) at a \textit{saturation density} $n_ {\rm _0} $ (=0.16
fm$^{-3}$), corresponding to the lowest bound state of SNM (the ground
state). On the contrary, the energy per particle in PNM is predicted
to be always positive, i.e. PNM as such does not exist in a bound
state and represents the highest excited state on nuclear
matter. These properties of SNM and PNM place powerful constraints on
the parameters of the Skyrme interaction, as will be discussed later.

Asymmetric nuclear matter that lies between the two extremes with the
proton/neutron ratio between zero and one has become a popular medium
for modeling large fractions of stellar matter in compact
objects, in particular in cold neutron stars.
Note that the term `compact object' in
astrophysics refers to an object whose gravity produces a deep
potential well. It is most commonly used to describe neutron stars and
black holes but is sometimes also used to refer to white dwarfs. Note
that the defining feature is the depth of the potential well and not
the strength of gravitational forces (which can be quite modest in the
case of high-mass black holes). The deep potential well means that
light coming from the surface to an outside observer is significantly
redshifted and light rays passing close to the object are
significantly bent.
The equation of state of nuclear matter is constructed from the
calculated density and composition dependence of energy per particle,
as discussed in Section~\ref{sec:relshf}. There is no direct
experimental information available on the behavior of the energy per
particle in nuclear matter. Theoretically it exhibits strong
dependence on the properties of the interaction between particles
present in the matter. In the investigation of this dependence for
different parameterizations of the Skyrme interaction, we use as a
benchmark the ab-initio calculation with the realistic A18+$\delta
v$+UIX$^*$ (APR) potential \cite{akm98}.  The APR model is based on
the Argonne A18 two-body interaction and includes three-body effects
through the Urbana UIX$^*$ potential and boost corrections to the
two-nucleon interaction which gives leading relativistic effects of
order $(v/c)^2$. This interaction is considered as one of the most
modern realistic potential used for description of nuclear matter at
present.

\subsection{Relevance of nuclear properties to physics of compact objects 
}
\label{sec:comp}

In order to construct compact star models, it is necessary to have an
equation of state (EoS) linking pressure with total energy-density
and, to obtain this, expressions must be supplied for the interaction
potentials or energy functionals of the particles concerned
\cite{hei00,lat00,hei00a}. The basic ingredient of the EoS, the
effective nucleon-nucleon interaction, is not very well known and
assumptions need to be made about its nature and form, particularly as
regards its behavior as a function of density. In this section we
concentrate on the role the non-relativistic phenomenological Skyrme
interaction plays in this context, and its limitations. Especially
relevant for our discussion are cold, non-rotational neutron stars as
naturally occurring compact objects. At around nuclear matter density,
neutron star matter can be well-represented as a homogeneous mixture
of neutrons, protons, electrons and muons ($n+p+e+\mu$ matter)
representing beta-equilibrium matter (BEM). These densities turn out
to be crucial for determining the properties of neutron star models
with masses near to the widely-used `canonical' value of 1.4
M$_{\odot}$. However, models of non-equilibrium systems like
core-collapse supernovae, both at sub- and super- nuclear densities,
also use EoS based on the Skyrme interaction \cite{sto06}.

We focus here on to what extent SHF fulfills the important requirement
that it describes not only finite nuclei but is also valid in infinite
nuclear matter. Almost 90 Skyrme functionals were recently tested in
nuclear matter and cold (T=0) non-rotational neutron star
models. Details can be found in \cite{rik03}. Here we state only the
main conclusion drawn in \cite{rik03} that Skyrme interactions, giving
broadly similar agreement with the experimental observables of nuclear
ground states as well as with properties of SNM at saturation density,
predict widely varying behavior for the observables of both symmetric
and asymmetric nuclear matter as a function of particle number
density. The underlying property, decisive for the validity of a
Skyrme interaction in nuclear matter models, is the density dependence
of the symmetry energy. Here we will demonstrate this result and its
implications for models of neutron stars on the same set of the Skyrme
parameterizations used in the previous sections, SkM$^*$, SLy6, SkP,
SKI3, SkI4 and BSk1 as examples.

The requirement that SHF remains applicable in various astrophysical
scenarios, utilizing the growing amount and precision of observational data,
in particular on neutron stars, places important constraints on the optimal
parameter sets. These will be discussed throughout the coming sections.

\subsection{Relation to SHF and the Equation of State}
\label{sec:relshf}      

 The key property of a system for use in EoS models is the total
energy per particle ${\mathcal E}=E/A$. As discussed above (see
Section~\ref{sec:infinite}), nuclear matter is a medium of constant
density.  This simplifies significantly the total energy defined for
finite nuclei by equation (\ref{eq:enfun}) in Section
\ref{sec:basicform}.  All terms depending on $\Delta\rho$, ${\bf j}$,
$\bsigma$, and ${\bf J}$ vanish. The assumption of absence of the
Coulomb force in nuclear matter is in line with the condition for
existence of stars, bound together by the gravitational force, many
orders or magnitude weaker than the electromagnetic force. The
c.m. correction approaches zero in infinite matter and pairing is
neglected as well. Thus the Skyrme energy functional in nuclear matter
becomes
\begin{subequations}
\begin{eqnarray}
  E
  &=&
  \int d^3\left\{{n \mathcal E}_{\rm kin}+{n \mathcal E}_{\rm Skyrme}\right\},
\label{eq:ennm}
\\
  {n \mathcal E}_{\rm Skyrme}
  &=&
  \frac{B_0+B_3n^\alpha}{2}n^{2} 
  - 
  \frac{B_0^{\prime}+B'_3n^\alpha}{2}\tilde n^2
  +B_1 n\tau
  -B_1^{\prime}\tilde n\tilde\tau
\label{eq:enfunsknm}
\end{eqnarray}
\end{subequations}
where ${\mathcal E}_{\rm kin}$ is taken in the same form as
(\ref{eq:enfun}).  The terms $\propto B_{\rm 0},B'_{\rm 0}$ correspond
to a zero-range attractive two-body force, terms $\propto B_3,B'_3$ to
a density-dependent repulsive force, and those $\propto B_1,B'_1$ are
related to the nucleon effective mass. The above expression for ${\cal
E}_{\rm {Skyrme}}$ is a function of 7 variable parameters $B_{\rm i},
B^\prime _ {\rm i}$ and $\alpha$. It is sometimes convenient to
express them in terms of 9 alternative Skyrme parameters $t_{\rm i}$,
$x_{\rm i}$ and $\alpha$ \cite{cha97} using relations (\ref{eq:bdef}).

The relation between finite nuclei and infinite matter is established
by considering a finite volume $V$ of homogenous matter containing $N$
neutrons and $Z$ protons, forming $A=N+Z$ nucleons. For matter with
N$\neq$Z the asymmetry parameter is defined as the ratio
\begin{equation}
  I
  =
  \frac{N-Z}{A}
  =
  \frac{n_n-n_p}{n}
  =
  \frac{\tilde n}{n}
  \quad.
\end{equation}
Then the proton and neutron number densities are given in terms of $I$
and $n$ as
 \begin{equation}
  n_p
  =
  \frac{Z}{V}
  = 
  \frac{n}{2}(1-I)
  \quad,\quad 
  n_n
  =
  \frac{N}{V}
  = 
  \frac{n}{2}(1+I)
  \quad,\quad 
  n
  =
  \frac{A}{V}
  =
  n_n+n_p
  \quad.
\label{eq:npna}
\end{equation}
Using the expression for the energy $E$ (\ref{eq:ennm}) we define the total
energy density of nuclear matter
\begin{equation}
  \epsilon(n,I)
  = 
  n( {\mathcal E} + mc^2)
\label{eq:endenstotal}
\end{equation}
where $\mathcal E=E/A$ is the binding energy per particle while
$n\mathcal E=E/V$ is the energy density, i.e. the energy per volume.

Nuclear matter obeys the standard thermodynamical relation between pressure
$P$ and Helmholtz free energy $F = E - TS$ where $E$ is the internal
energy (\ref{eq:ennm}) and $S$ is the entropy, namely
\begin{equation}
  P
  =
  \frac{\partial F}{\partial V}{\Big \vert}_{T,S}
  \quad.   
\end{equation}
This can be rewritten in terms of $n$ and $I$ in the form
\begin{equation}
  P
  =
  n^2\frac{\partial{\mathcal F}(n,I)}{\partial n}{\Big \vert}_{T,I} 
  \quad,\quad
  {\mathcal F}
  =
  \frac{F}{N}
  \quad.
\label{eq:press}
\end{equation}
where ${\mathcal F}$ is the free energy per particle.  For
systems at zero temperature ($T=0$) the equation ({\ref{eq:press})
reduces to
\begin{equation}
  P
  =
  n^2\frac{\partial{\mathcal E(n,I)}}{\partial n}{\Big \vert}_{I}
  \quad.
\label{eq:pressT0}
\end{equation}
The pressure is fully determined by the total energy per particle
${\cal E}(n,I)$ and its dependence on the particle number density $n$ and
composition $I$. This relation is often called the equation of state for
convenience - it can be very easily converted to the more customary form
relating pressure, temperature and volume as demonstrated above.
\begin{figure}
\begin{center}
{\includegraphics[angle=0,width=15cm]{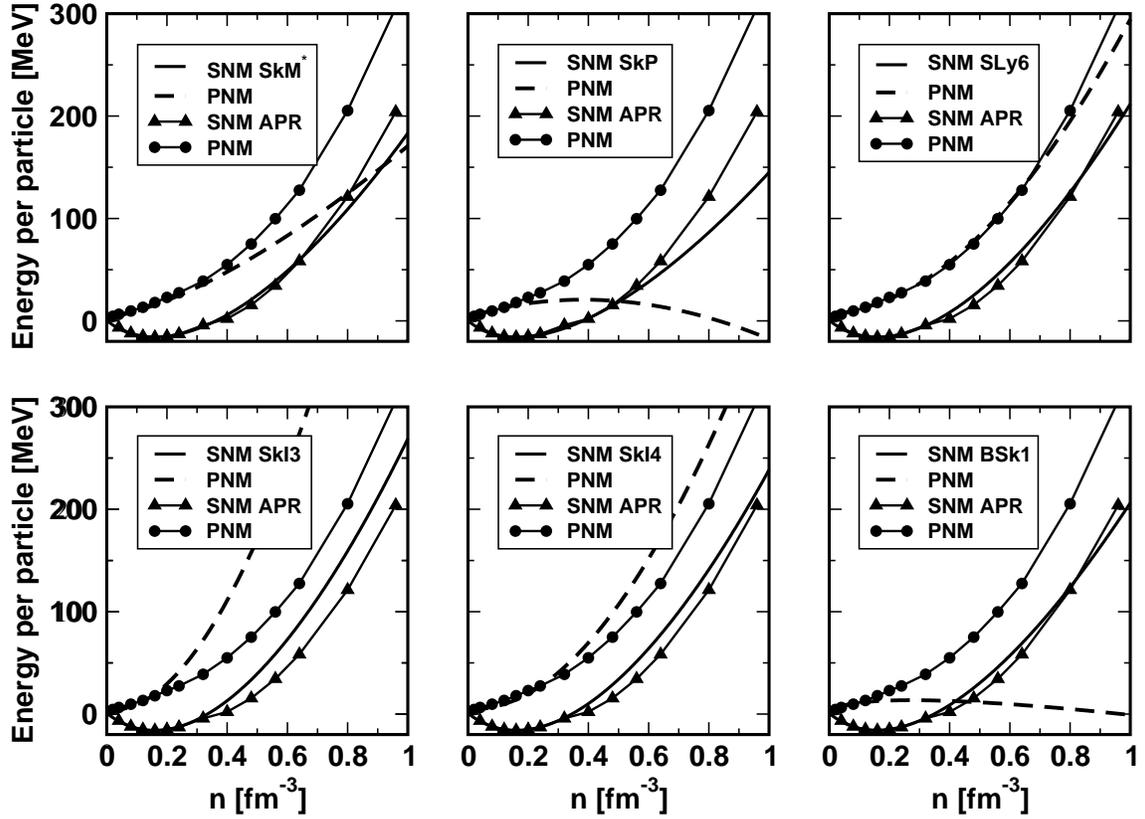}}
\caption{\label{fig:ea} 
The energy per particle for SNM and PNM is plotted as
a function of baryon number density $n$ for SkM$^*$, SkP, SLy6, SkI3, SkI4 and
BSk1 Skyrme parameterizations. For comparison, also shown are equivalent curves
for SNM (triangles) and PNM (circles) calculated using the APR potential
\protect\cite{akm98}. For more explanations, see text.}
\end{center}
\end{figure}
We illustrate, in Figure~\ref{fig:ea}, the energy per particle $\cal
E$ as a function of total nucleon number density $n$ for six different
parameterizations of the Skyrme interaction discussed above in
Section~\ref{sec:applfinite}. The calculation has been performed at
zero temperature for the two extreme phases of nuclear matter, SNM at
$I=0$ and PNM at $I=1$.  Examination of Figure~\ref{fig:ea} shows that
the density dependence of $\cal E$ for SNM is calculated to be rather
similar for all the selected parameterizations and compares reasonably
well with values yielded by the APR model. However this general
agreement does not carry over to the predictions for PNM for which the
density dependence of $\cal E$ varies significantly for different SHF
parameterizations at densities higher than about 0.2 fm$^{-3}$ and
agrees with the APR result only for SLy6.

\subsubsection{Key properties of nuclear matter} 
\label{sec:keyprop}

Although infinite nuclear matter is not directly observable, the
equilibrium parameters of SNM and some properties of asymmetric matter
provide a physically plausible and intuitive way to characterize the
bulk properties of a model. We consider in the following paragraphs
the saturation density of nuclear matter n$_{\rm 0}$, the binding
energy of SNM $\cal E$ at this density, the incompressibility modulus
$K$, the symmetry energy $a_{\rm sym}$ and the isoscalar and isovector
effective nucleon masses.  Currently accepted values are then compared
with predictions of the selected Skyrme parameterizations and with
selected interaction used in RMF models discussed in
Section~\ref{sec:functional}

The energy functional provides the binding as function of the
isoscalar and isovector densities, i.e. $\cal E$($n,\tilde n$).  In
SNM, the equilibrium energy and density are defined by the condition
\begin{equation}
  \frac{\partial{\mathcal E}}{\partial n}
  =
  0
\end{equation}
(the other condition $\partial{\mathcal E}/\partial \tilde{n}=0$ is
automatically satisfied for $\tilde n={n}_{\rm p}-{n}_{\rm n}=0$).
The best estimate of the density $n_{\rm 0}$ of symmetric nuclear
matter at saturation is based on the comparison of experimental and
calculated charge distribution in heavy nuclei.  The most precise
value $n_0 = 0.16 \pm 0.005$ fm$^{-3}$ is given in \cite{cha97}. A
more conservative value of $ 0.17 \pm 0.02$ fm$^{-3}$ is given in
Ref.\cite{mah89} where the error bar includes uncertainties in the
neutron density distribution and a correction for possible density
inhomogeneity in nuclear interior. The value of $0.16 \pm 0.02$
fm$^{-3}$ quoted in \cite{hei00} is of the same precision as that in
Ref. \cite{mah89}.

The accepted value of ${\mathcal E}_0 = \mathcal E (n_0,I=0)$ at
saturation (i.e. the minimum value) is taken to be equal to the
coefficient of the volume term $a_{\rm v}$ in the liquid-drop model,
obtained by fitting the binding energies of a large number of
nuclei. This procedure gives ${\cal{E}}_0 = -(16.0 \pm 0.2)$ MeV
\cite{cha97}.

The {\textit {incompressibility modulus}} $K$ is related to the
curvature of $E/A$ as \cite{cha97}
\begin{equation}
  K(n)
  =
  9n^2\frac{\partial^2{\cal{E}}}{\partial n^2}+18\frac{P(n)}{n}
  \quad.
\label{eq:comp}
\end{equation}
The value of $K_\infty = K(n_0)$ at saturation ($P=0$), represents an
important constraint on models of nuclear matter. However $K_\infty$
is a derived quantity and its `best' value is model dependent. Nix and
Moller estimate $K_{\infty} \simeq$~240 MeV \cite{nix95}, while
Hartree-Fock + RPA calculations of the giant isoscalar monopole
resonance (the breathing mode) \cite{bla80} imply K$_{\infty} = (210
\pm 20)$ MeV both with the use of Skyrme interactions \cite{gle90} and
with the Gogny potential \cite{bla95}. The generalized Skyrme
interactions \cite{far97}, fitted both to finite nuclei and the
breathing mode energies, give the best results for K$_{\infty} = (215
\pm 15)$ MeV.
In the non-relativistic approximation, the {\it{speed of ordinary sound}}
in the  nuclear medium is related to the incompressibility modulus $K$ by
\cite{hei00}
\begin{eqnarray}
  \frac{v_s}{c}
  = 
  \frac{dP(n)}{d\epsilon}
  =
  \sqrt{\frac{K(n)}{9(mc^2 + {\cal{E}} + \frac{P(n)}{n})}}
  \quad.
\label{vs}
\end{eqnarray}
It is desirable to check the density dependence of the speed of sound as it
may exceed the velocity of light at higher densities in non-relativistic
models \cite{hei00} and this unrealistic feature must be avoided.

One of the key properties of nuclear matter is the {\it{symmetry
energy}}, particularly important in the modeling of nuclear matter and
finite nuclei because it depends on the isospin part of the
interaction and is relevant for correctly describing nuclei with high
values of isospin far from stability. It reads
\begin{equation}
  \mathcal{S}(n)
  =
  \mathcal{E}(n,I=0)-\mathcal{E}(n,I=1) \quad.
\label{eq:symen}
\end{equation}
This quantity is related to the {\it{asymmetry coefficient}} $a_{\rm
sym}$ in the semiempirical mass formula under two assumptions, (i)
${\cal{E}}(n,I=0)$ is the minimum energy of the matter at given
density $n$ and thus in the expansion of ${\cal E}$(n,I) about this
value with respect to I (or equivalently $\tilde n)$ the leading
non-zero term is the second derivative term and (ii) all the other
derivatives in the expansion are negligible. Then \cite{Rei06a}
\begin{equation}
  a_{\rm sym}
  =
  \frac{1}{2}\left .\frac{\partial^2{\mathcal{E}}}{\partial I^2}\right|_{I=0}
  =
  \frac{n^2}{2}\left .\frac{\partial^2\mathcal E}{\partial \tilde{n}^2}\right|_{\tilde n=0}
  \quad.
\label{eq:asym}
\end{equation}
In practice, ${\cal E}$(n,I) contains always a small component
$\propto I^4$ \cite{Rei06a} and thus expression (\ref{eq:asym}) should
be treated as approximation.
A value a$_{\rm sym}=32.5 \pm 0.5$ MeV is found by fitting a large set
of experimental data in the Finite-Range-Droplet-Model (FRDM) \cite{mol95}.
The extrapolation of the various fits for the non-relativistic models (SHF and
Gogny) yield typical values of $a_{\rm sym}$ in the region of 27-38 MeV, see
\cite{hei00,ton00,Ben03aR,Rei06a} and Figure \ref{fig:nucmat_params}.  The
various RMF parameterization yield values in the 35--42 MeV range, see
\cite{bao97,Ben03aR,Rei06a} and Figure \ref{fig:nucmat_params}.  

The density variation of the asymmetry energy coefficient $a_{\rm
sym}$, defined by Eqs.~(\ref{eq:symen}) and illustrated in
Figure~\ref{fig:as} is determined by the density dependence of
${\cal{E}}$ in SNM and PNM.
\begin{figure}
\begin{center}
{\includegraphics[angle=0,width=15cm]{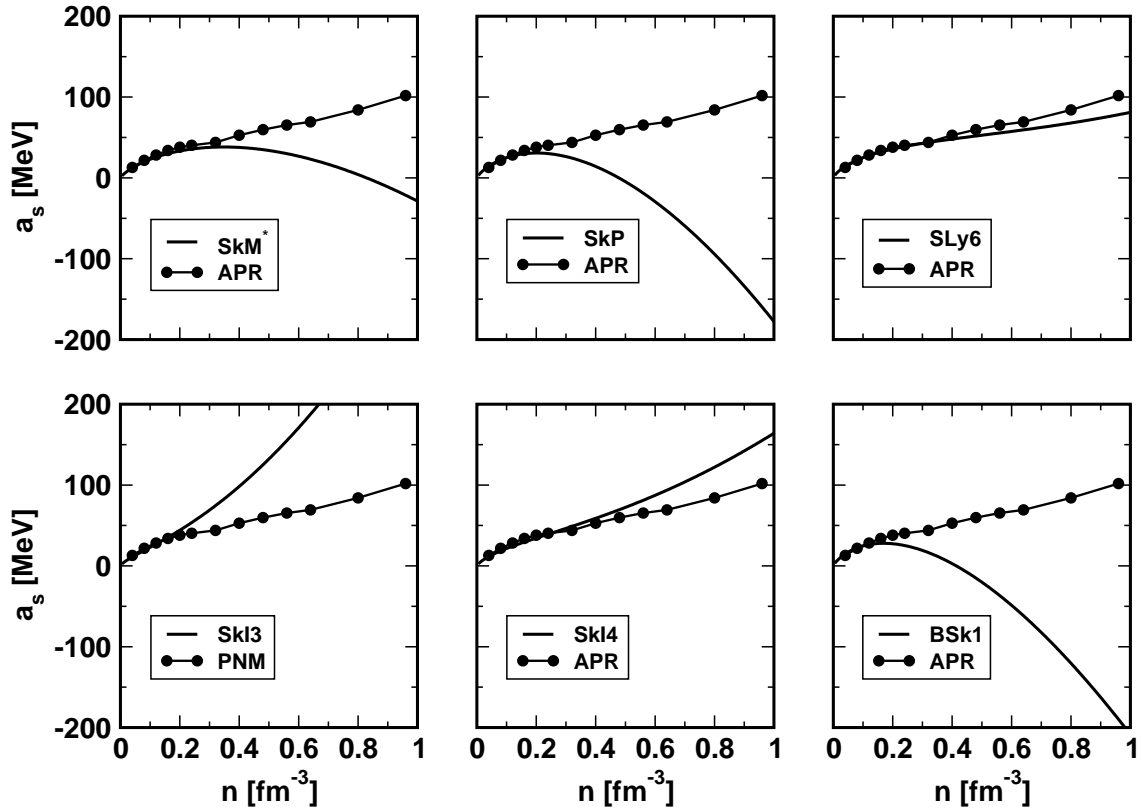}}
\caption{\label{fig:as}
The asymmetry coefficient 
$a_{\rm sym}$ as a function of particle number density $n$ for the six Skyrme parameterizations compared with the APR \protect\cite{akm98}.  
}
\end{center}
\end{figure}
We see two distinctly different patterns for the various Skyrme
parameterizations; $a_{\rm sym}$ is either growing or decreasing with
increasing particle number density. A more detailed analysis even
distinguishes between the rate of decrease of $a_{\rm sym}$ with
increasing $n$ \cite{rik03}. The transition to a negative value of
$a_{\rm sym}$ indicates that PNM becomes the ground state (which is
unphysical \cite{gle00}), the pressure becomes negative and the matter
collapses.  This is the case of the SkP and BSk1 parameterizations as
shown in Figure~\ref{fig:as}. Another (early) indicator of a possible
collapse of nuclear matter at high densities is too low a value for
the derivative of the symmetry energy with respect to particle number
density at $n_{\rm 0}$ (see Table~\ref{tab:star}). As we will show in
the next section that such parameterizations cannot produce
neutron-star models with masses as high as the `canonical' 1.4
M$_\odot$ and so they should be excluded from consideration for
astrophysical purposes. Of our sample group this applies to BSk1 and
SkP, but there are more examples of Skyrme parameterizations with the
same behavior as SkP and BSk1 in nuclear matter, identified as `group
III' in \cite{rik03}.

Finally, the {\textit{isoscalar}} and {\textit{isovector}} effective
nucleon masses in infinite nuclear matter (denoted by $m^*_{\rm s}/m$
and $m^*_{\rm v}/m$ respectively, measured in units of the vacuum
nucleon mass $m$) can be written as functions of the Skyrme parameters
and the density of the medium. The effective neutron mass in dense
asymmetric matter is then given by \cite{far01}
\begin{equation}
  \frac{\hbar^2}{m_{\rm n}^*}
  =
  (1+I)\frac{\hbar^2}{m_{\rm s}^*}-I\frac{\hbar^2}{m_{\rm v}^*}
\label{eq12}
\end{equation}
and similarly for the effective proton mass $m_{\rm p}^*$.
The isoscalar nucleon mass is deduced from the variation
$\partial_\tau E$ of the the energy functional and the isovector
effective mass comes from the variation $\partial_{\tilde\tau}E$. The
latter can be also formulated in terms of the \textit{isovector sum
rule enhancement factor} $\kappa$ \cite{Rin80aB} where
\begin{equation}
\frac{m_{\rm v}^*}{m}=(1+\kappa)
\end{equation}
A value of $\kappa=0$ means no enhancement, i.e. the isovector
effective mass equals the bare mass. Typical are slightly positive
$\kappa$ around 0.25 although little is certain about this parameter
(for details see \cite{Ben03aR}).

\begin{figure}
\centerline {\includegraphics[angle=0,width=15cm]{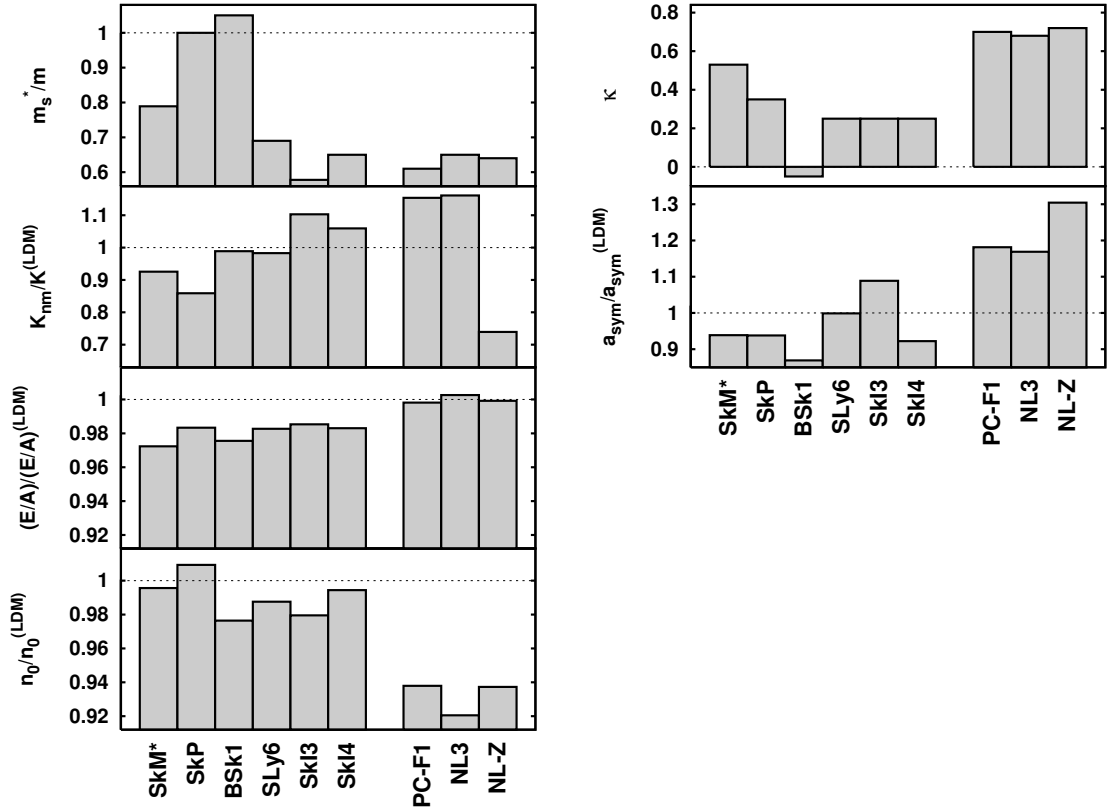}}
\caption{\label{fig:nucmat_params}
The key properties of nuclear matter:
${n}_{\rm 0}$ = saturation density of symmetric nuclear matter;
$E/A$ = binding energy per particle;
$K_{\rm nm}$ $\equiv$ $K_\infty$ = incompressibility;
$m_{\rm s}^*/m$ = isoscalar effective nucleon mass;
$a_{\rm sym}$ = asymmetry coefficient;
$\kappa$ = isovector sum-rule enhancement factor (related to isovector
effective mass;
Results for ${n}_{\rm 0}$, $E/A$ , $K_\infty$, $m_{\rm_s}^*/m$ and $a_{\rm sym}$  are shown for a variety of Skyrme and RMF parameterizations (see Table~\ref{tab:options}) as a ratio to the corresponding LDM values \cite{Mye98a,Mye98b}. $\kappa$ is given in absolute units as LDM value is not available for this quantity.
}
\end{figure}
Figure \ref{fig:nucmat_params} summarizes the predictions for these
basic properties of SNM at saturation density resulting from
SHF with our six parameterizations and in RMF using parameterizations
PC-F1, NL-3 and NL-Z in comparison with predictions of the Liquid Drop
Model (LDM). Binding energy and saturation density show little
variation within the SHF forces and agree well with the LDM
values. The RMF differs, slightly but significantly. The reasons are
not yet understood.
The {incompressibility} $K_\infty$ shows larger fluctuations, but they
gather nicely around a generally accepted value of $230\,{\rm MeV}$.
The isoscalar effective mass $m_{\rm s}^*/m$ tends to values below
one.  It is, however, a vaguely fixed property in SHF as one can see
from the large spread of results. Information on excitations is
required to better constrain the effective mass, see
e.g. \cite{Boh79aR,Cha97a} for the impact of giant resonances. The RMF
always prefers particularly low values which reflects the
counteracting interplay of the rather strong scalar and vector fields
\cite{Rei89aR}.

The isovector parameters $a_{\rm sym}$ and $\kappa$ are much less well
determined because of the limited extension of isotopic chains in the
nuclear landscape.  There are systematic differences between SHF and
RMF with RMF predicting generally larger values. The SHF values are
more consistent with the LDM which gives them somewhat better
credibility. Since the standard RMF model has much less flexibility in
the isovector channel, it is much harder to fit long isotopic
sequences and, in the extreme, sometimes impossible to produce the
equation of state for pure neutron matter \cite{wal74}. For that
reason, extensions of the RMF are being developed which provide more
isovector flexibility \cite{Typ99a,Hof01a}.

\subsubsection{$\beta$-equilibrium matter}
\label{sec:betaequil}
 
In cold neutron stars, as the density increases from about 0.75
$n_{\rm 0}$ up to 2--3 $n_{\rm 0}$, nuclear matter becomes a system of
unbound neutrons, protons and electrons and muons and, if enough time
is allowed, will develop equilibrium with respect to weak
interactions. All components that are present on a timescale longer
than the life-time of the system take part in equilibrium. For
example, neutrinos created in weak processes in a cold neutron star do
not contribute to the equilibrium conditions as they escape
rapidly. In this section we explore insights into the validity of the
selected Skyrme parameterizations obtained by investigating the
density dependence of the properties this homogenous phase.

BEM is characterized by the following processes:
\begin{displaymath}
  n\leftrightarrow p+e^- \leftrightarrow p + \mu^-
  \quad.
\end{displaymath}
Equilibrium implies that the chemical potentials should satisfy the conditions
\begin{equation} 
  \mu_{\rm n} 
  = 
  \mu_{\rm p}+\mu_{\rm e},\qquad \mu_{\rm \mu}=\mu_{\rm e}
  \quad,
\label{eq:equil}
\end{equation}
with each chemical potential $\mu_{\rm j}$ defined by
\begin{equation}
  \mu_{\rm j}
  =
  \frac{\partial \epsilon}{\partial n_{\rm j}}
\label{eq:chempot}
\end{equation}
where $\epsilon$ is the total energy density (\ref{eq:endenstotal})
and the $n_{\rm j}$'s are the particle number densities. Particle
fractions with respect to the total nucleon number density $n$ are
given as:
\begin{equation}
  y_{\rm j}
  =
  \frac{n_{\rm j}}{n}
  \quad.
\label{eq:yi}
\end{equation}
Charge neutrality of BEM implies $n_{\rm p}=n_{\rm e}+n_{\rm \mu}$.

The composition of BEM has significant impact on the properties of
neutron stars. In particular, the value taken by the proton fraction
$y_p$ has relevance for neutron star cooling, as discussed in
\cite{rik02}. The proton fraction is related to the asymmetry
parameter $I$, introduced earlier, by $I=1-2 y_{\rm p}$ and can also
be expressed in terms of the symmetry energy (\ref{eq:symen}) by
\cite{hei00}
\begin{equation}
  \hbar c (3 \pi^2 n y_{\rm p})^{1/3} 
  = 
  4 {\cal{S}}(n)(1-2y_{\rm p})
  \quad.
\label{eq:ypsym}
\end{equation}

An equation of state for zero-temperature beta-stable nucleon+lepton
matter can be constructed from SHF following a procedure described
previously, see e.g. \cite{rik03,cha97,rik02}. The total energy
density of the BEM is written as the sum of the nucleon and lepton
contributions\cite{cha97}:
\begin{eqnarray}
  \epsilon(n_{\rm p},n_{\rm n},n_{\rm e},n_{\rm \mu}) 
  &=& 
  \epsilon_N(n_{\rm p},n_{\rm n})+n_{\rm n}m_{\rm n}c^2 
  +n_{\rm p}m_{\rm p}c^2+ \epsilon_{\rm e}(n_{\rm e})
  +\epsilon_{\rm \mu}(n_{\rm \mu})+n_{\rm e}m_{\rm e}c^2 
  +n_{\rm \mu} m_{\rm \mu}c^2
\label{eq:totendens}
\end{eqnarray}
where $\epsilon_{\rm N}$ is the total nucleon energy density. The EoS for BEM is
determined by two expressions \cite{cha97}:
\begin{equation}
  \rho(n)
  =
  \frac{\epsilon(n)}{c^2}
  \quad,\quad 
  P(n) 
  = 
  n^2\frac{{\rm d}(\epsilon/n)}{{\rm d}n}
  \quad.
\label{eq:eos1}
\end{equation}
where $\rho$ is the mass density of the matter. The form used here for
the EoS is obtained by eliminating $n$ between equations (\ref{eq:eos1})
and evaluating the pressure as a function of the mass density $\rho$. This EoS forms input for neutron star models as discussed further in the following section.

For completeness we mention an important involvement of the
Skyrme-like interaction in construction of EoS for BEM based on
`realistic' potentials (see e.g. \cite{akm98} and references
therein). The APR model produces only a set of numerical results of
SNM and PNM energies at given densities but not as an analytic
function of densities. Therefore it is necessary to fit the energies
to a smooth function of density so that the derivatives needed to
obtain chemical potentials (\ref{eq:chempot}) and the pressure
(\ref{eq:press}) can be safely calculated. One then needs to
interpolate between $y_{\rm p}$ = 0.5 (SNM) and $y_{\rm p}$ = 0 (PNM)
results in order to find the value of $y_{\rm p}$ required by $\beta$
equilibrium. The smooth function of density used in this manipulation
is in the form of a generalized Skyrme interaction FPS especially
fitted for this purpose but also used elsewhere \cite{lor93}.

\begin{table} 
\caption{\label{tab:star} Observables of nuclear matter and cold (T=0)
non-rotating neutron stars: Calculated maximum mass with corresponding
radius and central density, and the radius and binding energy of a 1.4
M$_\odot$ star for Skyrme functionals SkM$^*$, SkP, SLy6, SkI3, SkI4
and BSk1. The symmetry energy at saturation density $n_{\rm 0}$,
3$n_{\rm 0}$ and the derivative of the symmetry energy at the
saturation density $\partial a_{\rm s}/\partial n(n_{\rm 0})$ are also
given. Results for four other EoS based on different models of the
nucleon-nucleon interaction, are added for comparison. See text for
more explanation.}
\vskip 15pt
\begin{tabular}{cccrlclllcc}
\hline
EoS &   $a_{\rm s}$($n_{\rm 0}$)  &  $a_{\rm s}$(3$n_{\rm 0}$) & $\frac{\partial a_{\rm s}}{\partial n}$($n_{\rm 0}$) &  M$_{\rm max}$  &  R$_{\rm max}$  &  $n^{\rm centr}_{\rm max}$  & $\rho^{\rm centr}_{\rm max}$  & R$_{1.4M_\odot}$  &  E$_{\rm bin}$  \\
       &    [MeV]  &  [MeV] &     & [M$_\odot$] & [km] &  fm$^{-3}$  & g/cm$^3$ & [km]  &  [10$^{53}$ ergs] \\
\hline

SkM$^*$           & 30.01   &  34.02  &  +95  & 1.62   & 8.94    & 1.66  & 3.80 &  10.54 &  3.09  \\         
SkP               & 29.71   & -0.43   &  +43  & 0.60   & 814   \\  
SLy6              & 32.09   &  51.84  &  +99  & 2.05   & 10.05   & 1.19  & 2.82 &  11.76 &  2.58  \\
SkI3              & 34.27   & 125.00  &  +212 & 2.19   & 11.19   & 0.98  & 2.33 &  13.56 &  2.18  \\
SkI4              & 29.50   &  98.85  &  +125 & 2.15   & 10.71   & 1.05  & 2.49 &  12.56 &  2.43  \\
BSk1	          & 27.86   & -15.06  &  +12  & 0.61   & 1089\\
FP                &  31     & 40      &       &  1.95  & 9.0     &  1.3 \\
APR      &  33.94  & 59.67   &       &  2.20  & 10.01   &  1.14 & 2.73 &  11.47 &  2.75 \\
BJ     &         &         &       &  1.85  & 9.90    &  1.31 & 3.06 &  11.86 &  2.51 \\
Hybrid            &         &         &       &  1.45  & 10.45   &  1.36 &
2.76 &  11.38 &  2.49 \\
\hline
\end{tabular}
\end{table}

\subsection{Neutron Stars}
\label{sec:neutronstars}

A neutron star is composed of matter at densities ranging from that of
terrestrial iron up to several times n$_{\rm 0}$ and, to describe this
theoretically, it is necessary to use a variety of models of atomic
and nuclear interactions. From the lowest densities up to
2.4$\times$10$^{-4}$ fm$^{-3}$ (1.5$\times$10$^{-3}$ $n_{\rm 0}$), the
neutron drip point \cite{bps71}), the matter is in the form of a
nuclear lattice with the nuclei going from those of the iron-group up
to progressively more neutron-rich ones as density increases. The
electrons are initially clustered around the nuclei but form an
increasingly uniform free electron gas with increasing density.
Beyond the neutron drip point, free neutrons appear. Above $\sim$ 0.1
fm$^{-3}$ (0.75 $n_{\rm 0}$), nuclei no longer exist and the matter
consists of homogeneous nucleon and electron fluids; with further
increases of density, muons appear in coexistence with the neutrons,
protons and electrons in beta equilibrium. At even higher densities,
heavier mesons and strange baryons are believed to play a role (see
e.g.\cite{arn77} and references therein,
\cite{pan71,bal99,wir93,hof01}). Ultimately, at the center of the
star, a quark matter phase may appear, either alone or coexisting with
hadronic matter \cite{gle00,iid98,bur02,men03}.

The SHF based models discussed in this section are used only for
modeling the nucleon+lepton phase of neutron-star matter. The lower
density region (the neutron star crust) will be discussed in more
detail in Section~\ref{sec:inhommatter}. Since only part of the star
is in the homogeneous phase, the calculated EoS needs to be matched,
at lower and (possibly) higher densities, onto other EoS reflecting
the composition of matter at those densities. For lower densities, the
Baym-Bethe-Pethick (BBP) \cite{bay71a} and Baym-Pethick-Sutherland
(BPS) EoS \cite{bay71b} is used, matching onto the Skyrme EoS at $n
\sim 0.1$ fm$^{-3}$ and going down to $n \sim 6.0 \times 10^{-12}$
fm$^{-3}$ ($3.75 \times 10^{-11} n_{\rm 0}$). At densities higher than
$\sim$ 3$n_{\rm 0}$, a hadronic EoS, for example the Bethe-Johnson
(BJ), \cite{rik02,bet74} is appropriate. Setting up an EoS over the
full density range allows calculation of one of the most important
observables of a neutron star, the maximum mass. However, this maximum
mass will be determined to large extent by the high density EoS and
will not be a test of SHF if another EoS, such as BJ, is used. It has
been argued that extrapolation of the Skyrme EoS to higher densities
is not unreasonable \cite{akm98,cha97} and that the error made by not
including the heavy baryons and possible quarks in the calculation may
not be significant. We will explore consequences of this extrapolation
and will discuss its validity in more detail in
Section~\ref{sec:validity}. We will also investigate star models at
around the `canonical' mass of $1.4 M_\odot$ and lower in which such
high densities are not reached and the applicability of SHF models is
more certain.

\subsubsection{Cold non-rotational neutron star models}
\label{sec:coldnonrot}

The basic characteristic of non-rotational cold neutron star models is
the relationship between the \textit{gravitational mass} M$_{\rm g}$
and radius $R$ of the neutron star. The most accurately measured
masses of neutron stars were, until very recently, consistent with the
range 1.26 to 1.45 M$_\odot$ \cite{lat05a}. However, Nice et
al. \cite{nic05} provided a dramatic result for the gravitational mass
of the PSR J0751+1807 millisecond pulsar M$_{\rm g}$~=~2.1$\pm $ 0.2
M$_\odot$ which makes it the most massive pulsar measured. This
observation offers one of the most stringent tests for EoS use in
calculation of cold neutron stars. It also sets an upper limit to the
mass density, or equivalently, the energy density, inside the star
\cite{lat05a}. A lower limit to mass density can be derived using the
latest data on the largest observed redshift from a neutron star
combined with its observational gravitational mass.

The gravitational mass and the radius are calculated using a tabulated
form of the composite EoS with the a chosen Skyrme interaction. The
Tolman-Oppenheimer-Volkov equation \cite{tol34,opp39}
\begin{equation}
  \frac{dP}{dr} 
  = 
  -\frac{Gm\rho}{r^2} \
  \frac{\left(1 + {P}/{\rho c^2} \right) 
  \left(1 + {4\pi r^3P}/{mc^2}\right)}{1-{2Gm}/{rc^2}}
\label{eq:TOV}
\end{equation}
 is integrated with
\begin{equation}
  m(r) 
  = 
  \int_0^r 4\pi {r^\prime}^2\rho (r^\prime)\,dr^\prime
\label{eq:relmas}
\end{equation} 
to obtain sequences of neutron-star models which for any specified
central density give directly the corresponding values for the total
gravitational mass $M_g$ and radius $R$ of the star (the surface being
at the location where the pressure vanishes).

The M$_{\rm g}$--R relation is shown in Figure~\ref{fig:ns} for the
SkM$^*$, SLy6, SkI3 and SkI4 parameterizations together with the data
for the APR model. Although each predicts a maximum mass it can be
seen that only the SLy6 parameterization gives a similar mass-radius
relation to APR.  The level of disagreement between the other Skyrme
based models and the APR is due to variation in the differences
between the density dependence of the symmetry energy (asymmetry
coefficient $a_{\rm sym}$) in these models as shown in
Figure~\ref{fig:as}. As discussed above and in detail in \cite{rik03},
the pressure in BEM is determined by the gradient of $a_{\rm sym}$ and
so it follows that parameterizations predicting slower growth of
$a_{\rm sym}$ with increasing density will give lower pressures for a
given density. The size of a neutron star depends on the internal
pressure generated by the star matter constituents that acts against
the gravitational pressure causing shrinkage of the star.If the
internal pressure vanishes (or becomes negative) the star starts to
collapse. This is the case of neutron star models based on the SkP and
BSk1 Skyrme parameterizations (see Table~\ref{tab:star}).

Maximum masses obtained from EoS based on SLy6, SkI3 and SkI4
parameterizations are well within the limits set by the new
observation reported in \cite{nic05}. On the other hand, the maximum
mass of the neutron star, calculated using the SkM$^*$
parameterization is considerably lower than that for the other models
shown here as expected considering the density dependence of $a_{sym}$
for this Skyrme model.

It is also important to calculate some other important properties of these
neutron star models. The total baryon number $A$ is given by
\begin{equation}
  A
  =
  \int_0^R \frac{4 \pi r^2 n(r) dr}
  {\left( 1 - {2Gm}/{r c^2}\right)^{1/2}} \quad .
\label{eq:totbarnum}
\end{equation}
The total baryon number $A$ multiplied by the atomic mass unit 931.50
MeV/c$^2$ defines the \textit{baryonic mass} M${\rm _0}$. The binding
energy released in a supernova core-collapse, forming eventually the
neutron star, is approximately
\begin{equation} 
  E_{bind}
  =
  (A m_{\rm 0} - M) c^2
  \quad,
\label{eq:binding}
\end{equation} 
where $m_{\rm 0}$ is the mass per baryon of $^{56}$Fe. Analysis of data from
supernova 1987A leads to an estimate of $E_{\rm bind} = 3.8 \pm 1.2 \times
10^{53}$ erg \cite{sch87}.

\begin{figure}
\begin{center}
{\includegraphics[angle=0,width=15cm]{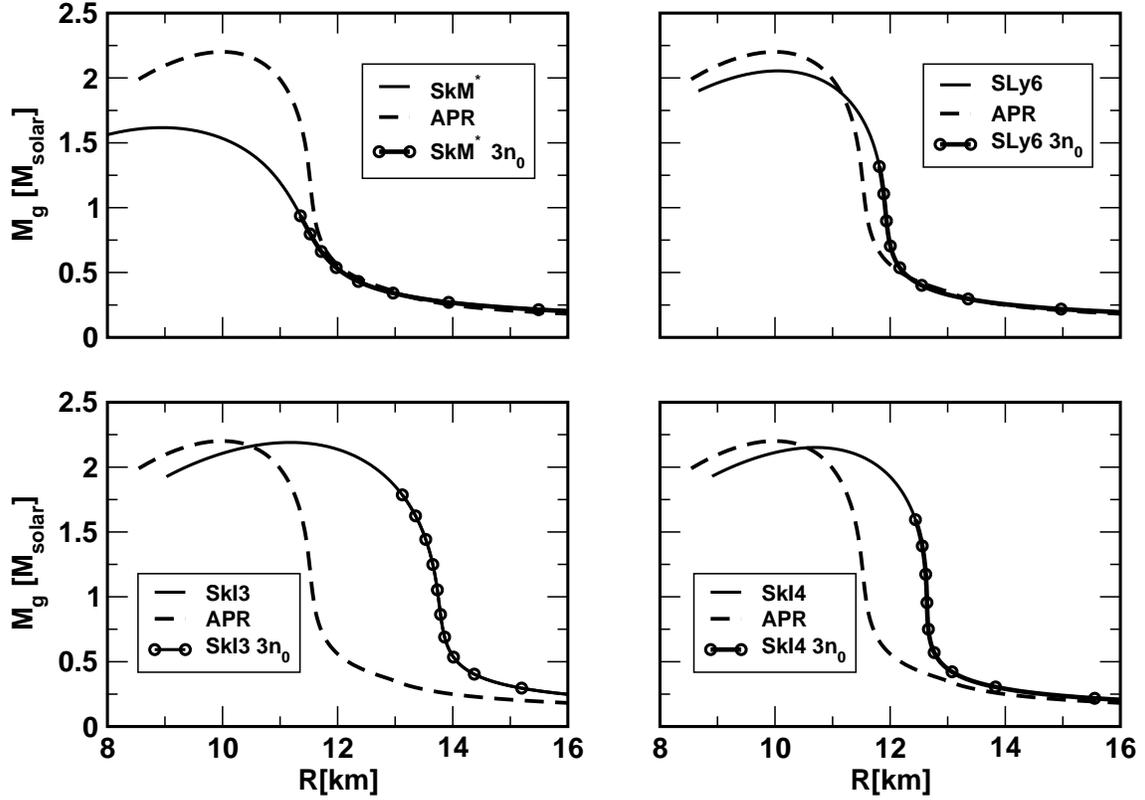}}
\caption{\label{fig:ns} The predicted gravitational masses of non-rotational
neutron-stars (measured in solar masses) plotted against radius (in
kilometers), calculated for $\beta$-stable nucleon+lepton matter using
selected Skyrme interactions, supplemented by BBP and BPS EoS's at low
densities. The circles indicate a sequence of neutron star models with central density lower than 3 $n_{\rm_0}$. For more explanation see text. }
\end{center}
\end{figure}
A summary of the majority of important parameters of nuclear matter
and of neutron star models is given in Table~\ref{tab:star} for the
six Skyrme parameterizations of our sample. The SHF based models are
compared with the APR model and some other established calculations by
Bethe-Johnson \cite{bet74}, Friedman and Pandharipande (FP)
\cite{fri81} and Glendenning \cite{gle00}. Details of these models are
beyond the scope of this work and can be found in the original
papers. We say here only that the EoS cover a great variety of
different physics including non-relativistic realistic potentials
\cite{fri81,akm98}, existence of strange hyperons in dense matter
\cite{bet74} and quark matter coexisting with relativistic nuclear
matter \cite{gle00}.  It is important to compare various models to
study the level of sensitivity of nuclear matter and neutron star
models to the physics underlying the EoS used. For example, the FP
model (still taken as a benchmark in some modern work \cite{Bro00b})
uses an old potential $v14$+TNI whilst the APR model is based on one
of the most modern Av18+$\delta v$+UIX$^\star$ realistic
potentials. The relative slope of EoS for the (SNM) and neutron (PNM)
with increasing baryon number density differs fundamentally in the two
models (see Table~\ref{tab:star}). At high densities, the FP model
predicts decreasing symmetry energy with increasing density in
contrast to the APR model. It can be seen that all models predict
central energy densities at maximum mass that are within the latest
limits set by Lattimer and Prakash \cite{lat05a}. As for the 1.4
M$_\odot$ models, we note that all predicted binding energies are
somewhat lower than the estimate based on the observation of the 1987A
SN \cite{sch87} and the corresponding radii are within the rather
broad observational limits. Neither binding energy nor radius
represent a strong constraint on the Skyrme parameterizations at
present but increased precision of observation, and, in particular,
measurement of the radius of a neutron star independently from its
mass, may change the situation.

Apart from the basic mass-radius relation, there are other features of
neutron stars that provide the intriguing possibility to constrain Eos
and thus the nucleon-nucleon interaction in star matter. With the fast
increasing quality and volume of observational data, these constraints
are of growing importance. We will discuss some of them in the rest of
this section.
   
An interesting possibility has been recently suggested by Lattimer and
Schultz \cite{lat05b}. Measurement of the moment of inertia of the
Pulsar A in the recently discovered double pulsar system PSR
J0737-3039 to $\sim$10\% accuracy would allow an accurate estimate of
the radius of the star and of the pressure in matter of density in the
vicinity of 1--2 $n_{\rm 0}$. This information would provide a strong
constraint on the EoS of neutron stars.

Another important constraint has been recently identified \cite{pod05}
in connection with the very precise determination of the gravitational
mass of Pulsar B in the system J0737-3039 M$_{\rm
g}$~=~1.249$\pm$0.001 M$_\odot$. Assuming that Pulsar B has formed in
an electron-capture supernova, a rather narrow range for the baryonic
mass M$_{\rm 0}$ between 1.366--1.375 M$_\odot$ of the pre-collapse
core can be determined. If there is no baryon loss during the
collapse, the newly born neutron star will have the same baryonic mass
as the progenitor star. For any given EoS for neutron star matter the
relation between the gravitational and baryonic mass is
calculated. The data for Pulsar B provide a very narrow window which
that relation has to satisfy.
\begin{SCfigure}
%\begin{center}
{\includegraphics[angle=0,width=10cm]{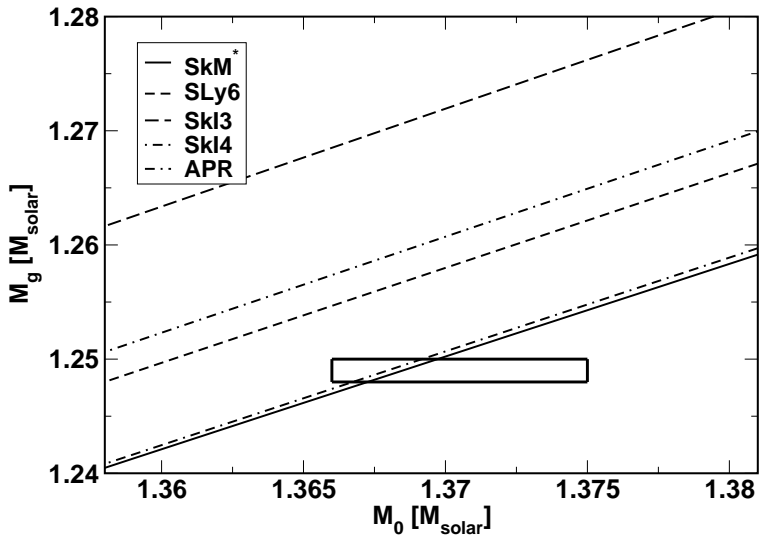}}
\caption{\label{fig:lowmass} Relation between the gravitational mass M$_{\rm g}$ of
neutron star models and their baryonic mass M$_{\rm 0}$ for SkM$^*$, SLy6, SkI3 and SkI4 Skyrme parameterizations. The APR
model is added for comparison.  }
%\end{center}
\end{SCfigure}
We show in Figure~\ref{fig:lowmass} an application of this constraint
to the Skyrme parameterizations discussed in this paper.  Of the
parameterizations studied in this work, only SkM$^*$ passes the test,
although it predicts a relatively low maximum mass of a neutron star
and yields only modestly growing symmetry energy with baryon number
density between n$_{\rm 0}$ and 3n$_{\rm 0}$. This finding seems to be
corroborated by the prediction of the neutron skin in $^{208}$Pb as
shown in Figure~\ref{fig:skin}. Both these results are obtained at
relatively low density (below and at nuclear saturation density). The
maximum mass of a neutron star is mainly determined by the high
density part of the EoS. The fact that the maximum mass of a neutron
star calculated using SkM$^*$ is rather low may be connected with a
the limited validity of the extrapolation procedure of the Skyrme EoS
to too high densities beyond 3 $n_{\rm 0}$ as will be discussed in
more detail in Section~\ref{sec:validity}. This constraint may be very
important for classification of the Skyrme parameterizations because
it concerns a relatively low density region where the applicability of
the Skyrme interaction is almost certainly valid. The constraint is
extremely stringent and may lead to an important restriction of Skyrme
parameterizations beyond finite nuclei.
  
The issue of neutron star cooling mechanisms can, in principle,
provide further important constraints on neutron-star models. However,
as discussed in more detail previously \cite{rik02}, the cooling
processes for both young and old neutron stars are not currently known
with any certainty, although several scenarios have been proposed
\cite{pag04,jon01}. We explore here the possible relevance of the
direct URCA processes (the name URCA stems from the name of a casino
in Rio de Janeiro; the neutrino energy loss was likened to losing
money at the casino by the authors of the original paper, see {\tt
http://en.wikipedia.org/wiki/Urca process}) given by
\begin{equation}
  n
  \longrightarrow 
  p + l + \bar{\nu}_{\rm l}, \hskip 1.5 truecm p + l 
  \longrightarrow 
  n + \nu_{\rm l},
\label{eq:urca}
\end{equation} 
(where $l$ stands for the leptons being considered here -- electrons and
muons) within the context of the Skyrme parameterizations studied in this
paper. For this direct URCA process to take place, the relative components
of the BEM must satisfy the appropriate conditions for conservation of
energy and momentum: 
$y^{1/3}_{\rm n} < y^{1/3}_{\rm p} + y^{1/3}_{\rm e}$ and 
$y^{1/3}_{\rm n} < y^{1/3}_{\rm p} + y^{1/3}_{\rm \mu}$.
It has been argued that this will be satisfied only at densities $n$
several times larger than the nuclear saturation density $n_{\rm 0}$
\cite{pet92,pra92,pra94} when the proton fraction in BEM reaches a
threshold value of $\sim$14\%. It follows from (\ref{eq:ypsym}) that
the steeper the increase of the symmetry energy with increasing number
density, the faster will be the growth of the proton fraction $y_{\rm
p}$. The results for the examples of the Skyrme parameterizations
investigated here are rather interesting. SkM$^*$ does not satisfy the
conditions for occurrence of the direct URCA process at any
density. This is in contrast to SkI3 and SkI4 for which the threshold
density (and corresponding mass) for the URCA process to take place is
rather low (n$_{\rm URCA}$ = 0.26 (M$_{\rm URCA}$ = 0.90 M$_\odot$)
and 0.50 fm$^{-3}$ (M$_{\rm URCA}$ = 1.59 M$_\odot$), respectively)
and SLy6 for which the direct URCA process can happen only at n$_{\rm
URCA}$ = 1.18 (M$_{\rm URCA}$ = 2.05 M$_\odot$). The latter critical
density and mass are equal to those calculated in maximum mass model
with the SLy6 EoS. However, we note that there are alternative direct
URCA processes involving hyperons and thus this constraint, although
interesting, is rather weak. Nevertheless an understanding of the
cooling mechanism of neutron stars is important and the Skyrme
interaction plays a significant role in these models \cite{pag04}. We
note that very recently a new constraint for the direct URCA cooling
mechanism has been suggested, based on analysis of population
synthesis scenarios \cite{pop06,kla06}. It finds that direct URCA
process (\ref{eq:urca}) should not happen in neutron stars with masses
lower than $\sim$1.5 M$\odot$ (see Table~\ref{tab:nsconstr})

\begin{SCfigure}
%\begin{center}
{\includegraphics[angle=0,width=10cm]{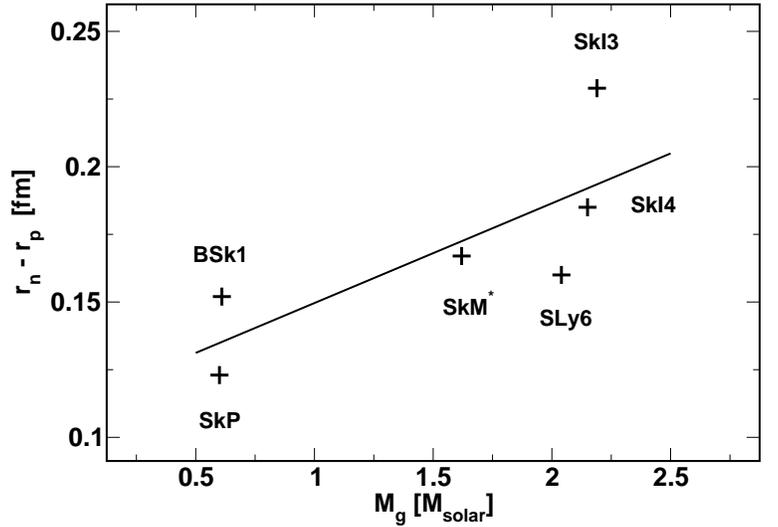}}
\caption{\label{fig:skin} Calculated neutron skin in $^{208}$Pb for
SkM$^*$, SLy6, SkI3 and SkI4 Skyrme parameterizations is displayed as
a function of the maximum mass of a neutron star, calculated using EoS
based on the same parameterization. The straight line shows the result
of linear regression.}
%\end{center}
\end{SCfigure}
It is intriguing to point out that the correlation between the
properties of neutron matter and the neutron skin in finite nuclei,
discussed generally in Section ~\ref{sec:neutronradii}, can be
extended to the correlation between maximum mass of a neutron star and
the neutron skin in neutron heavy nuclei. Models of finite nuclei
include surface and Coulomb effects, not present in calculation of the
symmetry energy in nuclear matter and yet the distinction between
Skyrme parameterizations on the basis of the \textit{density
dependence} of the symmetry energy is seen.  As demonstrated in
Figure~\ref{fig:skin}, the Skyrme parameterizations with the steepest
increase of the symmetry energy with growing density that predict the
highest maximum mass of the cold neutron star, also predict the
thickest neutron skin.  We show this effect for Pb$^{208}$ but it also
exists in $^{132}$Sn, $^{122}$Zr, $^{78}$Ni and $^{48}$Ca, as reported
in \cite{fle06}.

The Skyrme interaction has also been employed in a more subtle context
of nuclear matter and neutron star theory. The onset of ferromagnetic
instabilities in spin-polarized neutron matter at zero and finite
temperature has been investigated and identified for the SLy6 and SkI3
parameter sets \cite{rio05}. It has been shown that a new constraint
on the Skyrme parameters can be derived to avoid these
instabilities. The instability with respect to ferromagnetic collapse
was also studied in terms of Landau parameters in symmetric and
neutron matter and with respect to the constraint that the velocity of
sound has to remain smaller than the speed of light \cite{mar02}. A
set of 6 new Skyrme parameterizations has been developed using
parameters obeying the suggested constraints. It is interesting to
note that all these parameterizations predict decreasing symmetry
energy at high densities \cite{rik03}.

\subsubsection{Validity of the Skyrme model in high density matter: 
Hyperons and quark matter in neutron stars?}
\label{sec:validity}

The current trend in the development of modern effective nuclear
interactions is their applicability in finite nuclei as well as in
infinite nuclear matter. In particular, the study of asymmetric nuclear
matter (with neutron-proton ratio $0\le I\le1$) and pure neutron
matter are of importance as an extrapolation point connecting finite
nuclei at the line of stability, nuclei at the neutron drip line and
some density regions of neutron stars. One has to realize that all
effective nuclear interactions in use today have parameters fitted to
known properties of finite nuclei at or close to the stability line
and sometime properties of nuclear matter at saturation density. The
isospin and density dependence of these parameters is not
known. Nevertheless we routinely extend the use of these parameters up
to the drip lines and expect them to work. The validity of this
procedure is by no means guaranteed. Many aspects of nuclear
interactions (apart from surface effects) can be tested in infinite
nuclear matter and it is desirable to make such calculations. The
usual argument against this procedure is that models like SHF should
not be used beyond nuclear saturation density. This is a matter of
debate. The general consensus is that relativistic effects are small
up to 2--3 $n_{\rm 0}$ \cite{cha97} and the Skyrme interaction is well
justified for the description of nuclear matter consisting of nucleons
and leptons (no strange baryons or mesons)(see
e.g. \cite{cha97}). Indeed, in the degenerate fermi gas model, even at
density $\sim$0.5 fm$^{-3}$, the relativistic correction to the mean
energy of nucleons is about 10$\%$ which makes this density acceptable
from that point of view. Furthermore, the requirement of low momentum
approximation in the derivation of the Skyrme functional \cite{vau72}
sets a limit of about ($\pi k_F^{-1}>m_\pi c=1.4$ fm) which infers the
same limiting density of 0.5 fm$^{-3}$.

However, we reiterate that SHF describes only the {\textit
{nucleonic}} part of star matter. At high density and temperature it
is likely that heavy baryons and mesons will emerge at some point and
contribute to the total energy of dense matter. Simple estimates of
the threshold for $\Lambda$ and $\Sigma^{\pm,0}$ hyperons just based
on their rest-mass, were discussed in \cite{rik03} showing a
possibility that these particles will exist in matter at 3 $n_{\rm 0}$
and even below (1.5--2.0 $n_{\rm 0}$). This simple estimate neglects
the contribution to the threshold energy from the hyperon-nucleon and
hyperon-hyperon interaction, which is unknown at present and has been
modeled in a variety of ways, mainly in relativistic models (see
e.g. \cite{gle00,hof01,men03}), again predicting hyperon creation
close to, or below the 3 $n_{\rm 0}$.

Another issue that is of even more wide ranging consequence is the
question of the critical density when nucleons will start to loose
their identity and (partially) deconfined quark matter will emerge. At
densities of about 3 $n_{\rm 0}$ the nucleons will start to overlap
and formation of six-quark bags and increasing interaction strength
between quarks of different nucleons will become more likely. Again,
some calculations predict this critical density to be below $n_{\rm
0}$ \cite{gle00,men03}. Clearly no model based on nucleon degrees of
freedom should be used for the description of matter with a
significant deconfined quark component.

Non-nucleonic components in nuclear matter will affect the EoS, in
particular they will tend to decrease the slope of $\cal E$ as a
function of density (the EoS softens). As it was shown above, this
affects all the observables of dense matter and other quantities like
the maximum mass of cold neutron stars. Furthermore, it is important
to realize that if we limit the EoS only to hadronic degrees of
freedom, current neutron star models are not strictly speaking valid
at the baryon number densities of about 6--10 $n{\rm_0}$, needed to
calculate maximum mass models because at these densities nucleons and
hyperons are likely to loose their identity as composite particles and
make a partial or possibly complete transition to quark
matter. Following this logic, the neutron star models shown in
Figure~\ref{fig:ns} should be calculated only up to masses and radii
indicated by circles on the M$_{\rm g}$--R curves and the maximum mass
of cold neutron stars would not be obtainable from such models.
         
\subsubsection{Inhomogeneous matter.}
\label{sec:inhommatter}

In the previous sections, we discussed applications of SHF to models
of homogeneous nuclear matter consisting of unbound baryons and
leptons. However, there is a large inhomogeneous phase of matter at
lower densities that is present in crusts of neutron stars and forms a
part of the collapsing core at type II supernovae.
As already mentioned above, at densities below $\sim 6\times 10^{-11}$
fm$^{-3}$ the matter in this ground state consists of $^{56}$Fe nuclei
arranged in a lattice to minimize their Coulomb energy. A fraction of
electrons are bound to the $^{56}$Fe nuclei. In the density range
$\sim 6\times 10^{-11}\,{\rm fm}^{-3} < n <2.5\times 10^{-4}\,{\rm fm}^{-3}$ 
the equilibrium nuclei, now immersed in a relativistic electron gas
but still distributed in a lattice, become more neutron-rich as
electron capture and inverse $\beta$-decay take place while direct
$\beta$-decay is inhibited by the lack of unoccupied electron states.
At about $2.5 \times 10^{-4}$ fm$^{-3}$ (the neutron drip-line) finite
nuclei cannot support the neutron excess and start to emit neutrons to
continuum neutron states. From this density up to about 0.1 fm$^{-3}$,
where the transition to homogeneous nuclear matter occurs, there are
still bound nucleons, coexisting with a free neutron and electron
gases. The modeling of this most interesting region is difficult as
equilibrium emerges as result of a delicate balance between nuclear
surface energy (favoring large nuclei) and the Coulomb energy, the sum
of the positive nuclear self-energy and negative lattice-energy, which
favors small nuclei.  The current understanding is that the nuclei
form exotic phases (`the pasta phase') like rods, slabs, tubes,
inside-out bubbles, spaghetti or lasagna, stabilized by the Coulomb
interaction and characterized by various crystal lattice structures
\cite{oya93,lor93,mag02}.

The models used for the description of inhomogeneous matter phase fall
into two categories. One uses the semi-empirical mass formula or the
compressible liquid drop model (LDM) for the description of finite
nuclei and a separate model to calculate the free neutron gas energy
density (e.g. \cite{bet70,lam78,lat85,lat91}). The second category
includes self-consistent models that use the same interaction to
describe the nuclear matter inside and outside the finite nuclei
(e.g. \cite{bay71a,pet95,oya93,mag02}). The advantage of the second
class of models is obvious as they include in a consistent way the
effect of the external nucleons on nuclei which are not isolated but
embedded in the sea of neutrons. The main interest of these
calculations has focused on subnuclear densities, in particular the
`pasta' phase and the location and properties of the phase transition
between homogeneous and inhomogeneous matter. We note that there is no
need for a self-consistent calculation of homogeneous matter as in
this case the energy per particle $\cal E$ and related quantities can
be expressed analytically (\ref{eq:ennm}).

A major group of models use SHF to determine the ground state energy
of matter as a function of density, temperature and proton fraction by
minimization of the SHF energy density in coordinate space. In this
way the equilibrium energy per particle can be calculated as a
continuous function of density and the composition of nuclear matter
is determined naturally without {\it a priori} assumptions about
particular physical processes in certain density regions. One of the
first models of this kind was developed by Bonche and Vautherin
\cite{bon81} for inhomogeneous matter assuming spherical symmetry (1D
space) using the SkM and RPB Skyrme interactions. Later it was adopted
by Hillebrandt and Wolff \cite{hil85} with RATP, Ska and SkM Skyrme
parameterizations for tabulation of the EoS in a dense grid of energy
density, temperature (or entropy) and electron fraction for use in
hydrodynamical codes (for more details see section
\ref{sec:hotmatter}). Pethick et al. \cite{pet95} used the
Skyrme1$^\prime$, FPS and FPS21 parameterizations in their EoS. The
Gogny parameterization D1 has been also used to construct finite
temperature EoS \cite{hey88,ven92} based on the application of the
variational principle to the thermodynamical potential with special
attention to the liquid-gas phase transition. A semi-classical
approximation to SHF, the ETFSI method, with the parameterizations
RATP, SkM$^*$ and SkSC4--10 was also applied to derive EoS at finite
temperatures for inhomogeneous \cite{ons97} and homogeneous
\cite{ons94} nuclear matter under conditions appropriate to a
collapsing star assuming spherical symmetry. The EoS based on the RMF
theory with the TM1 parameter set for the RMF Lagrangian was used in
modeling supernova physics by Shen et al., \cite{she98}. Both,
homogeneous and inhomogeneous, matter are treated within the
model. However, the inhomogeneous matter is not calculated fully
self-consistently but is approximated as a b.c.c lattice of spherical
nuclei and the free energy is minimized with respect to cell volume,
proton and neutron radii and surface diffuseness of the nuclei and
densities of the outside neutron and electron gases. Menezes and
Provid\^{e}ncia \cite{men03} used RMF to construct EOS of mixed stars
formed by hadronic and quark matter in $\beta$ equilibrium at finite
temperature. The non-linear Walecka model and the MIT Bag and
Nambu-Jona-Lasinio models were used for the hadronic and quark matter,
respectively.

The major limitation of the models discussed so far, i.e. the
assumption of spherical symmetry for inhomogeneous matter, has been
removed in the recent work of Magierski and Heenen \cite{mag02} who
treated cold (T = 0) nuclear matter in the density region of the
`pasta' phase for the first time fully self-consistently in 3D
coordinate space (allowing for triaxial shapes) using SHF with SLy4
and SLy7 parameterizations. The SHF equations were solved in a
rectangular box with periodic boundary conditions.  This includes
naturally shell effects in neutron gas and pairing. In calculation of
the Coulomb energy, the model goes beyond the Wigner-Seitz
approximation of spherical cells (assumed in all the previously
mentioned models) and includes both interactions between nuclei on
different lattice sites and between protons and
electrons. Unfortunately no comparative study has been done as yet to
systematically explore the sensitivity of properties of inhomogeneous
matter to the choice of the SHF parameterization in comparison with
the other models.

\subsubsection{Hot matter and type II (core-collapse) supernova models.}
\label{sec:hotmatter}

The most sought after constraint on EoS in hot matter would be based
on success in modeling a type II core-collapse supernova. There is no
doubt that the supernova phenomenon exists in nature and is
accompanied by an explosion. Current models fail to reproduce this
explosion. One of the possible solutions of this problem lies in the
improvement of the EoS since the EoS at subnuclear densities controls
the rate of collapse, the amount of de-leptonization and thus the size
of the collapsing core and the bounce density. During the supernova
collapse and the birth of a proto-neutron star the matter can reach
temperatures up to 20 MeV and thus the finite temperature EoS is
used. The density at bounce of the collapsing core (for a detailed
description of the core collapse mechanism see e.g. \cite{mez05}) goes
up to 1.5--2.0 $n_{\rm 0}$ and thus includes both homogeneous and
inhomogeneous phases and the phase transition between the
two. Moreover, it may be just reaching the threshold for the
appearance of exotic components, like strange baryons and mesons, with
possibly a phase transition to quark matter at the upper end of the
density region. The timescale for supernova collapse is believed to be
of order of a second and the matter does have enough time to reach
$\beta$ equilibrium throughout the rapid changes. Thus the calculation
has to be performed for a fixed ratio of neutrons and protons, usually
y taken $y_{\rm p} \sim$ 0.3 \cite{mez05} and a relevant range of
densities and temperatures. There are so far two examples of a finite
temperature EoS, based on SHF, for use in supernova simulation
codes. The early calculations of Hillebrandt and Wolff, mentioned in
the previous section, used a spherically symmetrical SHF model (1D)
\cite{hil85}. As a new development, Newton and Stone
\cite{sto06,new06} extended the previous SHF 3D calculation
\cite{mag02} to finite temperature and wide range of densities for use
in supernova simulation models. The calculation scheme (using SkM$^*$
at present) treats the finite nuclei and the neutron gas in the
'pasta' phase as one entity and naturally allows effects such as (i)
neutron drip, which results in an external neutron gas, (ii) the
variety of exotic nuclear shapes expected for extremely neutron heavy
nuclei including, but not confined to, the `spaghetti' and `lasagne'
phases previously identified, and (iii) the subsequent dissolution of
these structures into homogeneous nuclear matter. The EoS is
calculated across phase transitions without recourse to interpolation
techniques between density regimes described by different physical
models \cite{new06}. The elimination (or significant smoothing) of the
phase transitions has serious consequences, e.g., for neutrino
transport in the `pasta' phase. It is clear that neutrino-nucleon and
neutrino-nucleus cross-sections cannot be considered just as
contributions from finite isolated nuclei and free neutron and
electron gas, but the complex density distribution of particles will
have to be taken into account.
\begin{SCfigure}
%\begin{center}
{\includegraphics[angle=0,width=8cm]{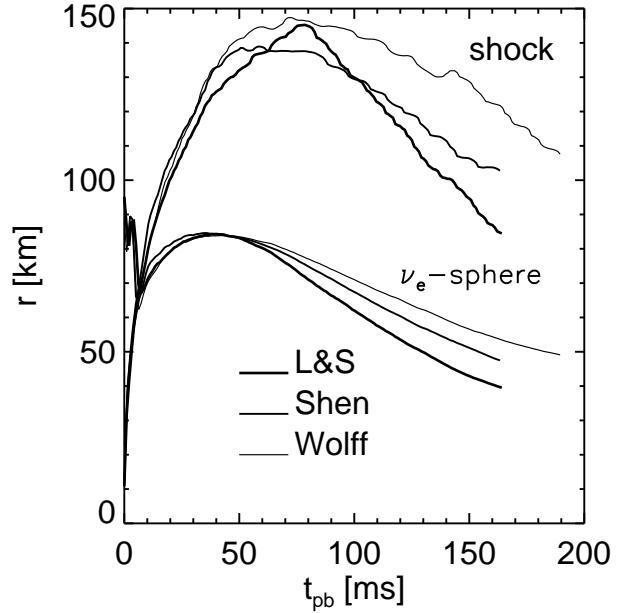}}
\caption{\label{fig:janka} Shock positions and neutrino sphere radii
of $\nu_{\rm e}$ as a function of time after bounce for collapse
simulations of a 15 M$_\odot$ progenitor star with three EoS. The
figure has adopted from \protect\cite{jan03}. See text for more
explanations.}
%\end{center}
\end{SCfigure}

An interesting comparison has been made recently by Janka {\it et al.}
\cite{jan03} who followed the development of the shock radius and
electron neutrino luminosity over the first 200 ms after core collapse
using three different EoS, Lattimer and Swesty \cite{lat91} based on
LDM, Shen et al. \cite{she98} using RMF and Hillebrandt and Wolff
\cite{hil85} utilizing SHF; the differences found may be seen as an
example of sensitivity of some predictions of supernova simulation
models to the choice of EoS.

% \section{Summarizing constraints on Skyrme parameterizations}
% \label{sec:summconstr}
% \subsection{Finite nuclei}
% \label{sec:finconstr}
% \subsection{Nuclear matter and Neutron star models}
% \label{sec:starconstr}
% \subsection{Supernova simulations}
% \label{sec:supernovaconstr}
%\include{sum-constr}
% this part --> in close cooperation
\section{Summarizing constraints on Skyrme parameterizations}
\label{sec:summconstr}

In this Section we draw together various constraints which are
applicable to confine the choice of SHF parameterizations and make some
suggestions concerning their implementation. This is a rather long
list which is, to some extent, still under development. In fact, all
practical adjustments of the Skyrme parameters up to now have started
from fitting a subset of these constraints, mainly the gross
properties of binding (Section \ref{sec:const-gross}), selected
information on spin-orbit splitting (Section \ref{sec:const-ls}), data
on odd-even staggering for pairing (Section \ref{sec:const-endif}),
and occasionally nuclear matter properties.  The other features
listed here are usually inspected a posteriori. The extensive list of
check-points presented here has never been utilized in fitting
procedures in full before.

All observables in finite systems (see Section \ref{sec:finconstr})
have to checked for possible correlation effects from soft modes (see
Section \ref{sec:beyond}). Gross properties are generally robust
while details as, e.g., isotopic and isotonic differences are more
likely to be affected.

\subsection{Finite nuclei}
\label{sec:finconstr}

\subsubsection{Gross properties: energy, charge radii}
\label{sec:const-gross}

The most important features to be properly reproduced are the nuclear
bulk properties:\\

\noindent
\hspace*{1em} binding energy $E_{\rm B}$,\\
\hspace*{1em} charge r.m.s. radius $r_{\rm ch}$,\\
\hspace*{1em} charge diffraction radius $R_{\rm ch}$, and\\
\hspace*{1em} charge surface thickness $\sigma_{\rm ch}$. \\ 

These are
usually considered in a large selection of spherical (doubly magic and
semi-magic) nuclei. Thus nuclei which are likely to carry sizeable
correlations should be excluded.  This holds, e.g., for nuclei with
$N=Z$ because of a rather large Wigner energy. Light nuclei should
also be scrutinized for correlation effects. The best possible
selection of experimental data is still a question of debate.

\subsubsection{Spin-orbit splitting}
\label{sec:const-ls}

The spin-orbit splitting is automatically provided in RMF but has to
be adjusted `by hand' in SHF. Experimental data for this adjustment
are taken from the spectra of doubly-magic nuclei, for levels which
are close to the Fermi energy and thus rather free from polarization
and correlation effects. Up to now, almost all fits have considered
spin-orbit splittings only in $^{16}$O and $^{208}$Pb. Other options
for which good data exist, such as $^{40}$Ca, $^{48}$Ca, $^{90}$Zr, or
$^{132}$Sn, should be exploited.

\subsubsection{Neutron radii}

Information on neutron radii $r_{\rm n}$ and/or neutron skin $r_{\rm
n}-r_{\rm p}$ are most useful to pin down the isovector properties of
the parameterization, particularly the symmetry energy $a_{\rm
sym}$. The problem here is that $r_{\rm n}$ is deduced from
model-dependent analysis of scattering experiments. To overcome this
obstacle, direct modeling of the scattering process by SHF should be
considered.

\subsubsection{Energy differences}
\label{sec:const-endif}

The odd-even staggering of binding energies in medium heavy and heavy
nuclei is used to adjust the pairing strength. It is recommended to
use the staggering from the five-point difference formula to minimize
perturbation from competing polarization (Jahn-Teller effect)
\cite{Ben00c}. 

Two-proton and two-neutron separation energies can supply information
on shell structure, particularly on the magnitude of shell gaps.  A
proper description is crucial for the predictive value of SHF in
application to the r-process (see Section~\ref{sec:nuclo}).
However, one has to be aware of possible correlation effects
on this quantity (see Section \ref{sec:collcorr}).

\subsubsection{Isotopic shifts}

The isotope shift in radii of Pb can be used a test of the strength
of the isovector spin-orbit force \cite{Rei95a}.

Much information is contained in other chains to, such as the
negligible difference in charge radii between $^{40}$Ca and $^{48}$Ca,
or the large shift in certain regions of, e.g., Sr or Hg isotopes.
However, these chains are likely to include sizeable correlation
effects.

\subsubsection{Deformations}

More advanced test cases can be found in properties of deformed
nuclei, their ground state deformations and rotational bands.
A possible selection may be:\\

\noindent
\hspace*{1em} deformation in heaviest nuclei ($^{264}$Hs),\\
\hspace*{1em} triaxial deformation ($^{138}$Sm, $^{188}$Os),\\
\hspace*{1em} octupole deformation ($^{222,224}$Th, $^{146}$Ba),\\
\hspace*{1em} super-deformed states ($^{192,194}$Hg, $^{194}$Pb,$^{238}$U),\\
\hspace*{1em} hyper-deformed states ($^{232}$Sm)         ,\\
\hspace*{1em} prolate-oblate competition ($^{184,186}$Pt, $^{184,186}$Hg),\\
\hspace*{1em} spherical-to-deformed transition in heavy Zr 
              isotopes $^{96,98,100}$Zr, and/or fission barriers in actinides.\\

Deformation properties are often sensitive to details of the shell
structure and constitute extremely critical tests. Nuclei with well
developed stable deformation are generally well described within a
mean-field model.  Transitional regime, although very interesting,
has to be investigated for possible influence of correlation effects.

\subsubsection{Excitation properties}

Excitation properties provide a rich pool of information. Giant
resonances in heavy nuclei are closely related to key properties of
nuclear matter such as incompressibility (isoscalar monopole),
symmetry energy (isovector dipole), or effective mass (isoscalar
quadrupole). The isovector dipole resonance in light nuclei is not yet
under control and presents a big puzzle (see Section
\ref{sec:giantres}).

Excitation with unnatural parity as, e g., the Gamow-Teller
resonances can serve to fix the not so well known properties of SHF in
the spin channel (see Section \ref{sec:gamow-teller}). There is still a long
road ahead in exploiting these aspects.

Low-energy quadrupole excitations carry combined information on shell
structure (near doubly magic nuclei) and on pairing.  However, the
modeling has yet to be improved in order to truly optimize the large
amplitude collective path (see Section \ref{sec:lampl}).

\subsection{Nuclear matter and neutron star models}
\label{sec:starconstr}

The constraints one nuclear matter properties are of two types. The
first concerns expected values of some empirical quantities at nuclear
saturation density n$_{\rm 0}$. These values and/or their lower and upper
limits, coming from merging data obtained from LDM or indirectly from
experiments as discussed in Section~\ref{sec:keyprop}, are summarized
in Table~\ref{tab:nmconstr}. The LDM values are an extrapolation from
finite nuclei data, similarly as the various SHF values. They serve
merely as an orientation to the extend that models which deviate too
much from these `commonly accepted values' are to be questioned.

\begin{table}[h] 
\caption{\label{tab:nmconstr} 
Summary of constraints from nuclear
matter at saturation density. 
The ``commonly accepted values'' are taken from the LDM as
published in \cite{Mye98a,Mye98b}.
The calculated values for the Skyrme
parameterizations used in the review are given for illustration. All
symbols are explained in the text.  }
\vskip 15pt
\begin{center}
\begin{tabular}{l|rl|rrrrrrrr}
\hline
  Constraint&\multicolumn{2}{c|}{LDM}   &     SkM$^*$     &  SkP     &  SLy6    & SkI3     &   SkI4    & BSk1 \\
\hline
$n_{\rm 0}$ [fm$^{-3}$]      & 0.16 & $\pm$ 0.01   &     0.160       &  0.163   &  0.159   & 0.158    &   0.160   &  0.157 \\
${\cal E}_{\rm 0}$ [MeV]     & -16.1 &$\pm$ 0.1 &    -15.77       & -15.95   &  -15.92  & -15.98   &   -15.94  &  -15.80 \\
$K_\infty$ [MeV]      & 230 & $\pm$ 20       &     217.5       &  201.9   &  230.8   &  259.0   &   249.1   &  232.1 \\
$a_{\rm sym}$($n_0$) [MeV] & 31 &$\pm$ 1.5           &     30.06       &  30.02   &  32.00   &  34.89   &   29.54   &  27.82 \\
$m_{\rm s}^*/m$              &    &     1.0       &     0.79        &  1.00    &  0.69    &   0.58   &    0.65   &  1.05 \\
\hline
\end{tabular}
\end{center}
\end{table}

The second type of constraint is given by the $\textit{density
dependence}$ of energy per particle ${\cal E}_{\rm 0}$, the asymmetry
coefficient $a_{\rm sym}$(n$_{\rm 0}$), the isoscalar effective nucleon mass
$m_{\rm s}^*/m$ and derived quantities like the pressure, incompressibility
and the velocity of sound in SNM, PNM and BEM. There is no firmly
based set of values of these quantities and their various
dependences. The best that can be done is to compare their predicted
values to a `realistic' \textit{ab initio} theoretical model. However, they
have a profound effect on the EoS used in neutron star models which
can indeed in turn can be tested against observational data. Although
the data are still lacking desired accuracy and are related to each
other (usually as a function of the star mass), they have already
proven to be powerful enough to eliminate some EoS (and thus Skyrme
parameterizations) from use in neutron star models \cite{rik03} although
they perform well in finite nuclei. Considerable progress in
increasing the quality of data on cold neutron stars is expected in
the near future when, for example, there is hope to have independent
values of neutron star radii rather than values determined as a
function of the gravitational mass. The recently proposed method of
its independent determination could be realized in a few years
\cite{lat05b}. We list in table~\ref{tab:nsconstr} the main
constraints based on neutron star models.
\begin{table} 
\caption{\label{tab:nsconstr} Summary of constraints from properties
of cold neutron stars. Fullfilment of some of these constraints for
the sample Skyrme parameterizations is indicated by `+' (fulfilled),
`(+)' (fulfilled but the constraint is weak), `-' (failed). `0' stays
for undecided as either the constraints is uncertain at present or the
property was not tested for the parameterization in question. Symbols
are explained in the text.}
\vskip 15 pt
\begin{center}
\begin{tabular}{l|r|cccccc}
\hline
  Constraint &     Expected value &               SkM$^*$ &    SkP   &  SLy6   & SkI3  &   SkI4  & BSk1 \\
\hline
$M_{\rm max}$ [M$_\odot$]     & 2.1$\pm$ 0.2                         0   &     -    &    +    &   +   &    +    &   -   \\
$R_{\rm max}$ [km]            & 8--15          &                     +   &     -    &    +    &   +   &    +    &   -   \\
$R_{1.4_\odot}$ [km]      & 9--16          &                     +   &     -    &    +    &   +   &    +    &   -   \\
$E_{\rm bin}$ [10$^{53}$ ergs] & 3.8 $\pm$ 1.2   &                    0   &     -    &    0    &   0   &    0    &   -   \\
$\rho^{\rm centr}_{\rm max}$ [10$^{15}$ gcm$^{-3}$]& 0.5--4.0&  (+)  &     -    &   (+)   &  (+)  &   (+)   &   -   \\
$M_{\rm 0}$ [M$_\odot$]         & 1.356 $<$ 1.375&                     +   &     -    &    -    &   -   &    -    &   -   \\    
$ M_{\rm URCA}$ [M$_\odot$]   & $>$ 1.5         &                     0   &     -    &    +    &   0   &    +    &   -   \\     \hline
\end{tabular}
\end{center}
\end{table}

Certain other constraints, such as those connected with polarized
nuclear matter and its ferromagnetic collapse, have not been
quantified as yet for the majority for SHF models and are yet to be
explored. We mention that very recently \cite{kla06} a similar set of
constraints has been suggested for testing effective interactions used
in EoS based on RMF models.

\subsection{Supernova simulations}
\label{sec:supernovaconstr}

The most powerful constraint on the SHF equation of state would be
achieving the explosion of the core-collapse supernova model. It is
important to reiterate that the density region at which this collapse
is expected to happen (1.5--2.0 $n_0$) is well within the expected
window of validity of the SHF model beyond the saturation
density. However these tests are rather complicated as there are
certain other factors in the supernova process that may affect the
occurrence of the explosion and their relation with the EoS are not
fully established. However, supernova physics offers a new ground for
testing the SHF models which is just starting to be explored
\cite{sto05}.

% this part --> in close cooperation
% \section{Conclusions}
% \label{sec:concl}
%\include{concl}
% this part --> in close cooperation
\section{Conclusions}
\label{sec:concl}

The application of the Skyrme Hartree-Fock model (SHF) to finite
nuclei, nuclear matter and in astrophysical contexts has been
critically surveyed, with occasional consideration links of its
relativistic cousin, the relativistic mean-field model, and to
ab-initio models. SHF provides an effective energy-functional which is
motivated by many-body theory and leaves about 6--12 free parameters
which are adjusted phenomenologically. The functional is universal in
the sense that it, in principle, applicable to all nuclei (except the
very light) irrespective to their shape or shell structure. The
current models usually calculate static and dynamical properties of
even-even nuclei exactly and approximate models for odd-A and odd-odd
nuclei in some fashion, although an exact approach is in principle
possible and is waiting to be implemented in SHF codes. Static
observables include nuclear ground state binding energies, profile and
shape of charge and neutron density distributions, nuclear radii and
fission barriers. The SHF models provide, in the small-amplitude
limit of TDHF, useful information on the excitation spectrum,
particularly on the various giant resonances which form important
features of the nuclear excitation spectrum. Moreover, the large
amplitude limit of TDHF covers a wide range of phenomena, low-lying
collective modes, fission, rotational bands, fusion and nuclear
collisions.

The overall global agreement of the SHF model with existing data is
overwhelmingly good. Local discrepancies in SHF predictions, however,
exist which is fully understandable because the effective functionals
has a simple analytical form and is preferably appropriate for
reproducing smooth trends. Moreover, the isospin and density
dependence of these parameters is not yet well known. It is often
observed that local deviation between SHF calculation and experiment
is not statistical but exhibits systematic trends which indicate
aspects for future improvement (see Section~\ref{sec:etrends}). As an
example, discrepancies between experimental and calculated
two-neutron-separation energies may originate in detailed, but locally
important, physics like shell effects, shape fluctuations and level
densities at the Fermi surface to which global observables have
limited sensitivity. Similarly, the whole range of dynamical nuclear
properties is reproduced well by the SHF model, again with exception
which constitute major challenges yet to be addressed (see
Section~\ref{sec:dynamic}). For example, no SHF parameterization can
simultaneously reproduce the average peak position of the GDR in heavy
and light nuclei.
 
Discrepancies between results of SHF calculation and experimental data
often originate from correlations beyond mean field (see
Section~\ref{sec:beyond}). There is a large potential for systematic
investigation of these effects that may lead to better understanding
of the limitations of the SHF and similar self-consistent mean-field
models. Another major improvement can be achieved by gradual
elimination of imperfections in fitting of the Skyrme energy
functional. It is well known that due to mutual correlation between
the variable parameters of the Skyrme functional, there is, in
principle, an infinite number of combinations of the parameters which
could be fitted to experimental data and produce results with an
equally satisfactory agreement with experiment. There exists already a
great manifold of such sets in the
literature and it is very difficult to make a well founded choice
between them, in particular in calculations of properties of finite
nuclei. This situation is not unique to SHF but a generic problem of
all types of nuclear effective energy functionals. It is caused by the
fact that an ab-initio derivation of the functionals is hindered by
the uncertainty in the true nature and form of the genuine
nucleon-nucleon interaction in the nuclear environment.

One way forward in confining the manifold of permissible
parameterizations is to systematically increase the amount and variety
of experimental data used in the fitting process. Attempts to use only
one selected type of observable in fitting, like nuclear masses, and
to ignore all the other experimental data, did not meet with real
success. The common practice adopted in fitting of
most parameterizations is to use ground state properties of magic or
semi-magic nuclei in the hope that these nuclei will be least affected
by correlations beyond the mean field and that many effects like shape
fluctuation, shell effects and pairing may be neglected. The
increasing amount of experimental information on nuclei far from
stability allows to include more stringent information, particularly
concerning isovector properties of the effective interactions. This
should improve steadily the extrapolative power of SHF which is at
present still rather unsatisfactory (i.e. the deviation between
predictions of different SHF parameterizations is increasing with
growing distance from the stability line which causes extrapolation of
SHF models to the drip-line regions lacking credibility).  Establishing
more reliable extrapolations is the major challenge for future
development because that is desperately needed for a variety of
astrophysical applications, like understanding the r-process. The same
holds for the construction of SHF based EoS for description of nuclear
matter under extreme conditions of high density and temperature that
exists in neutron star and supernovae. Similarly, in terrestrial
`extreme' conditions, reliable EoS are desirable for modeling of
heavy-ion collision processes and investigations of super-heavy
nuclei.

We believe it would be very beneficial to develop modern Skyrme
parameter sets that would satisfy simultaneously the extended set of
constraints summarized in Section~\ref{sec:summconstr}. Although this
will not create a {\em unique} set of the Skyrme parameters, it will
give the best chance to create a model that would be applicable on a
global scale.  It has been shown in Section~\ref{sec:infinite} that
properties of nuclear matter even at densities below 3 $n_{\rm 0}$ offer a
sensitive test of the Skyrme parameterizations. It is plausible that
extension of the SHF models to systems with high values of isospin,
like cold neutron stars, would provide a distant extrapolation point
with the closed-shell stable nuclei at the other extreme and the
neutron-drip-line nuclei between the two. This development may even
give us a better insight into the physical connection between Skyrme
parameters (or their selected combinations) and some nuclear
properties and thus give more control over the model. As long as a
quantitative performance of more fundamental models of nuclear
interactions, perhaps based on subnucleon degrees of freedom, is still
ahead in the future, up-to-date SHF models are likely to remain an
important tool in low-energy nuclear physics and astrophysics for 
some time to come.

% \section{Acknowledgements}
%\label{sec:ackno}
\section{Aknowledgements}
\label{sec:ackno}
We wish to thank N.~J.~Stone for reading of the manuscript and many
constructive suggestions. 
Helpful comments of D.~M.~Brink, D.~Dean, P.~Guichon, W.~R.~Hix,
P.~Kl\"upfel, K.-L.~Kratz, J.~C.~Miller and B.~Pfeiffer are gratefully
acknowledged.
This work has been supported by US DOE grant
DE-FG02-94ER40834, from DOE Scientific Discovery through Advanced
Computing Grant, and from the BMBF (project 06 ER 124).

\appendix
%\section{Parameters of the sample group of Skyrme parametrisations used in this review in nuclear matter calculations}
%\label{app1}
\newpage
\section{Appendix 1}
\label{app1}
\subsection{Parameters of the sample group of Skyrme parametrisations in nuclear matter calculations}
Parameters of the Skyrme parametrizations SkM$*$, SkP, SLy6, SkI3, SkI4 and BSk1 are listed in Table~\ref{tab:param}.    
\begin{table}[h] 
\caption{\label{tab:param} 
Parameters of the sample Skyrme interactions. For definitions see (\ref{eq:enfun},\ref{eq:bdef}).
}
\vskip 15pt
\begin{center}
\begin{tabular}{l|rrrrrrrrr}
\hline
Skyrme    &      $B_0$  &  $B^\prime_0$ & $B_1$ &  $B^\prime_1$ & $B_3$  &  $B^\prime_3$  & $\alpha$ \\ \hline
  SkM$^*$ &  -1983.75 & -780.28  & 2924.06  &  974.69  &   34.688   &  34.063   &   0.166667 \\
  SkP     &  -2198.78 & -1161.17 &  3507.93 &  1592.67 &    -0.000  &   44.642  &    0.166667 \\
  SLy6    &  -1859.63 & -1642.67 &  2563.69 &  3170.43 &    58.621  &  -26.016  &    0.166667 \\
  SkI3    &  -1322.16 &  -712.47 &  1519.91 &  1816.40 &    96.258  &  -63.957  &    0.25 \\
  SkI4    &  -1391.87 &  -839.84 &  1819.43 &  1995.55 &    69.927  &  -37.655  &    0.25 \\
  BSk1    &  -1372.84 & -1006.74 &  2520.88 &  2223.54 &    -6.277  &    0.000  &    0.333333 \\  
\hline
\hline
Skyrme   &   $t_0$  &   $t_1$  &  $t_2$  &  $t_3$  &  $x_0$  &   $x_1$  &   $x_2$  &   $x_3$  & $\alpha$ \\ \hline
SkM$^*$  &  -2645.00 &   410.00 &  -135.00 & 15595.00  &  0.090  &   0.000  &   0.000  &   0.000  & 0.166667 \\
 SkP     & -2931.70  &  320.62  & -337.41  & 18708.96  &  0.292  &   0.653  &  -0.537  &   0.181  & 0.166667 \\
  SLy6   & -2479.50  &   462.18 &  -448.61 & 13673.00  &  0.825  &  -0.465  &  -1.000  &   1.355  & 0.166667 \\
  SkI3   & -1762.88  &  561.61  & -227.09  & 8106.20   &  0.308  &  -1.172  &  -1.091  &  1.293   &  0.25 \\
  SkI4   &  -1855.83 &   473.83 &  1006.86 &  9703.61  &  0.405  &  -2.889  &  -1.325  &  1.145   &  0.25 \\
  BSk1   &  -1830.45 &   262.97 &  -296.45 & 13444.70  &  0.600  &  -0.500  &  -0.500  &   0.823  &  0.333333 \\
\hline
\end{tabular}
\end{center}
\end{table}

%\begin{thebibliography}{00}


% notes:

% \bibitem{label2}
% If there is a note, it should come last:
% \bibitem{label3} \note
% \end{subbibitems}

%\bibitem{}

%\end{thebibliography}

\bibliographystyle{amsunsrt}

\bibliography{pgr,js}
%\bibliography{nucl/biblio.bib,nucl/exp.bib,nucl/reviews.bib,nucl/reviewsexp.bib,nucl/books.bib}  % set a link to that site

\end{document}